# Projective Limits of State Spaces
# I. Classical Formalism

Suzanne Lanéry[1,2] and Thomas Thiemann[1]

[1] Institute for Quantum Gravity, Friedrich-Alexander University Erlangen-Nürnberg, Germany

[2] Mathematics and Theoretical Physics Laboratory, François-Rabelais University of Tours, France

## Abstract

In this series of papers, we investigate the projective framework initiated by Jerzy Kijowski [6] and Andrzej Okołów [14, 15], which describes the states of a quantum (field) theory as projective families of density matrices. The present first paper aims at clarifying the classical structures that underlies this formalism, namely projective limits of symplectic manifolds. In particular, this allows us to discuss accurately the issues hindering an easy implementation of the dynamics in this context, and to formulate a strategy for overcoming them.

## Contents

# 1 Introduction

An important step toward the quantization of a classical theory is the choice of a home for the kinematical quantum states: typically, we look for an Hilbert space supporting a representation of an algebra of selected kinematical observables. As long as we only deal with finitely many degrees of freedom, a comprehensive survey of the available options might still be within reach. But the extent and implications of this initial choice tend to get dramatically more involved in the case of field theory, where the huge algebra of kinematical observables can give rise to an intricate forest of representations. Unfortunately, it is hard to concisely formalize which requirements the elected representation should satisfy. In the worst case, we are left with the 'trial and error' method: pick some representation with attractive properties and check whether the next steps of the quantization program work well on it or not.

These next steps can fail for various reasons, one of them being that we committed ourselves to a space of kinematical states that, at a closer look, does not support the states we are really interested in. In particular, the space of physical quantum states, solving the dynamical constraints of the theory, should be rich enough. Refined Algebraic Quantization [4] is a way to look for physical states out of the representation we initially choose as the space of kinematical states; however, there are unfavorable situations, where we do not really know how to construct the additional input it requires.

Also, our space of states should contain the 'coherent states' needed to explore the semi-classical limit of the theory: we would like to associate to any point in the classical phase space a corresponding quantum state, suitably peaked around that point (see [8, 3] for a discussion of this problem in the case of Loop Quantum Gravity, together with possible ways to circumvent it).

These issues motivate the search for alternative ways of building the space of kinematical states. Here, we will focus on a formalism first introduced by Jerzy Kijowski in the late '70s [6] and further developed by Andrzej Okołów recently [14, 15]. The idea is to work in a setting that is more general than Hilbert spaces, and allows us to rely more heavily on the physical interpretation of the kinematical observables, namely how they are measured in practice. This tends to give state spaces that are bigger, but nevertheless technically easier to handle. In particular, we thus start with better chances to find the particular states we are looking for.

In the present work, we try to develop this formalism at a fairly general level, going beyond the extensive studies that have been carried out in special cases so far. To this intent, we will start by a detailed exposition of projective limits of symplectic manifolds, that build the natural classical counterpart of the quantum state spaces we want to discuss. An important observation is that such projective limits admit, at least locally, a preferred factorized description (prop. 2.10). Therefore, we will look more closely at those projective systems where the factorization holds globally: not only they are often more convenient, they also reflect the core properties of the structures we are considering, so they are well-suited to get a first hold of complex questions. This will be in particular comfortable when turning to the quantum formalism in [9], but we will always try to sketch some ideas on how to strengthen those of our results that make explicit use of such a global factorization.

Specific difficulties arise when trying to deal with constraints in this approach to (quantum) field theory. In section 3, we will take advantage of having at our disposal a classical precursor of the formalism to analyze this question without having to deal at the same time with the inherent subtleties of the quantum dynamics. We will outline a suitable strategy, with the aim of doing justice both to the deep physical meaning of the issues at hand and to their practical significance for computations. This strategy will be applied to two simple toy models in [10].

Unless otherwise stated, all symplectic manifolds will be smooth manifolds with smooth symplectic structures, and all maps between them will be smooth. Where infinite dimensional manifolds are considered, these are Banach-modeled smooth manifolds, and symplectic structures on them are always strong symplectic structures [2, chap. VII].

# 2 Projective limits of classical phase spaces

The aim of this section is to describe the classical structures that, while underlying the constructions considered in previous works [6, 14, 15], have not been explicitly analyzed so far. The discussion of the physical interpretation will follow closely the one that has been given in these references.

The idea of the projective framework is to assemble a complicated classical theory (typically a field theory) from a collection of easier, smaller, truncated classical theories, by appropriately sewing them together. The motivation for this is twofold.

From the physical point of view, even when considering a theory with an infinite number of degrees of freedom, any given realistic experiment will involve only a finite number of observables, since measuring an infinite number of observables would require infinite time as well as infinite memory space (in fact, this means that any experiment can only measure a finite number of *boolean* observables, but we will not be that radical here, and will satisfy us with small truncated theories that are described by finite dimensional phase spaces). We will therefore think of the small partial theories as spanned by a finite number of elementary degrees of freedom. By "elementary", we mean those that can be measured in one experimental step, hence the justification for the choice of a collection of truncations should ultimately come from a careful analysis of what concrete experiments actually measure.

From a technical point of view, the smaller and easier theories are meant to be a convenient arena to develop systematic ways of calculating physical predictions. Indeed, a theoretical model will then be optimally useful if it comes with finite algorithms prescribing how to compute, at a given precision, the outcome of any arbitrary experiment.

Note however that the intuitive understanding just sketched has some weak points. One of them is that, even if we are considering only finitely many observables, it might occur that the Poisson-algebra they are generating cannot live on a finite dimensional symplectic manifold. Another problem is related to the formulation of deterministic predictions while considering only finitely many degrees of freedom out of a field theory. Our viewpoint here is that these problems should not be relevant for the *kinematical* observables (these are supposed to build an easy algebra, and the question of writing down predictions does not belong to the kinematical level). Therefore, we postpone this discussion to section 3, where we will refine the present framework to take into account the dynamics.

## 2.1 Projective systems of classical phase spaces

Having a collection of partial theories is not enough, we need to say how to connect them together in a consistent way (ie. we do not want our physical predictions to depend on the particular partial theory in which we computed them). To look at this question, we consider two partial theories $\mathcal{M}$ and $\mathcal{N}$, where $\mathcal{M}$ is a more detailed description of the physical system at hand, in the sense that all degrees of freedom that are retained by $\mathcal{N}$ are also retained by $\mathcal{M}$. The link between them has then two dual aspects. On the one side, we want to associate, with any state in $\mathcal{M}$, a state in $\mathcal{N}$, by forgetting the details we presently do not need. On the other side, we want to identify the observables that can be defined on $\mathcal{N}$ with a subalgebra of the ones that can be defined on $\mathcal{M}$.

Given a specific experiment, any partial theory big enough to describe that experiment (ie hosting at least all the observables involved in it) should lead to the same predictions. In other words, the two identifications mentioned above (downward identification of the states and upward identification of the observables) should intertwine the evaluation of an observable on a state.

These considerations lead to the following formulation of how some degrees of freedom, spanning a symplectic manifold $\mathcal{N}$, can be seen as being extracted out of a bigger symplectic manifold $\mathcal{M}$: what we need is a projection $\pi : \mathcal{M} \to \mathcal{N}$, and we will mount observables on $\mathcal{N}$ to observables on $\mathcal{M}$ by taking their pullback. We impose a compatibility condition between the projection $\pi$ and the symplectic structures of $\mathcal{M}$ and $\mathcal{N}$ to ensure that the Poisson bracket computed between two observables in $\mathcal{N}$ is identified with the one computed between the corresponding observables mounted in $\mathcal{M}$.

**Definition 2.1** A smooth, surjective map $\pi : \mathcal{M} \to \mathcal{N}$ between two smooth (possibly infinite dimensional) symplectic manifolds $\mathcal{M}, \Omega_\mathcal{M}$ and $\mathcal{N}, \Omega_\mathcal{N}$ is said to be compatible with the symplectic structures iff:

$$\forall x \in \mathcal{M}, \forall \upsilon \in T'_{\pi(x)}(\mathcal{N}), \quad \underline{\upsilon} = T_x \pi \left( \underline{\pi^* \upsilon} \right) \tag{2.1.1}$$

where $T'_{\pi(x)}(\mathcal{N})$ is the topological dual of $T_{\pi(x)}(\mathcal{N})$, $T_x \pi$ is the differential of $\pi$ at $x$, and $\underline{\upsilon}$ (resp. $\underline{\pi^* \upsilon}$) is the unique vector in $T_{\pi(x)}(\mathcal{N})$ (resp. $T_x(\mathcal{M})$) such that $\upsilon = \Omega_{\mathcal{N},\pi(x)}(\underline{\upsilon}, \cdot)$ (resp. $\pi^* \upsilon = \Omega_{\mathcal{M},x}(\underline{\pi^* \upsilon}, \cdot)$).

**Proposition 2.2** If $\pi : \mathcal{M} \to \mathcal{N}$ satisfies def. 2.1 and $f, g : \mathcal{N} \to \mathbb{R}$ are smooth maps on $\mathcal{N}$, then $\{f, g\}_\mathcal{N} \circ \pi = \{f \circ \pi, g \circ \pi\}_\mathcal{M}$ where $\{\cdot, \cdot\}_\mathcal{N}$ (resp. $\{\cdot, \cdot\}_\mathcal{M}$) denotes the Poisson brackets on $\mathcal{N}$ (resp. $\mathcal{M}$).

**Proof** Eq. (2.1.1) is equivalent to:

$$\forall x \in \mathcal{M}, \forall \mu, \upsilon \in T^*_{\pi(x)}(\mathcal{N}), \mu \left( \underline{\upsilon} \right) = \pi^* \mu \left( \underline{\pi^* \upsilon} \right).$$

Using the definition of the Poisson brackets, we therefore have:

$$\forall x \in \mathcal{M}, \{f, g\}_\mathcal{N} \circ \pi(x) = dg_{\pi(x)} \left( \underline{df_{\pi(x)}} \right) = d(g \circ \pi)_x \left( \underline{d(f \circ \pi)_x} \right) = \{f \circ \pi, g \circ \pi\}_\mathcal{M}(x).$$

$\square$

Next, the collection of partial theories, together with the projections between them, can be arranged into a structure of projective limit. Such a construction has been considered for example in

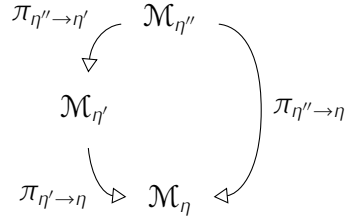

Figure 2.1 – Three-spaces consistency for projective systems of phase spaces

[18].

That the label set $\mathcal{L}$ indexing the partial theories should be directed is manifest if we go back to the interpretation of these small theories as the arenas to describe specific experiments: if we want to describe an elaborate experimental protocol, combining two sub-experiments, that can be described respectively in $\mathcal{M}_\eta$ and $\mathcal{M}_{\eta'}$, we need a symplectic manifold $\mathcal{M}_{\eta''}$, containing the degrees of freedom in $\mathcal{M}_\eta$ as well as the ones in $\mathcal{M}_{\eta'}$, in order to model the full experiment. And the three-spaces consistency condition (fig. 2.1) ensures that the connection between a bigger partial theory $\mathcal{M}_{\eta''}$ and a smaller one $\mathcal{M}_\eta$ is unambiguous, namely that it coincides with the identification we get if we perform the truncation in two successive steps, going first from $\mathcal{M}_{\eta''}$ to an intermediary $\mathcal{M}_{\eta'}$ and then from $\mathcal{M}_{\eta'}$ to $\mathcal{M}_\eta$.

With this structure for the state space, the observables naturally build an inductive limit, which is consistent with the discussion above regarding the mounting of observables and indeed corresponds to the standard construction when looking at functions on a projective limit.

**Definition 2.3** A projective system of phase spaces is a triple $\left(\mathcal{L}, \left(\mathcal{M}_\eta\right)_{\eta \in \mathcal{L}}, \left(\pi_{\eta' \to \eta}\right)_{\eta \preccurlyeq \eta'}\right)$ where:

1. $\mathcal{L}$ is a preordered, directed set (we denote the pre-order, ie. a reflexive and transitive binary relation, by $\preccurlyeq$, its inverse by $\succcurlyeq$);

2. $\left(\mathcal{M}_\eta\right)_{\eta \in \mathcal{L}}$ is a family of symplectic manifolds indexed by $\mathcal{L}$;

3. $\left(\pi_{\eta' \to \eta}\right)_{\eta \preccurlyeq \eta'}$ is a family of surjective maps $\pi_{\eta' \to \eta} : \mathcal{M}_{\eta'} \to \mathcal{M}_\eta$ indexed by $\{\eta, \eta' \in \mathcal{L} \mid \eta \preccurlyeq \eta'\}$ such that $\pi_{\eta' \to \eta}$ is compatible with the symplectic structures, $\pi_{\eta \to \eta} = \mathrm{id}_{\mathcal{M}_\eta}$ and $\forall \eta, \eta', \eta'' \in \mathcal{L}, \eta \preccurlyeq \eta' \preccurlyeq \eta'' \Rightarrow \pi_{\eta'' \to \eta} = \pi_{\eta' \to \eta} \circ \pi_{\eta'' \to \eta'}$.

Whenever possible, we will use the shortened notation $(\mathcal{L}, \mathcal{M}, \pi)^\downarrow$ instead of $\left(\mathcal{L}, \left(\mathcal{M}_\eta\right)_{\eta \in \mathcal{L}}, \left(\pi_{\eta' \to \eta}\right)_{\eta \preccurlyeq \eta'}\right)$.

The projective limit of $(\mathcal{L}, \mathcal{M}, \pi)^\downarrow$, denoted by $\mathcal{S}^\downarrow_{(\mathcal{L},\mathcal{M},\pi)}$, is the space:

$$\mathcal{S}^\downarrow_{(\mathcal{L},\mathcal{M},\pi)} := \left\{ (x_\eta)_{\eta \in \mathcal{L}} \in \prod_{\eta \in \mathcal{L}} \mathcal{M}_\eta \ \middle|\ \forall \eta \preccurlyeq \eta',\ \pi_{\eta' \to \eta}(x_{\eta'}) = x_\eta \right\}.$$

On $\mathcal{S}^\downarrow_{(\mathcal{L},\mathcal{M},\pi)}$ we put the initial topology with respect to the family of projections $\left(\pi_\eta\right)_{\eta \in \mathcal{L}}$ where:

$$\pi_\eta : \mathcal{S}^\downarrow_{(\mathcal{L},\mathcal{M},\pi)} \to \mathcal{M}_\eta$$
$$(x_{\eta'})_{\eta' \in \mathcal{L}} \mapsto x_\eta =: \left[(x_{\eta'})_{\eta' \in \mathcal{L}}\right]_\eta.$$

**Definition 2.4** An observable over a projective limit of phase spaces $\mathcal{S}^{\downarrow}_{(\mathcal{L},\mathcal{M},\pi)}$ is an equivalence class in $\bigcup_{\eta \in \mathcal{L}} C^{\infty}(\mathcal{M}_\eta, \mathbb{R})$ for the equivalence relation defined by:

$$\forall \eta, \eta' \in \mathcal{L}, \forall f_\eta \in C^{\infty}(\mathcal{M}_\eta, \mathbb{R}), \forall f_{\eta'} \in C^{\infty}(\mathcal{M}_{\eta'}, \mathbb{R}),$$
$$f_\eta \sim f_{\eta'} \Leftrightarrow \left(\exists \eta'' \in \mathcal{L} \mid \eta \preccurlyeq \eta'', \eta' \preccurlyeq \eta'' \ \&\ f_\eta \circ \pi_{\eta'' \to \eta} = f_{\eta'} \circ \pi_{\eta'' \to \eta'}\right) \quad (2.4.1)$$

The space of observables over $\mathcal{S}^{\downarrow}_{(\mathcal{L},\mathcal{M},\pi)}$ will be denoted by $\mathcal{O}^{\downarrow}_{(\mathcal{L},\mathcal{M},\pi)}$. The definition of the equivalence relation ensures that the evaluation $f(x) = f_\eta(x_\eta)$ of an element of $f = [f_\eta]_\sim$ of $\mathcal{O}^{\downarrow}_{(\mathcal{L},\mathcal{M},\pi)}$ on a point $x = (x_\eta)_{\eta \in \mathcal{L}}$ in $\mathcal{S}^{\downarrow}_{(\mathcal{L},\mathcal{M},\pi)}$ is well-defined. From prop. 2.2 the Poisson bracket of two elements of $\mathcal{O}^{\downarrow}_{(\mathcal{L},\mathcal{M},\pi)}$ is well-defined as an element of $\mathcal{O}^{\downarrow}_{(\mathcal{L},\mathcal{M},\pi)}$ ($\forall \eta' \succcurlyeq \eta$, $f_\eta \circ \pi_{\eta' \to \eta} \in [f_\eta]_\sim$, hence, $\mathcal{L}$ being directed, we can find a common label to compute the Poisson bracket).

## 2.2 Maps between classical state spaces

A question that occurs frequently when working with the structure introduced above, is to ask what happens if we restrict ourselves to a directed subset $\mathcal{L}'$ of the label set $\mathcal{L}$. It is immediate that a state $(x_\eta)_{\eta \in \mathcal{L}}$ in the projective structure based on $\mathcal{L}$ defines a state $(x_\eta)_{\eta \in \mathcal{L}'}$ in the one based on $\mathcal{L}'$, simply by throwing away all the $x_\eta$ for $\eta \in \mathcal{L} \setminus \mathcal{L}'$. But this map from $\mathcal{S}^{\downarrow}_{(\mathcal{L},\mathcal{M},\pi)}$ into $\mathcal{S}^{\downarrow}_{(\mathcal{L}',\mathcal{M},\pi)}$ will in general neither be injective nor surjective.

The injectivity might fail because the structure based on $\mathcal{L}'$ retains less observables than the structure based on $\mathcal{L}$, and states that can, thanks to these additional observables, be distinguished in the latter may be indistinguishable in the former. That also the surjectivity might fails is more subtle: it can occur if $\mathcal{L}$ has a label $\eta$ that is above an infinite number of labels in $\mathcal{L}'$. Then, given a state $(x_\eta)_{\eta \in \mathcal{L}'}$ in $\mathcal{S}^{\downarrow}_{(\mathcal{L}',\mathcal{M},\pi)}$, it may indeed not be possible to find an $x_\eta$ that will project correctly on all the $x_{\eta'}$ for $\eta' \in \mathcal{L}$ with $\eta' \preccurlyeq \eta$.

In the particular case of $\mathcal{L}'$ being cofinal in $\mathcal{L}$, we can however completely identify the two projective structure, since we can reconstruct any thrown away $x_\eta$ for $\eta \in \mathcal{L} \setminus \mathcal{L}'$ by projecting down from some $\eta' \in \mathcal{L}'$ above $\eta$.

**Proposition 2.5** Let $(\mathcal{L}, \mathcal{M}, \pi)^{\downarrow}$ be a projective system of phase spaces and let $\mathcal{L}'$ be a directed subset of $\mathcal{L}$. We define the map:

$$\sigma : \mathcal{S}^{\downarrow}_{(\mathcal{L},\mathcal{M},\pi)} \to \mathcal{S}^{\downarrow}_{(\mathcal{L}',\mathcal{M},\pi)}$$
$$(x_\eta)_{\eta \in \mathcal{L}} \mapsto (x_\eta)_{\eta \in \mathcal{L}'}.$$

Then, we have a map $\alpha : \mathcal{O}^{\downarrow}_{(\mathcal{L}',\mathcal{M},\pi)} \to \mathcal{O}^{\downarrow}_{(\mathcal{L},\mathcal{M},\pi)}$ such that:

$$\forall x \in \mathcal{S}^{\downarrow}_{(\mathcal{L},\mathcal{M},\pi)}, \forall f \in \mathcal{O}^{\downarrow}_{(\mathcal{L}',\mathcal{M},\pi)}, \ \alpha(f)(x) = f(\sigma(x)), \quad (2.5.1)$$

and:

$$\forall f, g \in \mathcal{O}^{\downarrow}_{(\mathcal{L}',\mathcal{M},\pi)}, \ \{\alpha(f), \alpha(g)\} = \alpha\left(\{f, g\}\right). \tag{2.5.2}$$

If $\mathcal{L}'$ is cofinal in $\mathcal{L}$, we have in addition that $\sigma$ and $\alpha$ are bijective maps.

**Proof** It's an immediate check that $\sigma$ is indeed valued in $\mathcal{S}^{\downarrow}_{(\mathcal{L}',\mathcal{M},\pi)}$.

Then, for $f = [f_\eta]_{\sim,\mathcal{L}'} \in \mathcal{O}^{\downarrow}_{(\mathcal{L}',\mathcal{M},\pi)}$, we define:

$$\alpha(f) = [f_\eta]_{\sim,\mathcal{L}} \text{ for } f_\eta \text{ a representative of } f.$$

We have:

$$\forall \eta, \eta' \in \mathcal{L}', \forall f_\eta \in C^\infty(\mathcal{M}_\eta, \mathbb{R}), \forall f_{\eta'} \in C^\infty(\mathcal{M}_{\eta'}, \mathbb{R}), \ f_\eta \sim_{\mathcal{L}'} f_{\eta'} \ \Rightarrow \ f_\eta \sim_{\mathcal{L}} f_{\eta'},$$

hence $\alpha$ is well-defined as a map $\mathcal{O}^{\downarrow}_{(\mathcal{L}',\mathcal{M},\pi)} \to \mathcal{O}^{\downarrow}_{(\mathcal{L},\mathcal{M},\pi)}$.

Eq. (2.5.1) and eq. (2.5.2) hold because we can choose any representative we want to carry out the evaluation or to compute the Poisson brackets.

We now suppose that $\mathcal{L}$ is cofinal. Then, we can define:

$$\widetilde{\sigma}: \mathcal{S}^{\downarrow}_{(\mathcal{L}',\mathcal{M},\pi)} \to \mathcal{S}^{\downarrow}_{(\mathcal{L},\mathcal{M},\pi)}$$
$$(x_\eta)_{\eta \in \mathcal{L}'} \mapsto (\widetilde{x}_\eta)_{\eta \in \mathcal{L}},$$

where for $\eta \in \mathcal{L}$, $\widetilde{x}_\eta = \pi_{\eta' \to \eta} x_{\eta'}$, with $\eta' \in \mathcal{L}'$ and $\eta' \succcurlyeq \eta$. If $\eta''$ is an other element of $\mathcal{L}'$ such that $\eta'' \succcurlyeq \eta$, there exists $\eta''' \in \mathcal{L}' / \eta' \preccurlyeq \eta''' \ \& \ \eta'' \preccurlyeq \eta'''$ ($\mathcal{L}'$ is directed by hypothesis), hence:

$$\pi_{\eta' \to \eta} x_{\eta'} = \pi_{\eta' \to \eta} \pi_{\eta''' \to \eta'} x_{\eta'''} = \pi_{\eta''' \to \eta} x_{\eta'''} = \pi_{\eta'' \to \eta} \pi_{\eta''' \to \eta''} x_{\eta'''} = \pi_{\eta'' \to \eta} x_{\eta''}.$$

If $\eta \in \mathcal{L}'$, we can choose $\eta = \eta'$, so that $\widetilde{x}_\eta = x_\eta$, therefore $\sigma \circ \widetilde{\sigma} = \mathrm{id}_{\mathcal{S}^{\downarrow}_{(\mathcal{L}',\mathcal{M},\pi)}}$. On the other hand, if there exists an element $(\overline{x}_\eta)_{\eta \in \mathcal{L}} \in \mathcal{S}^{\downarrow}_{(\mathcal{L},\mathcal{M},\pi)}$, such that $\forall \eta' \in \mathcal{L}', \ x_{\eta'} = \overline{x}_{\eta'}$, then $\widetilde{x}_\eta = \pi_{\eta' \to \eta} \overline{x}_{\eta'} = \overline{x}_\eta$, therefore $\widetilde{\sigma} \circ \sigma = \mathrm{id}_{\mathcal{S}^{\downarrow}_{(\mathcal{L},\mathcal{M},\pi)}}$.

Then, for $f = [f_\eta]_{\sim,\mathcal{L}} \in \mathcal{O}^{\downarrow}_{(\mathcal{L},\mathcal{M},\pi)}$, we define:

$$\widetilde{\alpha}(f) = [f_\eta \circ \pi_{\eta' \to \eta}]_{\sim,\mathcal{L}'},$$

for $f_\eta$ a representative of $f$ and $\eta' \in \mathcal{L}'$ such that $\eta' \succcurlyeq \eta$. If $\eta''$ is an other element of $\mathcal{L}'$ such that $\eta'' \succcurlyeq \eta$, there exists $\eta''' \in \mathcal{L}' / \eta' \preccurlyeq \eta''' \ \& \ \eta'' \preccurlyeq \eta'''$ ($\mathcal{L}'$ is directed by hypothesis), hence:

$$(f_\eta \circ \pi_{\eta' \to \eta}) \circ \pi_{\eta''' \to \eta'} = f_\eta \circ \pi_{\eta''' \to \eta} = (f_\eta \circ \pi_{\eta'' \to \eta}) \circ \pi_{\eta''' \to \eta''},$$

so that $f_\eta \circ \pi_{\eta' \to \eta} \sim_{\mathcal{L}'} f_\eta \circ \pi_{\eta'' \to \eta}$.

If $f_\kappa$ is an other representative of $f$, there exists $\mu \in \mathcal{L} / \mu \succcurlyeq \eta \ \& \ \mu \succcurlyeq \kappa$ such that $f_\eta \circ \pi_{\mu \to \eta} = f_\kappa \circ \pi_{\mu \to \kappa}$. Since $\mathcal{L}'$ is cofinal in $\mathcal{L}$, we can choose $\mu' \in \mathcal{L}'$ such that $\mu' \succcurlyeq \mu$, and we have:

$$f_\eta \circ \pi_{\mu' \to \eta} = f_\eta \circ \pi_{\mu \to \eta} \circ \pi_{\mu' \to \mu} = f_\kappa \circ \pi_{\mu \to \kappa} \circ \pi_{\mu' \to \mu} = f_\kappa \circ \pi_{\mu' \to \kappa},$$

hence $\widetilde{\alpha}$ is well-defined as a map $\mathcal{O}^{\downarrow}_{(\mathcal{L},\mathcal{M},\pi)} \to \mathcal{O}^{\downarrow}_{(\mathcal{L}',\mathcal{M},\pi)}$.

If $f_\eta$ is a representative of $f$ with $\eta \in \mathcal{L}'$, we can choose $\eta' = \eta$, so that $\widetilde{\alpha}(f) = [f_\eta]_{\sim,\mathcal{L}'}$, therefore $\widetilde{\alpha} \circ \alpha = \mathrm{id}_{\mathcal{O}^{\downarrow}_{(\mathcal{L}',\mathcal{M},\pi)}}$. On the other hand, we have for all $\eta \in \mathcal{L}$ and all $\eta' \in \mathcal{L}$ with $\eta' \succcurlyeq \eta$, $[f_\eta \circ \pi_{\eta' \to \eta}]_{\sim,\mathcal{L}} = [f_\eta]_{\sim,\mathcal{L}}$, therefore $\alpha \circ \widetilde{\alpha} = \mathrm{id}_{\mathcal{O}^\downarrow_{(\mathcal{L},\mathcal{M},\pi)}}$. □

We can now rewrite in terms of the concepts we have introduced the program that has been followed in [6, 14, 15]. When considering a field theory constructed on an infinite dimensional manifold $\mathcal{M}_\infty$, we will first, relying on our understanding of how physical effects are measured in practice, undertake to identify what the elementary observables should be, and try to construct a corresponding collection of interconnected partial theories, where each partial theory $\mathcal{M}_\eta$ will be associated to a finite subset of elementary observables (those that can be defined on $\mathcal{M}_\eta$, ie. that only depend on the degrees of freedom retained by $\mathcal{M}_\eta$). Since we have naturally a projection from $\mathcal{M}_\infty$ into each partial theory $\mathcal{M}_\eta$, we also immediately have a map from $\mathcal{M}_\infty$ into $\mathcal{S}^\downarrow_{(\mathcal{L},\mathcal{M},\pi)}$.

If the set of all elementary observables separate the points of $\mathcal{M}_\infty$, then this map will be injective. Moreover, the projective limit of phase spaces will provide an extension of $\mathcal{M}_\infty$ in the sense that $\mathcal{M}_\infty$ can be identified with a dense subspace of $\mathcal{S}^\downarrow_{(\mathcal{L},\mathcal{M},\pi)}$.

Note that choosing a collection of elementary observables is not the same as choosing preferred coordinates on $\mathcal{M}_\infty$, for the construction here does not require the elementary observables to be *independent*, they can form an overdetermined system. This can make physically a crucial difference: to illustrate this point, one can think at the set of elementary observables as analogous to the set of all the linear forms on a vector space, while preferred coordinates would correspond to the choice of a basis (compare the two examples given in [10] for examples implementing a projective structure along these lines; the model in [10, section 2] relies on a choice of basis, while the one in [10, section 3] does not). The set of all linear forms encodes nothing less but nothing more than the linear structure of the vector space, and this structure might indeed have a deep physical relevance, while we probably want to avoid relying on a preferred basis, in order not to break the invariance under isomorphisms.

**Definition 2.6** We say that a (possibly infinite dimensional) symplectic manifold $\mathcal{M}_\infty$ is rendered by a projective system of phase spaces $(\mathcal{L},\mathcal{M},\pi)^\downarrow$ if for all $\eta \in \mathcal{L}$ there exists an application $\pi_{\infty \to \eta} : \mathcal{M}_\infty \to \mathcal{M}_\eta$ such that:

1. $\forall \eta \in \mathcal{L}$, $\pi_{\infty \to \eta}$ is surjective and compatible with the symplectic structures;
2. $\forall \eta \preccurlyeq \eta' \in \mathcal{L}$, $\pi_{\infty \to \eta} = \pi_{\eta' \to \eta} \circ \pi_{\infty \to \eta'}$.

Hence, we have a projective system of phase spaces $(\mathcal{L} \sqcup \{\infty\}, \mathcal{M}, \pi)^\downarrow$, where we extend the preorder of $\mathcal{L}$ to $\mathcal{L} \sqcup \{\infty\}$ by requiring $\forall \eta \in \mathcal{L}, \infty \succ \eta$. From prop. 2.5, we have maps $\sigma_\searrow : \mathcal{S}^\downarrow_{(\mathcal{L} \sqcup \{\infty\}, \mathcal{M}, \pi)} \to \mathcal{S}^\downarrow_{(\mathcal{L},\mathcal{M},\pi)}$ and $\sigma_\infty^{-1} : \mathcal{S}^\downarrow_{(\{\infty\},\mathcal{M},\pi)} \to \mathcal{S}^\downarrow_{(\mathcal{L} \sqcup \{\infty\},\mathcal{M},\pi)}$ (since $\{\infty\}$ is cofinal in $\mathcal{L} \sqcup \{\infty\}$), so by identifying $\mathcal{S}^\downarrow_{(\{\infty\},\mathcal{M},\pi)}$ with $\mathcal{M}_\infty$, we define:

$$\sigma_\downarrow := \sigma_\searrow \circ \sigma_\infty^{-1} : \mathcal{M}_\infty \to \mathcal{S}^\downarrow_{(\mathcal{L},\mathcal{M},\pi)}.$$

Similarly, we have $\alpha_\searrow : \mathcal{O}^\downarrow_{(\mathcal{L},\mathcal{M},\pi)} \to \mathcal{O}^\downarrow_{(\mathcal{L} \sqcup \{\infty\},\mathcal{M},\pi)}$ and $\alpha_\infty^{-1} : \mathcal{O}^\downarrow_{(\mathcal{L} \sqcup \{\infty\},\mathcal{M},\pi)} \to \mathcal{O}^\downarrow_{(\{\infty\},\mathcal{M},\pi)}$, so by identifying $\mathcal{O}^\downarrow_{(\{\infty\},\mathcal{M},\pi)}$ with $C^\infty(\mathcal{M}_\infty, \mathbb{R})$, we define:

$$\alpha_\uparrow := \alpha_\infty^{-1} \circ \alpha_\searrow : \mathcal{O}^\downarrow_{(\mathcal{L},\mathcal{M},\pi)} \to C^\infty(\mathcal{M}_\infty, \mathbb{R}).$$

**Proposition 2.7** With the notations of def. 2.6, $\sigma_\downarrow \langle \mathcal{M}_\infty \rangle$ is dense in $\mathcal{S}^\downarrow_{(\mathcal{L},\mathcal{M},\pi)}$.

**Proof** Let $(x_\eta)_{\eta \in \mathcal{L}} \in \mathcal{S}^\downarrow_{(\mathcal{L},\mathcal{M},\pi)}$. For $\eta \in \mathcal{L}$, we choose $y^\eta \in \mathcal{M}_\infty$ such that $\pi_{\infty \to \eta}(y^\eta) = x_\eta$ (this is possible, since $\pi_{\infty \to \eta}$ is surjective). We have:

$$\forall \eta \in \mathcal{L}, \forall \eta' \succcurlyeq \eta, \left[\sigma_\downarrow\left(y^{\eta'}\right)\right]_\eta = \pi_{\infty \to \eta}\left(y^{\eta'}\right) = \pi_{\eta' \to \eta} \circ \pi_{\infty \to \eta'}\left(y^{\eta'}\right) = \pi_{\eta' \to \eta}\left(x_{\eta'}\right) = x_\eta.$$

Hence, the net $\left(\sigma_\downarrow\left(y^{\eta'}\right)\right)_{\eta' \in \mathcal{L}}$ converges in $\mathcal{S}^\downarrow_{(\mathcal{L},\mathcal{M},\pi)}$ to $(x_\eta)_{\eta \in \mathcal{L}}$, therefore $(x_\eta)_{\eta \in \mathcal{L}} \in \overline{\mathrm{Im}\, \sigma_\downarrow}$. □

We close this subsection by mentioning the construction of a different kind of maps between projective systems of phase spaces, that will be of interest when dealing with concrete examples. Indeed, we will often encounter the situation of having a projective system that has been originally constructed over a very large and complicated label set (in particular this can be a side-effect of the way we will handle constraints, as exhibited in section 3), but whose structure happens to be considerably simpler, because we can group the labels into classes partitioning $\mathcal{L}$, in such a way that the projective consistency conditions force a symplectomorphic identification between the manifolds $\mathcal{M}_\eta$ for all $\eta$ belonging to the same class. Then, we probably want to define a label set $\mathcal{L}^\bullet$ by quotienting $\mathcal{L}$ according to those classes, and to identify the original projective system on $\mathcal{L}$ with an easier one built on $\mathcal{L}^\bullet$.

For example, suppose that the elements of $\mathcal{L}$ are pairs $(\varepsilon, \theta)$, ordered in the product order (aka. $(\varepsilon, \theta) \preccurlyeq (\varepsilon', \theta') \Leftrightarrow \varepsilon \preccurlyeq \varepsilon'$ & $\theta \preccurlyeq \theta'$). Now, if it turns out that $\mathcal{M}_{(\varepsilon,\theta)}$ only depends on $\varepsilon$ and $\pi_{(\varepsilon',\theta') \to (\varepsilon,\theta)}$ only on $\varepsilon$ and $\varepsilon'$, then the projective condition on the states will actually impose $x_{(\varepsilon,\theta_1)} = x_{(\varepsilon,\theta_2)}$. Thus this projective limit is in reality just a projective limit on the set of all $\varepsilon$.

This is a tool that we will use repeatedly in [10] (and also when proceeding to applications in quantum gravity).

**Proposition 2.8** Let $\mathcal{L}$ and $\mathcal{L}^\bullet$ be directed preordered sets and assume that we are given:

1. a surjective map $\ell : \mathcal{L} \to \mathcal{L}^\bullet$ such that $\forall \eta \preccurlyeq \eta' \in \mathcal{L}$, $\ell(\eta) \preccurlyeq \ell(\eta')$;

2. a projective system of phase spaces $(\mathcal{L}^\bullet, \mathcal{M}^\bullet, \pi^\bullet)^\downarrow$ on $\mathcal{L}^\bullet$;

3. and for all $\eta \in \mathcal{L}$, a symplectic manifold $\mathcal{M}_\eta$ together with a symplectomorphism $\mu_\eta : \mathcal{M}_\eta \to \mathcal{M}^\bullet_{\ell(\eta)}$.

Then, defining for all $\eta \preccurlyeq \eta' \in \mathcal{L}$ the projection:

$$\pi_{\eta' \to \eta} := \mu_\eta^{-1} \circ \pi^\bullet_{\ell(\eta') \to \ell(\eta)} \circ \mu_{\eta'}, \qquad (2.8.1)$$

$(\mathcal{L}, \mathcal{M}, \pi)^\downarrow$ is a projective system of phase spaces and the map:

$$\begin{aligned} \kappa : \mathcal{S}^\downarrow_{(\mathcal{L}^\bullet,\mathcal{M}^\bullet,\pi^\bullet)} &\to \mathcal{S}^\downarrow_{(\mathcal{L},\mathcal{M},\pi)} \\ (x^\bullet_{\eta^\bullet})_{\eta^\bullet \in \mathcal{L}^\bullet} &\mapsto \left(\mu_\eta^{-1}\left(x^\bullet_{\ell(\eta)}\right)\right)_{\eta \in \mathcal{L}} \end{aligned}, \qquad (2.8.2)$$

is bijective. Moreover, there exists a bijective map $\lambda : \mathcal{O}^\downarrow_{(\mathcal{L},\mathcal{M},\pi)} \to \mathcal{O}^\downarrow_{(\mathcal{L}^\bullet,\mathcal{M}^\bullet,\pi^\bullet)}$ such that:

$$\forall x^\bullet \in \mathcal{S}^\downarrow_{(\mathcal{L}^\bullet,\mathcal{M}^\bullet,\pi^\bullet)}, \forall f \in \mathcal{O}^\downarrow_{(\mathcal{L},\mathcal{M},\pi)},\ \lambda(f)(x^\bullet) = f(\kappa(x^\bullet)). \tag{2.8.3}$$

**Proof** First, we check that $\kappa$ is well-defined. Let $\left(x^\bullet_{\eta^\bullet}\right)_{\eta^\bullet \in \mathcal{L}^\bullet} \in \mathcal{S}^\downarrow_{(\mathcal{L}^\bullet,\mathcal{M}^\bullet,\pi^\bullet)}$ and let $\eta \preccurlyeq \eta' \in \mathcal{L}$. We have $\ell(\eta) \preccurlyeq \ell(\eta')$ and from eq. (2.8.1):

$$\pi_{\eta' \to \eta}\left(\mu_{\eta'}^{-1}\left(x^\bullet_{\ell(\eta')}\right)\right) = \mu_\eta^{-1} \circ \pi^\bullet_{\ell(\eta') \to \ell(\eta)}\left(x^\bullet_{\ell(\eta')}\right) = \mu_\eta^{-1}\left(x^\bullet_{\ell(\eta)}\right),$$

hence $\left(\mu_\eta^{-1}\left(x^\bullet_{\ell(\eta)}\right)\right)_{\eta \in \mathcal{L}} \in \mathcal{S}^\downarrow_{(\mathcal{L},\mathcal{M},\pi)}$.

To prove that $\kappa$ is bijective, we define:

$$\begin{aligned}\widetilde{\kappa} : \mathcal{S}^\downarrow_{(\mathcal{L},\mathcal{M},\pi)} &\to \mathcal{S}^\downarrow_{(\mathcal{L}^\bullet,\mathcal{M}^\bullet,\pi^\bullet)} \\ (x_\eta)_{\eta \in \mathcal{L}} &\mapsto (x^\bullet_{\eta^\bullet})_{\eta^\bullet \in \mathcal{L}^\bullet}\end{aligned},$$

where $\forall \eta^\bullet \in \mathcal{L}^\bullet$, $x^\bullet_{\eta^\bullet} := \mu_\eta(x_\eta)$ for any $\eta$ such that $\ell(\eta) = \eta^\bullet$ (making use of the surjectivity of $\ell$). $x^\bullet_{\eta^\bullet}$ does not depend on the choice of $\eta \in \ell^{-1}\langle \eta^\bullet \rangle$; indeed, if $\ell(\eta) = \ell(\eta')$, there exists $\eta'' \in \mathcal{L}$ such that $\eta'' \succcurlyeq \eta, \eta'$, hence:

$$\mu_\eta(x_\eta) = \mu_\eta \circ \pi_{\eta'' \to \eta}(x_{\eta''}) = \pi^\bullet_{\ell(\eta'') \to \ell(\eta)} \circ \mu_{\eta''}(x_{\eta''})$$

$$= \pi^\bullet_{\ell(\eta'') \to \ell(\eta')} \circ \mu_{\eta''}(x_{\eta''}) = \mu_{\eta'} \circ \pi_{\eta'' \to \eta'}(x_{\eta''}) = \mu_{\eta'}(x_{\eta'}).$$

And by construction of $\widetilde{\kappa}$, we have $\kappa \circ \widetilde{\kappa} = \mathrm{id}_{\mathcal{S}^\downarrow_{(\mathcal{L},\mathcal{M},\pi)}}$ as well as $\widetilde{\kappa} \circ \kappa = \mathrm{id}_{\mathcal{S}^\downarrow_{(\mathcal{L}^\bullet,\mathcal{M}^\bullet,\pi^\bullet)}}$.

Now, we define $\lambda$ by:

$$\begin{aligned}\lambda : \mathcal{O}^\downarrow_{(\mathcal{L},\mathcal{M},\pi)} &\to \mathcal{O}^\downarrow_{(\mathcal{L}^\bullet,\mathcal{M}^\bullet,\pi^\bullet)} \\ [f_\eta]_\sim &\mapsto [f_\eta \circ \mu_\eta^{-1}]_\sim\end{aligned}.$$

$\lambda$ is well-defined, for we have:

$$\forall \eta, \eta' \in \mathcal{L},\ f_\eta \sim f_{\eta'} \Leftrightarrow \left(\exists \eta'' \succcurlyeq \eta', \eta\ /\ f_\eta \circ \pi_{\eta'' \to \eta} = f_{\eta'} \circ \pi_{\eta'' \to \eta'}\right)$$

$$\Leftrightarrow \left(\exists \eta'' \succcurlyeq \eta', \eta\ /\ f_\eta \circ \mu_\eta^{-1} \circ \pi^\bullet_{\ell(\eta'') \to \ell(\eta)} = f_{\eta'} \circ \mu_{\eta'}^{-1} \circ \pi^\bullet_{\ell(\eta'') \to \ell(\eta')}\right)$$

$$\Rightarrow \left(f_\eta \circ \mu_\eta^{-1} \sim f_{\eta'} \circ \mu_{\eta'}^{-1}\right).$$

And by construction of $\lambda$, eq. (2.8.3) is fulfilled.

Finally, to prove that $\lambda$ is bijective, we construct a map $\widetilde{\lambda}$ by:

$$\begin{aligned}\widetilde{\lambda} : \mathcal{O}^\downarrow_{(\mathcal{L}^\bullet,\mathcal{M}^\bullet,\pi^\bullet)} &\to \mathcal{O}^\downarrow_{(\mathcal{L},\mathcal{M},\pi)} \\ [f^\bullet_{\eta^\bullet}]_\sim &\mapsto [f_\eta]_\sim\end{aligned},$$

where $f_\eta$ is defined for any $\eta$ such that $\ell(\eta) = \eta^\bullet$ by $f_\eta = f^\bullet_{\eta^\bullet} \circ \mu_\eta$. To check that $\widetilde{\lambda}$ is well-defined, let $\eta, \eta' \in \mathcal{L}$ such that there exist $f^\bullet_{\ell(\eta)}, f^\bullet_{\ell(\eta')} \in [f^\bullet_{\eta^\bullet}]_\sim$ (note that this also covers the case $\ell(\eta) = \ell(\eta') = \eta^\bullet$). Then, there exists $\eta^{\bullet\prime\prime}$ such that:

$$f^\bullet_{\ell(\eta)} \circ \pi^\bullet_{\eta^{\bullet\prime\prime} \to \ell(\eta)} = f^\bullet_{\ell(\eta')} \circ \pi^\bullet_{\eta^{\bullet\prime\prime} \to \ell(\eta')},$$

and, since $\ell$ is surjective, there exists $\eta'' \in \mathcal{L}$ such that $\ell(\eta'') = \eta^{\bullet\prime\prime}$. Next, using that $\mathcal{L}$ is a directed set, there exists $\eta''' \in \mathcal{L}$ with $\eta''' \succcurlyeq \eta, \eta', \eta''$. Therefore, we have:

$$f^\bullet_{\ell(\eta)} \circ \pi^\bullet_{\ell(\eta''') \to \ell(\eta)} = f^\bullet_{\ell(\eta')} \circ \pi^\bullet_{\ell(\eta''') \to \ell(\eta')}$$

$$f^\bullet_{\ell(\eta)} \circ \mu_\eta \circ \pi_{\eta''' \to \eta} = f^\bullet_{\ell(\eta')} \circ \mu_{\eta'} \circ \pi_{\eta''' \to \eta'}$$

$$f^\bullet_{\ell(\eta)} \circ \mu_\eta \sim f^\bullet_{\ell(\eta')} \circ \mu_{\eta'}.$$

And by construction of $\widetilde{\lambda}$, we have $\widetilde{\lambda} \circ \lambda = \mathrm{id}_{\mathcal{O}^\downarrow_{(\mathcal{L}, \mathcal{M}, \pi)}}$ as well as $\lambda \circ \widetilde{\lambda} = \mathrm{id}_{\mathcal{O}^\downarrow_{(\mathcal{L}^\bullet, \mathcal{M}^\bullet, \pi^\bullet)}}$. □

**Proposition 2.9** The previous result still holds if, instead of requiring $\ell$ to be surjective, we simply require $\ell \langle \mathcal{L} \rangle$ to be a cofinal part of $\mathcal{L}^\bullet$.

**Proof** This follows by combining prop. 2.8 with prop. 2.5. □

## 2.3 Factorizing systems

For technical convenience, we will often specialize to a particular class of projective systems of phase phases, namely the situation where for any $\eta \preccurlyeq \eta'$, the symplectic manifold $\mathcal{M}_{\eta'}$ can be identified with the Cartesian product of $\mathcal{M}_\eta$ with a symplectic manifold $\mathcal{M}_{\eta' \to \eta}$ (in other words the discarded degrees of freedom can be collected into a phase space $\mathcal{M}_{\eta' \to \eta}$).

This restriction is in fact not as radical as one could first think, for given a projection $\pi : \mathcal{M} \to \mathcal{N}$ as in def. 2.1, $\mathcal{M}$ can always be *locally* written as a Cartesian product of symplectic manifolds in such a way that $\pi$ correspond to the projection map on one factor of the product. Moreover, there is only one (local) decomposition having this property.

At the level of observables, writing $\mathcal{M}$ as a Cartesian product $\mathcal{N}^\perp \times \mathcal{N}$ implies that the algebra $\mathcal{O}'$ of all observables over $\mathcal{M}$ is generated by $\mathcal{O} \cup \mathcal{O}^\perp$, with $\mathcal{O}$ the subalgebra of $\mathcal{O}'$ defined by the observables over $\mathcal{N}$ and $\mathcal{O}^\perp$ by the ones over $\mathcal{N}^\perp$. And asking the symplectic structure on $\mathcal{M}$ to agree with the symplectic structure on the Cartesian product moreover requires that any observable in $\mathcal{O}$ Poisson-commutes with any observable in $\mathcal{O}^\perp$. This is the reason why, at least locally, the symplectic structure on $\mathcal{M}$ prescribes how to choose a subalgebra $\mathcal{O}^\perp$ completing $\mathcal{O}$: $\mathcal{O}^\perp$ has to be the set of all observables having vanishing Poisson brackets with any observable in $\mathcal{O}$.

To understand better why this factorization of $\mathcal{M}$ will not always hold globally, we can examine how the proof of prop. 2.10 below is done: what we have is a foliation of $\mathcal{M}$, of which each leaf is locally diffeomorphic to $\mathcal{N}$ via $\pi$. It is precisely when this local diffeomorphic identification fails to be a global one, that we will not get a global factorization. This can happen at two different levels. First, the restriction of $\pi$ to a given leaf is not necessarily a covering map, although it is locally diffeomorphic: there can be 'completeness' issues, as exemplified by the ad hoc situation where $\mathcal{M} = \{(x_1, x_2; p_1, p_2) \mid |x_2| < \exp(x_1)\} \subset T^*(\mathbb{R}^2)$ and $\mathcal{N} = \{(x_1; p_1)\} \subset T^*(\mathbb{R})$. Second, a covering map need not be bijective, unless $\mathcal{N}$ is simply-connected: for example, $\mathcal{M}$ being a symplectomorphic covering of $\mathcal{N}$ provides a projection that is compatible with the symplectic structures in the sense of def. 2.1, but there is no corresponding factorization.

Unless otherwise stated, all manifolds considered in the present subsection will be *finite* dimensional manifolds.

**Proposition 2.10** Let $\mathcal{M}$, $\mathcal{N}$ be finite dimensional symplectic manifolds and suppose that there exists $\pi : \mathcal{M} \to \mathcal{N}$ satisfying def. 2.1.

Then, for $x \in \mathcal{M}$, there exist an open neighborhood $\mathcal{U}$ of $x$ in $\mathcal{M}$, an open neighborhood $\mathcal{V}$ of $\pi(x)$ in $\mathcal{N}$, a manifold $\mathcal{W}$ and a symplectic structure $\Omega_\mathcal{W}$ on $\mathcal{W}$ such that there exists a diffeomorphism $\Phi : \mathcal{V} \times \mathcal{W} \to \mathcal{U}$ satisfying $\forall y \in \mathcal{V}, \forall w \in \mathcal{W}$, $\pi \circ \Phi(y, w) = y$ and $\Phi^*\Omega_\mathcal{M} = \Omega_\mathcal{N} \times \Omega_\mathcal{W}$.

Moreover, $\Phi$ is unique in the following sense: if $\mathcal{U}'$ is an open subset of $\mathcal{U}$, $\mathcal{V}'$ is a connected open subspace of $\mathcal{V}$, $\mathcal{W}'$ is a symplectic manifold and $\Phi' : \mathcal{V}' \times \mathcal{W}' \to \mathcal{U}'$ is a symplectomorphism such that $\forall y \in \mathcal{V}', \forall z \in \mathcal{W}'$, $\pi \circ \Phi'(y, z) = y$, then there exists a symplectomorphism $\psi : \mathcal{W}' \to \mathcal{W}''$ (with $\mathcal{W}''$ an open subset of $\mathcal{W}$) such that $\forall y \in \mathcal{V}', \forall z \in \mathcal{W}'$, $\Phi'(y, z) = \Phi(y, \psi(z))$.

**Proof** *Existence.* We call $D = \dim(\mathcal{M})$, $d = \dim(\mathcal{N})$ and $n = D - d$. For all $x \in \mathcal{M}$, we define:

$$W_x = \{w \in T_x(\mathcal{M}) \mid T_x\pi(w) = 0\} \text{ and } V_x = \{v \in T_x(\mathcal{M}) \mid \forall w \in W_x, \Omega_{\mathcal{M},x}(v, w) = 0\}.$$

We have $\forall v \in T^*_{\pi(x)}(\mathcal{N}), \forall w \in W_x$, $\Omega_{\mathcal{M},x}(\underline{\pi^*v}, w) = v \circ T_x\pi(w) = 0$, hence $\forall v \in T^*_{\pi(x)}(\mathcal{N})$, $\underline{\pi^*v} \in V_x$. For $u \in T_x(\mathcal{M})$, we define $v_u = \Omega_{\mathcal{N},\pi(x)}(T_x\pi(u), \cdot)$. Using eq. (2.1.1), we get $\forall u \in T_x(\mathcal{M})$, $T_x\pi\left(\underline{\pi^*v_u}\right) = \underline{v_u} = T_x\pi(u)$. So we can write $u \in T_x(\mathcal{M})$ as $u = (u - \underline{\pi^*v_u}) + \underline{\pi^*v_u}$ with $u - \underline{\pi^*v_u} \in W_x$ and $\underline{\pi^*v_u} \in V_x$.

Hence, we have $W_x + V_x = T_x(\mathcal{M})$, and therefore $W_x \oplus V_x = T_x(\mathcal{M})$, since $\dim(V_x) = \dim(\mathcal{M}) - \dim(W_x)$. Moreover, since eq. (2.1.1) implies that $T_x\pi$ is surjective, we have $\dim(W_x) = D - d$ and $\dim(V_x) = D - (D - d) = d$.

Now we choose $x \in \mathcal{M}$ and we consider a coordinate patch $\mathcal{V}_1$ on $\mathcal{N}$ containing $\pi(x)$, with coordinates $y_1, \ldots, y_d$. We define:

$$X_{i,x'} := \underline{\pi^*dy_{i,\pi(x')}} = \underline{d(y_i \circ \pi)_{x'}} \text{ for all } x' \in \mathcal{U}_1 := \pi^{-1}\langle \mathcal{V}_1 \rangle.$$

$X_1, \ldots, X_d$ are vector fields on $\mathcal{U}_1$ such that $\forall x' \in \mathcal{U}_1, (X_{1,x'}, \ldots, X_{d,x'})$ is a basis of $V_{x'}$. We calculate the Lie brackets between two of these vector fields:

$$[X_i, X_j] = \left[\underline{dy_i \circ \pi}, \underline{dy_j \circ \pi}\right] = \underline{d\left(\{y_i \circ \pi, y_j \circ \pi\}\right)} = \underline{d\left(\{y_i, y_j\} \circ \pi\right)} = \underline{\pi^*d\left(\{y_i, y_j\}\right)}$$

where the second equality expresses the Lie brackets of two Hamiltonian vector fields and the third equality comes from prop. 2.2.

Therefore, we have $\forall x' \in \mathcal{U}_1, [X_i, X_j]_{x'} \in V_{x'}$. From Frobenius theorem [11, theorem 14.5], there exist an open neighborhood $\mathcal{U}_2$ of $x$ in $\mathcal{U}_1$ and coordinates $x_1, \ldots, x_d, x_{d+1}, \ldots, x_D$ over $\mathcal{U}_2$ such that $\forall x' \in \mathcal{U}_2, \partial_{x_1,x'}, \ldots, \partial_{x_d,x'}$ is a basis of $V_{x'}$. We define:

$$\widetilde{\Phi} : \mathcal{U}_2 \to \mathcal{N} \times \mathbb{R}^n$$
$$x' \mapsto \pi(x'), (x_{d+1}(x'), \ldots, x_D(x')).$$

We can now show that $T_x\widetilde{\Phi} : T_x(\mathcal{M}) \to T_{\pi(x)}(\mathcal{N}) \times \mathbb{R}^n$ is bijective. Indeed, let $u \in T_x(\mathcal{M})$ such that $T_x\widetilde{\Phi}(u) = 0$. Then, in particular, we have $T_x\pi(u) = 0$, so $u \in W_x$. On the other hand, we have $dx_{k,x}(u) = 0$ for $k = d + 1, \ldots, D$, so $u$ is a linear combination of $\partial_{x_1,x}, \ldots, \partial_{x_d,x}$, hence $u \in V_x$. From $W_x \oplus V_x = T_x(\mathcal{M})$, $u = 0$. Therefore $T_x\widetilde{\Phi}$ is injective, thus bijective, for

$\dim(T_x(\mathcal{M})) = \dim\left(T_{\pi(x)}(\mathcal{N}) \times \mathbb{R}^n\right).$

From the inverse function theorem [11, theorem 5.11], there exists an open neighborhood $\mathcal{U}_3$ of $x$ in $\mathcal{U}_2$ such that $\widetilde{\Phi}\big|_{\mathcal{U}_3} : \mathcal{U}_3 \to \widetilde{\Phi}\langle\mathcal{U}_3\rangle$ is a diffeomorphism. Hence there exist an open connected neighborhood $\mathcal{V}$ of $\pi(x)$ in $\mathcal{N}$, an open subset $\mathcal{W}$ of $\mathbb{R}^n$, an open neighborhood $\mathcal{U}$ of $x$ in $\mathcal{U}_3$ and a diffeomorphism $\Phi : \mathcal{V} \times \mathcal{W} \to \mathcal{U}$ such that $\forall y \in \mathcal{V}, \forall z \in \mathcal{W}, \widetilde{\Phi}(\Phi(y,z)) = (y,z)$. In particular, $\forall y \in \mathcal{V}, \forall z \in \mathcal{W}, \pi \circ \Phi(y,z) = y$.

At every point $x' \in \mathcal{M}$ and for every vector $v, w \in T_{\Phi^{-1}(x')}(\mathcal{V} \times \mathcal{W})$, $T_{\Phi^{-1}(x')}\Phi(v,0) \in V_{x'}$ and $T_{\Phi^{-1}(x')}\Phi(0,w) \in W_{x'}$. In particular, we have $\Omega_{\mathcal{M},x'}\left(T_{\Phi^{-1}(x')}\Phi(v,0), T_{\Phi^{-1}(x')}\Phi(0,w)\right) = 0$.

We now consider $z \in \mathcal{W}$, $w, w' \in T_z(\mathcal{W})$ and we define for all $y \in \mathcal{V}$,

$$\Omega^y_{\mathcal{W},z}(w, w') = \Omega_{\mathcal{M},\Phi(y,z)}(T_{(y,z)}\Phi(0,w), T_{(y,z)}\Phi(0,w')).$$

Let $\widetilde{Y}$ be a vector field on $\mathcal{V}$, let $\widetilde{Z}, \widetilde{Z}'$ be vector fields on $\mathcal{W}$ such that $\widetilde{Z}_z = w$, $\widetilde{Z}'_z = w'$. We define the vector fields $Y = \Phi_*\left(\widetilde{Y},0\right)$, $Z = \Phi_*\left(0,\widetilde{Z}\right)$ and $Z' = \Phi_*\left(0,\widetilde{Z}'\right)$ on $\mathcal{M}$. From $[Y,Z] = [Y,Z'] = 0$ and $d\Omega_{\mathcal{M}} = 0$, we have:

$$Y\left(\Omega_{\mathcal{M}}\left(Z, Z'\right)\right) = Z\left(\Omega_{\mathcal{M}}\left(Y, Z'\right)\right) - Z'\left(\Omega_{\mathcal{M}}\left(Y, Z\right)\right) + \Omega_{\mathcal{M}}\left([Z, Z'], Y\right)$$

since $Y_{x'} \in V_{x'}$ and $Z_{x'}, Z'_{x'}, [Z, Z']_{x'} \in W_{x'}$, we have $Y\left(\Omega_{\mathcal{M}}\left(Z, Z'\right)\right) = 0$. Therefore the differential of $y \mapsto \Omega^y_{\mathcal{W},z}(w, w')$ is zero at every point $y \in \mathcal{V}$, and $\mathcal{V}$ being connected, $\Omega^y_{\mathcal{W},z}(w, w')$ does not depend on $y$. So, we define $\Omega_{\mathcal{W},z}(w,w') = \Omega^y_{\mathcal{W},z}(w,w')$.

We now can check using eq. (2.1.1) and the definition of $\Omega_{\mathcal{W}}$ that $\Phi^*\Omega_{\mathcal{M}} = \Omega_{\mathcal{N}} \times \Omega_{\mathcal{W}}$. Therefore $\Omega_{\mathcal{W}}$ is a symplectic structure on $\mathcal{W}$ and $\Phi : \mathcal{V} \times \mathcal{W} \to \mathcal{U}$ is a symplectomorphism.

*Uniqueness.* We consider symplectic manifolds $\mathcal{V}$, $\mathcal{W}$, and $\mathcal{W}'$, a connected open subset $\mathcal{V}'$ of $\mathcal{V}$, and an application $\widetilde{\psi} : \mathcal{V}' \times \mathcal{W}' \to \mathcal{W}$ such that:

$$\Psi : \mathcal{V}' \times \mathcal{W}' \to \mathcal{V} \times \mathcal{W}$$
$$y, z \mapsto y, \widetilde{\psi}(y,z)$$

induces a symplectomorphism $\mathcal{V}' \times \mathcal{W}' \to \Psi\langle\mathcal{V}' \times \mathcal{W}'\rangle$.

For $y \in \mathcal{V}', z \in \mathcal{W}', v \in T_y(\mathcal{V}'), w \in T_z(\mathcal{W}')$, we then have:

$$0 = \Omega_{\mathcal{V} \times \mathcal{W},\Psi(y,z)}\left(T_{(y,z)}\Psi(v,0), T_{(y,z)}\Psi(0,w)\right) = \Omega_{\mathcal{W},\psi(y,z)}\left(T_{(y,z)}\widetilde{\psi}(v,0), T_{(y,z)}\widetilde{\psi}(0,w)\right).$$

However, for $\Psi$ to be a diffeomorphism, $T_{(y,z)}\widetilde{\psi}(0,w)$ should run through $T_{\psi(y,z)}(\mathcal{W})$ when $w$ runs through $T_z(\mathcal{W})$. Therefore, we should have $T_{(y,z)}\widetilde{\psi}(v,0) = 0$. Hence, $\mathcal{V}'$ being connected, $\widetilde{\psi}(y,z)$ cannot depend upon $y$. Accordingly, we define $\psi(z) := \widetilde{\psi}(y,z)$, and $\Psi|_{\mathcal{V}' \times \mathcal{W}' \to \Psi\langle\mathcal{V}' \times \mathcal{W}'\rangle}$ being a symplectomorphism requires that $\psi|_{\mathcal{W}' \to \psi\langle\mathcal{W}'\rangle}$ should be a symplectomorphism.

*Note.* A more concise (albeit less instructive) proof of this result can be achieved by considering the closed 2-form $\sigma := \Omega_{\mathcal{M}} - \pi^*\Omega_{\mathcal{N}}$ and applying a standard result of symplectic geometry [17, § 5.24], telling us that the kernel of $\sigma$ is an involutive distribution, and that $\sigma$ defines a symplectic form on the quotient. □

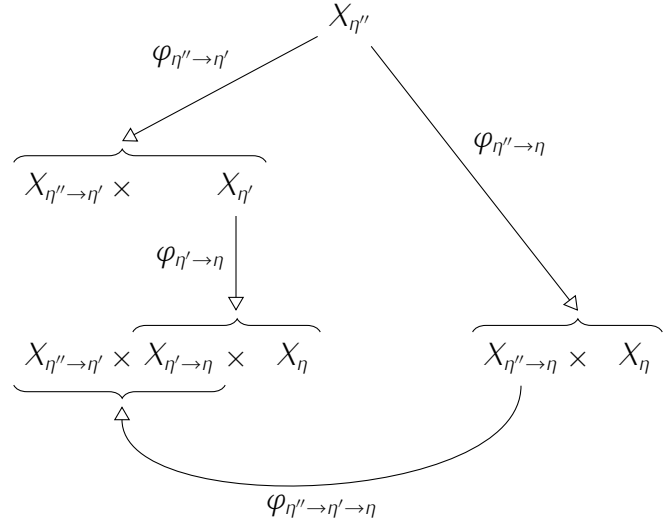

Figure 2.2 – Three-spaces consistency for factorizing systems

In order to build a structure describing a collection of interconnected partial theories, where the relation between a more detailed partial theory $\mathcal{M}_{\eta'}$ and a less detailed one $\mathcal{M}_\eta$ is given by a factorization of $\mathcal{M}_{\eta'}$ as $\mathcal{M}_{\eta' \to \eta} \times \mathcal{M}_\eta$, we also need to reformulate the three-spaces consistency condition that we had for a projective system (fig. 2.1) in terms of a factorization requirement. For this, we ask for the symplectic manifold $\mathcal{M}_{\eta'' \to \eta}$, that holds the degrees of freedom discarded when going directly from $\mathcal{M}_{\eta''}$ to $\mathcal{M}_\eta$, to decompose as the Cartesian product of $\mathcal{M}_{\eta'' \to \eta'}$ with $\mathcal{M}_{\eta' \to \eta}$, where $\mathcal{M}_{\eta'' \to \eta'}$ holds the degrees of freedom discarded when going as a first step from $\mathcal{M}_{\eta''}$ to $\mathcal{M}_{\eta'}$, and $\mathcal{M}_{\eta' \to \eta}$ holds the ones discarded when going as a second step from $\mathcal{M}_{\eta'}$ to $\mathcal{M}_\eta$ (fig. 2.2).

Having a factorizing system defined this way then provides us immediately with a projective system as above. Reciprocally, if we give us a projective system of phase spaces in which any projection $\pi_{\eta' \to \eta}$ can be understood as projecting on a factor of a Cartesian product (that is, if the result of prop. 2.10 happens to hold globally and not just locally) and if moreover all the $\mathcal{M}_\eta$ are connected (which sounds physically sensible when speaking of phases spaces), we can construct a corresponding factorizing system of phase spaces.

**Definition 2.11** A factorizing system is a quintuple:
$$\left( \mathcal{L}, \left( X_\eta \right)_{\eta \in \mathcal{L}}, \left( X_{\eta' \to \eta} \right)_{\eta \preccurlyeq \eta'}, \left( \varphi_{\eta' \to \eta} \right)_{\eta \preccurlyeq \eta'}, \left( \varphi_{\eta'' \to \eta' \to \eta} \right)_{\eta \preccurlyeq \eta' \preccurlyeq \eta''} \right)$$
where:

1. $\mathcal{L}$ is a preordered, directed set;

2. $\left( X_\eta \right)_{\eta \in \mathcal{L}}$ is a family of spaces indexed by $\mathcal{L}$;

3. $\left( X_{\eta' \to \eta} \right)_{\eta \preccurlyeq \eta'}$ is a family of spaces indexed by $\{ \eta, \eta' \in \mathcal{L} \mid \eta \preccurlyeq \eta' \}$, such that, for all $\eta \in \mathcal{L}$, $X_{\eta \to \eta}$ has only one element;

4. $\left( \varphi_{\eta' \to \eta} \right)_{\eta \preccurlyeq \eta'}$ is a family of bijective maps $\varphi_{\eta' \to \eta} : X_{\eta'} \to X_{\eta' \to \eta} \times X_\eta$ indexed by $\{ \eta, \eta' \in \mathcal{L} \mid \eta \preccurlyeq \eta' \}$ such that $\varphi_{\eta \to \eta}$ is trivial;

5. $(\varphi_{\eta''\to\eta'\to\eta})_{\eta\preccurlyeq\eta'\preccurlyeq\eta''}$ is a family of bijective maps $\varphi_{\eta''\to\eta'\to\eta} : X_{\eta''\to\eta} \to X_{\eta''\to\eta'} \times X_{\eta'\to\eta}$ indexed by $\{\eta, \eta', \eta'' \in \mathcal{L} \mid \eta \preccurlyeq \eta' \preccurlyeq \eta''\}$ such that $\varphi_{\eta''\to\eta'\to\eta}$ is trivial whenever two labels among $\eta, \eta', \eta''$ are equal and:

$$\forall \eta, \eta', \eta'' \in \mathcal{L} \mid \eta \preccurlyeq \eta' \preccurlyeq \eta'', \quad (\varphi_{\eta''\to\eta'\to\eta} \times \mathrm{id}_{X_\eta}) \circ \varphi_{\eta''\to\eta} = (\mathrm{id}_{X_{\eta''\to\eta'}} \times \varphi_{\eta'\to\eta}) \circ \varphi_{\eta''\to\eta'}. \quad (2.11.1)$$

Whenever possible, we will use the shortened notation $(\mathcal{L}, X, \varphi)^\times$ instead of $\left(\mathcal{L}, (X_\eta)_{\eta\in\mathcal{L}}, (X_{\eta'\to\eta})_{\eta\preccurlyeq\eta'}, (\varphi_{\eta'\to\eta})_{\eta\preccurlyeq\eta'}, (\varphi_{\eta''\to\eta'\to\eta})_{\eta\preccurlyeq\eta'\preccurlyeq\eta''}\right)$.

**Definition 2.12** A factorizing system of phase spaces is a factorizing system $(\mathcal{L}, \mathcal{M}, \varphi)^\times$ where:

1. for all $\eta \in \mathcal{L}$, $\mathcal{M}_\eta$ is a symplectic manifold, and for all $\eta \preccurlyeq \eta' \in \mathcal{L}$, $\mathcal{M}_{\eta'\to\eta}$ is a symplectic manifold, except if $\eta' = \eta$ in which case $\mathcal{M}_{\eta\to\eta}$ is a set with just one element;

2. for all $\eta \preccurlyeq \eta' \in \mathcal{L}$, $\varphi_{\eta'\to\eta}$ is a symplectomorphism, and for all $\eta \preccurlyeq \eta' \preccurlyeq \eta'' \in \mathcal{L}$, $\varphi_{\eta''\to\eta'\to\eta}$ is a symplectomorphism.

**Proposition 2.13** If $(\mathcal{L}, \mathcal{M}, \varphi)^\times$ fulfills def. 2.12 and if, for $\eta \preccurlyeq \eta' \in \mathcal{L}$, we define:

$$s_{\eta'\to\eta} : \mathcal{M}_{\eta'\to\eta} \times \mathcal{M}_\eta \to \mathcal{M}_\eta \qquad \text{and} \qquad \pi_{\eta'\to\eta} = s_{\eta'\to\eta} \circ \varphi_{\eta'\to\eta} \qquad (2.13.1)$$
$$(y, x) \mapsto x$$

then $(\mathcal{L}, \mathcal{M}, \pi)^\downarrow$ is a projective system of phase spaces.

Accordingly, we define the space of states by $\mathcal{S}^\times_{(\mathcal{L},\mathcal{M},\varphi)} := \mathcal{S}^\downarrow_{(\mathcal{L},\mathcal{M},\pi)}$ (def. 2.3) and the space of observables by $\mathcal{O}^\times_{(\mathcal{L},\mathcal{M},\varphi)} := \mathcal{O}^\downarrow_{(\mathcal{L},\mathcal{M},\pi)}$ (def. 2.4).

**Proof** We need to prove that $\forall \eta \preccurlyeq \eta' \in \mathcal{L}$, $\pi_{\eta'\to\eta}$ is a surjective map compatible with the symplectic structures, that $\forall \eta \in \mathcal{L}$, $\pi_{\eta\to\eta} = \mathrm{id}_{\mathcal{M}_\eta}$, and that $\forall \eta \preccurlyeq \eta' \preccurlyeq \eta'' \in \mathcal{L}$, $\pi_{\eta''\to\eta} = \pi_{\eta'\to\eta} \circ \pi_{\eta''\to\eta'}$.

For $\eta \in \mathcal{L}$, we have $\pi_{\eta\to\eta} = \mathrm{id}_{\mathcal{M}_\eta}$ (identifying $\mathcal{M}_\eta$ and its trivial Cartesian product with a one-element set), so in particular it is a surjective map compatible with the symplectic structures.

Let $\eta \prec \eta' \in \mathcal{L}$. $\mathcal{M}_{\eta'\to\eta} \neq \varnothing$ (as a manifold), hence $s_{\eta'\to\eta}$ is surjective, therefore $\pi_{\eta'\to\eta}$ is a surjective map.

Let $(y, x) \in \mathcal{M}_{\eta'\to\eta} \times \mathcal{M}_\eta$ and let $\upsilon \in T_x^*(\mathcal{M}_\eta)$. We have:

$\forall w, v \in T_{(y,x)}(\mathcal{M}_{\eta'\to\eta} \times \mathcal{M}_\eta),$

$$\upsilon \circ \left[T_{(y,x)} s_{\eta'\to\eta}\right](w, v) = \upsilon(v) = \Omega_{\mathcal{M}_\eta, x}(\underline{\upsilon}, v) = \Omega_{\mathcal{M}_{\eta'\to\eta} \times \mathcal{M}_\eta, (y,x)}((0, \underline{\upsilon}), (w, v)),$$

so that $\underline{s^*_{\eta'\to\eta}\upsilon} = (0, \underline{\upsilon})$, hence $\left[T_{(y,x)} s_{\eta'\to\eta}\right](\underline{s^*_{\eta'\to\eta}\upsilon}) = \underline{\upsilon}$. Therefore $s_{\eta'\to\eta}$ is compatible with the symplectic structures, and since $\varphi_{\eta'\to\eta}$ is a symplectomorphism, $\pi_{\eta'\to\eta}$ is compatible with the symplectic structures.

Let $\eta \preccurlyeq \eta' \preccurlyeq \eta'' \in \mathcal{L}$ and define:

$$s_{\eta''\to\eta'\to\eta} : \mathcal{M}_{\eta''\to\eta'} \times \mathcal{M}_{\eta'\to\eta} \times \mathcal{M}_\eta \to \mathcal{M}_\eta \quad .$$
$$(z, y, x) \mapsto x$$

We have:

$$s_{\eta''\to\eta'\to\eta} \circ \left(\text{id}_{\mathcal{M}_{\eta''\to\eta'}} \times \varphi_{\eta'\to\eta}\right) = s_{\eta'\to\eta} \circ \varphi_{\eta'\to\eta} \circ s_{\eta''\to\eta'},$$

and $\quad s_{\eta''\to\eta'\to\eta} \circ \left(\varphi_{\eta''\to\eta'\to\eta} \times \text{id}_{\mathcal{M}_\eta}\right) = s_{\eta''\to\eta}.$

Hence, composing eq. (2.11.1) to the right with $s_{\eta''\to\eta'\to\eta}$ gives:

$$s_{\eta''\to\eta} \circ \varphi_{\eta''\to\eta} = s_{\eta'\to\eta} \circ \varphi_{\eta'\to\eta} \circ s_{\eta''\to\eta'} \circ \varphi_{\eta''\to\eta'},$$

so that we have $\pi_{\eta''\to\eta} = \pi_{\eta'\to\eta} \circ \pi_{\eta''\to\eta'}$. $\square$

**Proposition 2.14** Let $(\mathcal{L}, \mathcal{M}, \pi)^{\downarrow}$ be a projective system of phase spaces and suppose that:

1. for all $\eta \in \mathcal{L}$, $\mathcal{M}_\eta$ is connected;

2. for all $\eta \prec \eta' \in \mathcal{L}$, there exist a symplectic manifold $\mathcal{M}_{\eta'\to\eta}$ and a symplectomorphism $\varphi_{\eta'\to\eta} : \mathcal{M}_{\eta'} \to \mathcal{M}_{\eta'\to\eta} \times \mathcal{M}_\eta$ such that $\pi_{\eta'\to\eta} = s_{\eta'\to\eta} \circ \varphi_{\eta'\to\eta}$.

Then, we can complete this input into a factorizing system $(\mathcal{L}, \mathcal{M}, \varphi)^{\times}$.

**Proof** For $\eta \in \mathcal{L}$, we define $\mathcal{M}_{\eta\to\eta}$ to be a space with one element and $\varphi_{\eta\to\eta}$ to be the trivial identification.

Let $\eta \preccurlyeq \eta' \preccurlyeq \eta'' \in \mathcal{L}$. What we need to show is that there exists a symplectomorphism $\varphi_{\eta''\to\eta'\to\eta} : \mathcal{M}_{\eta''\to\eta} \to \mathcal{M}_{\eta''\to\eta'} \times \mathcal{M}_{\eta'\to\eta}$ such that eq. (2.11.1) is fulfilled. If two labels among $\eta, \eta', \eta''$ are equal, we can choose $\varphi_{\eta''\to\eta'\to\eta}$ to be the trivial identification, so we now consider the case $\eta \prec \eta' \prec \eta''$.

We define:

$$\Psi := \varphi_{\eta''\to\eta} \circ \varphi_{\eta''\to\eta'}^{-1} \circ \left(\text{id}_{\eta''\to\eta'} \times \varphi_{\eta'\to\eta}\right)^{-1} : \mathcal{M}_{\eta''\to\eta'} \times \mathcal{M}_{\eta'\to\eta} \times \mathcal{M}_\eta \to \mathcal{M}_{\eta''\to\eta} \times \mathcal{M}_\eta.$$

$\Psi$ is a symplectomorphism and satisfies:

$$\forall (z, y, x) \in \mathcal{M}_{\eta''\to\eta'} \times \mathcal{M}_{\eta'\to\eta} \times \mathcal{M}_\eta, \quad s_{\eta''\to\eta} \circ \Psi(z, y, x) = x.$$

Hence, applying the uniqueness part from the proof of prop. 2.10 (with $\mathcal{V} = \mathcal{V}' = \mathcal{M}_\eta$, $\mathcal{W} = \mathcal{M}_{\eta''\to\eta}$ and $\mathcal{W}' = \mathcal{M}_{\eta''\to\eta'} \times \mathcal{M}_{\eta'\to\eta}$, using that $\mathcal{M}_\eta$ is connected, as $\mathcal{V}'$ must be), there exists a symplectomorphism $\psi : \mathcal{M}_{\eta''\to\eta'} \times \mathcal{M}_{\eta'\to\eta} \to \mathcal{M}_{\eta''\to\eta}$ such that $\Psi = \psi \times \text{id}_{\mathcal{M}_\eta}$. Thus we define $\varphi_{\eta''\to\eta'\to\eta} = \psi^{-1}$. $\square$

If we have a family of finite dimensional symplectic manifolds, where each $\mathcal{M}_\eta$ modeling a partial theory can be written as a cotangent bundle on a configuration space $\mathcal{C}_\eta$, then a factorizing system built over the family $(\mathcal{C}_\eta)_{\eta\in\mathcal{L}}$ can automatically be lifted as a factorizing system over the family $(\mathcal{M}_\eta)_{\eta\in\mathcal{L}}$. Reciprocally, if we build a projective system of symplectic manifolds over this family, such that each projection can be understood as arising from a factorization of the underlying configuration spaces, and if additionally all the configuration spaces are connected, then not only can the projective system of symplectic manifolds be put into a factorizing form (as follows from prop. 2.14), but this factorizing form goes down to a factorizing system of the configuration spaces.

It is important to note that, at the level of configuration spaces, a factorizing system contains much more input than a projective system does. The situation here is different than what we have at the level of phase spaces, where projective and factorizing systems can, let aside global considerations, be matched unambiguously. The reason for this disparity is that the symplectic structure on the phase spaces played a crucial role in the proof of prop. 2.10: when looking at a projection between configuration spaces, that retains only a subset of the configuration variables, we have no additional structure that would allows us to select a preferred complementary set of discarded variables.

**Definition 2.15** A factorizing system of smooth manifolds is a factorizing system $(\mathcal{L}, \mathcal{C}, \varphi)^\times$ (def. 2.11; in particular eq. (2.11.1) holds) where:

1. for all $\eta \in \mathcal{L}$, $\mathcal{C}_\eta$ is a smooth manifold, and for all $\eta \preccurlyeq \eta' \in \mathcal{L}$, $\mathcal{C}_{\eta' \to \eta}$ is a smooth manifold, except if $\eta' = \eta$ in which case $\mathcal{C}_{\eta \to \eta}$ is a set with just one element;

2. for all $\eta \preccurlyeq \eta' \in \mathcal{L}$, $\varphi_{\eta' \to \eta}$ is a diffeomorphism, and for all $\eta \preccurlyeq \eta' \preccurlyeq \eta'' \in \mathcal{L}$, $\varphi_{\eta'' \to \eta' \to \eta}$ is a diffeomorphism.

**Proposition 2.16** If $(\mathcal{L}, \mathcal{C}, \varphi)^\times$ fulfills def. 2.15 and if:

1. for all $\eta \in \mathcal{L}$ (resp. all $\eta, \eta' \in \mathcal{L}$ with $\eta \prec \eta'$), we define $\mathcal{M}_\eta := T^*(\mathcal{C}_\eta)$ (resp. $\mathcal{M}_{\eta' \to \eta} := T^*(\mathcal{C}_{\eta' \to \eta})$), equipped with the canonical symplectic structure on a cotangent bundle;

2. for all $\eta, \eta' \in \mathcal{L}$ with $\eta \prec \eta'$ (resp. all $\eta, \eta', \eta'' \in \mathcal{L}$ with $\eta \prec \eta' \prec \eta''$), we naturally lift $\varphi_{\eta' \to \eta} : \mathcal{C}_{\eta'} \to \mathcal{C}_{\eta' \to \eta} \times \mathcal{C}_\eta$ (resp. $\varphi_{\eta'' \to \eta' \to \eta} : \mathcal{C}_{\eta'' \to \eta} \to \mathcal{C}_{\eta'' \to \eta'} \times \mathcal{C}_{\eta' \to \eta}$) to a map $\widetilde{\varphi}_{\eta' \to \eta} : \mathcal{M}_{\eta'} \to \mathcal{M}_{\eta' \to \eta} \times \mathcal{M}_\eta$ (resp. $\widetilde{\varphi}_{\eta'' \to \eta' \to \eta} : \mathcal{M}_{\eta'' \to \eta} \to \mathcal{M}_{\eta'' \to \eta'} \times \mathcal{M}_{\eta' \to \eta}$) between the cotangent bundles;

3. for all $\eta \in \mathcal{L}$, we define $\mathcal{M}_{\eta \to \eta}$ to be a set with one element, and for all $\eta \in \mathcal{L}$ (resp. all $\eta, \eta', \eta'' \in \mathcal{L}$ with $\eta \preccurlyeq \eta' \preccurlyeq \eta''$ and at least two labels equals) we define $\widetilde{\varphi}_{\eta \to \eta}$ (resp. $\widetilde{\varphi}_{\eta'' \to \eta' \to \eta}$) to be the trivial identification;

then $(\mathcal{L}, \mathcal{M}, \widetilde{\varphi})^\times$ is a factorizing system of phase spaces.

**Proof** We need to prove that $\forall \eta \prec \eta'$, $\widetilde{\varphi}_{\eta' \to \eta}$ is a symplectomorphism, that $\forall \eta \prec \eta' \prec \eta''$, $\widetilde{\varphi}_{\eta'' \to \eta' \to \eta}$ is a symplectomorphism and that eq. (2.11.1) for the maps $\varphi$ is lifted up to the corresponding equation for the maps $\widetilde{\varphi}$.

For $\eta \in \mathcal{L}$, the symplectic structure on $\mathcal{M}_\eta = T^*(\mathcal{C}_\eta)$ is defined by:

$$\forall (x, p) \in \mathcal{M}_\eta, \forall w, w' \in T_{(x,p)}(\mathcal{M}_\eta), \quad \Omega_{\mathcal{M}_\eta, (x,p)}(w, w') := w'_{\text{VER}}(w_{\text{HOR}}) - w_{\text{VER}}(w'_{\text{HOR}}), \quad (2.16.1)$$

where we define for $w \in T_{(x,p)}(\mathcal{M}_\eta)$, $w_{\text{HOR}} \in T_x(\mathcal{C}_\eta)$ to be the horizontal projection of $w$, and $w_{\text{VER}} \in T^*_x(\mathcal{C}_\eta)$ to be the vertical part of $w$ defined using some local coordinate system around $x$ (the map $w \mapsto w_{\text{VER}}$ depends on this choice of local coordinates, however the anti-symmetrization in eq. (2.16.1) ensures that the definition of $\Omega_{\mathcal{M}_\eta, (x,p)}$ is independent of this choice).

For $\eta \prec \eta' \in \mathcal{L}$, the map $\widetilde{\varphi}_{\eta' \to \eta} : \mathcal{M}_{\eta'} \to \mathcal{M}_{\eta' \to \eta} \times \mathcal{M}_\eta$ is defined by:

$$\forall (x', p') \in \mathcal{M}_{\eta'}, \widetilde{\varphi}_{\eta' \to \eta}(x', p') := \left( \left( f_{\eta' \to \eta} \circ \varphi_{\eta' \to \eta}(x'), p' \circ \left[ T_{\varphi_{\eta' \to \eta}(x')} \varphi^{-1}_{\eta' \to \eta} \right] (\cdot, 0) \right), \right.$$

$$\left. \left( s_{\eta' \to \eta} \circ \varphi_{\eta' \to \eta}(x'), p' \circ \left[ T_{\varphi_{\eta' \to \eta}(x')} \varphi^{-1}_{\eta' \to \eta} \right] (0, \cdot) \right) \right),$$

where $f_{\eta' \to \eta} : \mathcal{C}_{\eta' \to \eta} \times \mathcal{C}_\eta \to \mathcal{C}_{\eta' \to \eta}$ and $s_{\eta' \to \eta} : \mathcal{C}_{\eta' \to \eta} \times \mathcal{C}_\eta \to \mathcal{C}_\eta$ are the projection maps of the Cartesian product. This map is bijective, because $\varphi_{\eta'\to\eta}$ and $\left[T_{\varphi_{\eta'\to\eta}(x')}\varphi_{\eta'\to\eta}^{-1}\right]$ are.

Let $(x, p) \in \mathcal{M}_\eta$, $(y, q) \in \mathcal{M}_{\eta'\to\eta}$ and $(x', p') = \widetilde{\varphi}_{\eta'\to\eta}\left((y, q), (x, p)\right)$. From the definition of $\widetilde{\varphi}_{\eta'\to\eta}$, we have for all $w \in T_{(x',p')}(\mathcal{M}_{\eta'})$:

$$\left([T_{x',p'}\widetilde{\varphi}_{\eta'\to\eta}](w)\right)_{\text{HOR}} = \left([T_{x'} f_{\eta'\to\eta} \circ \varphi_{\eta'\to\eta}](w_{\text{HOR}}), [T_{x'} s_{\eta'\to\eta} \circ \varphi_{\eta'\to\eta}](w_{\text{HOR}})\right).$$

Now, we choose local coordinates around $x$ in $\mathcal{C}_\eta$ and around $y$ in $\mathcal{C}_{\eta'\to\eta}$, so we have local coordinates around in $(y, x)$ in $\mathcal{C}_{\eta'\to\eta} \times \mathcal{C}_\eta$ that we can transport through $\varphi_{\eta'\to\eta}^{-1}$ as local coordinates around $x' = \varphi_{\eta'\to\eta}^{-1}(y, x)$ in $\mathcal{C}_{\eta'}$. Using these to define $(\cdot)_{\text{VER}}$ in $T_{(x,p)}(\mathcal{M}_\eta)$, $T_{(y,q)}(\mathcal{M}_{\eta'\to\eta})$ and $T_{(x',p')}(\mathcal{M}_{\eta'})$, we have for all $w \in T_{(x',p')}(\mathcal{M}_{\eta'})$:

$$\left([T_{x',p'}\widetilde{\varphi}_{\eta'\to\eta}](w)\right)_{\text{VER}} = \left(w_{\text{VER}} \circ [T_{y,x}\varphi_{\eta'\to\eta}^{-1}](\cdot, 0),\ w_{\text{VER}} \circ [T_{y,x}\varphi_{\eta'\to\eta}^{-1}](0, \cdot)\right).$$

Therefore:

$$\forall w, w' \in T_{(x',p')}(\mathcal{M}_{\eta'}),\quad \Omega_{\mathcal{M}_{\eta'\to\eta}\times\mathcal{M}_\eta,((y,x),(q,p))}\left([T_{x',p'}\widetilde{\varphi}_{\eta'\to\eta}](w), [T_{x',p'}\widetilde{\varphi}_{\eta'\to\eta}](w')\right) =$$

$$= w'_{\text{VER}} \circ [T_{y,x}\varphi_{\eta'\to\eta}^{-1}]\left([T_{x'}f_{\eta'\to\eta}\circ\varphi_{\eta'\to\eta}]w_{\text{HOR}},\ 0\right) +$$

$$+ w'_{\text{VER}} \circ [T_{y,x}\varphi_{\eta'\to\eta}^{-1}]\left(0,\ [T_{x'}s_{\eta'\to\eta}\circ\varphi_{\eta'\to\eta}](w_{\text{HOR}})\right) - \left(w \leftrightarrow w'\right)$$

$$= w'_{\text{VER}} \circ [T_{y,x}\varphi_{\eta'\to\eta}^{-1}] \circ [T_{x'}\varphi_{\eta'\to\eta}](w_{\text{HOR}}) - \left(w \leftrightarrow w'\right)$$

$$= \Omega_{\mathcal{M}_{\eta'},(x',p')}(w, w').$$

So $\widetilde{\varphi}_{\eta'\to\eta}$ is a symplectomorphism, and in the same way we prove that for all $\eta \prec \eta' \prec \eta''$, $\widetilde{\varphi}_{\eta''\to\eta'\to\eta}$ is a symplectomorphism.

Let $\eta \prec \eta' \prec \eta'' \in \mathcal{L}$, eq. (2.11.1) for the maps $\varphi$ implies:

$$f_{\eta''\to\eta'\to\eta} \circ \varphi_{\eta''\to\eta'\to\eta} \circ f_{\eta''\to\eta} \circ \varphi_{\eta''\to\eta} = f_{\eta''\to\eta'} \circ \varphi_{\eta''\to\eta'},$$

$$s_{\eta''\to\eta'\to\eta} \circ \varphi_{\eta''\to\eta'\to\eta} \circ f_{\eta''\to\eta} \circ \varphi_{\eta''\to\eta} = f_{\eta'\to\eta} \circ \varphi_{\eta'\to\eta} \circ s_{\eta''\to\eta'} \circ \varphi_{\eta''\to\eta'},$$

$$\&\quad s_{\eta''\to\eta} \circ \varphi_{\eta''\to\eta} = s_{\eta'\to\eta} \circ \varphi_{\eta'\to\eta} \circ s_{\eta''\to\eta'} \circ \varphi_{\eta''\to\eta'},$$

(where $f_{\eta''\to\eta'\to\eta} : \mathcal{C}_{\eta''\to\eta'} \times \mathcal{C}_{\eta'\to\eta} \to \mathcal{C}_{\eta''\to\eta}$ and $s_{\eta''\to\eta'\to\eta} : \mathcal{C}_{\eta''\to\eta'} \times \mathcal{C}_{\eta'\to\eta} \to \mathcal{C}_{\eta''\to\eta}$ are the projection maps of the Cartesian product), and, for all $z, y, x \in \mathcal{C}_{\eta''\to\eta'} \times \mathcal{C}_{\eta'\to\eta} \times \mathcal{C}_\eta$:

$$\left[T_{\varphi_{\eta''\to\eta'\to\eta}^{-1}(z,y),x}\varphi_{\eta''\to\eta}^{-1}\right](\cdot, 0) \circ \left[T_{z,y}\varphi_{\eta''\to\eta'\to\eta}^{-1}\right](\cdot, 0) = \left[T_{z,\varphi_{\eta'\to\eta}^{-1}(y,x)}\varphi_{\eta''\to\eta'}^{-1}\right](\cdot, 0),$$

$$\left[T_{\varphi_{\eta''\to\eta'\to\eta}^{-1}(z,y),x}\varphi_{\eta''\to\eta}^{-1}\right](\cdot, 0) \circ \left[T_{z,y}\varphi_{\eta''\to\eta'\to\eta}^{-1}\right](0, \cdot) = \left[T_{z,\varphi_{\eta'\to\eta}^{-1}(y,x)}\varphi_{\eta''\to\eta'}^{-1}\right](0, \cdot) \circ \left[T_{y,x}\varphi_{\eta'\to\eta}^{-1}\right](\cdot, 0),$$

$$\&\quad \left[T_{\varphi_{\eta''\to\eta'\to\eta}^{-1}(z,y),x}\varphi_{\eta''\to\eta}^{-1}\right](0, \cdot) = \left[T_{z,\varphi_{\eta'\to\eta}^{-1}(y,x)}\varphi_{\eta''\to\eta'}^{-1}\right](0, \cdot) \circ \left[T_{y,x}\varphi_{\eta'\to\eta}^{-1}\right](0, \cdot),$$

therefore eq. (2.11.1) is fulfilled for the maps $\widetilde{\varphi}$. $\qquad\square$

**Proposition 2.17** Let $(\mathcal{L}, \mathcal{M}, \pi)^\downarrow$ be a projective system of phase spaces and suppose that:
1. $\forall \eta \in \mathcal{L}$, $\mathcal{M}_\eta = T^*(\mathcal{C}_\eta)$ where $\mathcal{C}_\eta$ is a smooth connected manifold;

2. $\forall \eta \prec \eta' \in \mathcal{L}$, there exist a smooth manifold $\mathcal{C}_{\eta' \to \eta}$ and a diffeomorphism $\varphi_{\eta' \to \eta} : \mathcal{C}_{\eta'} \to \mathcal{C}_{\eta' \to \eta} \times \mathcal{C}_{\eta}$ such that $\pi_{\eta' \to \eta} = \widetilde{s}_{\eta' \to \eta} \circ \widetilde{\varphi}_{\eta' \to \eta}$, where $\widetilde{s}_{\eta' \to \eta} : T^* \left( \mathcal{C}_{\eta' \to \eta} \times \mathcal{C}_{\eta} \right) \simeq T^* \left( \mathcal{C}_{\eta' \to \eta} \right) \times T^* \left( \mathcal{C}_{\eta} \right) \to T^* \left( \mathcal{C}_{\eta} \right)$ is the projection on the second Cartesian factor and $\widetilde{\varphi}_{\eta' \to \eta} : T^* \left( \mathcal{C}_{\eta'} \right) \to T^* \left( \mathcal{C}_{\eta' \to \eta} \times \mathcal{C}_{\eta} \right)$ is the cotangent lift of $\varphi_{\eta' \to \eta}$.

Then, we can complete this input into a factorizing system $(\mathcal{L}, \mathcal{C}, \varphi)^\times$.

**Proof** For $\eta \in \mathcal{L}$, we define $\mathcal{C}_{\eta \to \eta}$ to be a space with one element and $\varphi_{\eta \to \eta}$ to be the trivial identification.

Let $\eta \preccurlyeq \eta' \preccurlyeq \eta'' \in \mathcal{L}$. What we need to show is that there exists a diffeomorphism $\varphi_{\eta'' \to \eta' \to \eta} : \mathcal{C}_{\eta'' \to \eta} \to \mathcal{C}_{\eta'' \to \eta'} \times \mathcal{C}_{\eta' \to \eta}$ such that eq. (2.11.1) is fulfilled. If two labels among $\eta, \eta', \eta''$ are equal, we can choose $\varphi_{\eta'' \to \eta' \to \eta}$ to be the trivial identification, so we now consider the case $\eta \prec \eta' \prec \eta''$.

Let $x'', p'' \in T^*(\mathcal{C}_{\eta'})$ and define:

$$(y', q'; x', p') = \widetilde{\varphi}_{\eta'' \to \eta'}(x'', p''),$$

$$(y, q; x, p) = \widetilde{\varphi}_{\eta' \to \eta}(x', p'),$$

and $(z, r; x^\bullet, p^\bullet) = \widetilde{\varphi}_{\eta'' \to \eta}(x'', p'')$.

Now, from $\pi_{\eta'' \to \eta} = \pi_{\eta' \to \eta} \circ \pi_{\eta'' \to \eta'}$, we have $x = x^\bullet$ and $p = p^\bullet$, hence:

$$s_{\eta' \to \eta} \circ \varphi_{\eta' \to \eta} \circ s_{\eta'' \to \eta'} \circ \varphi_{\eta'' \to \eta'}(x'') = s_{\eta'' \to \eta} \circ \varphi_{\eta'' \to \eta}(x''),$$

and $p'' \circ \left[ T_{y', x'} \varphi_{\eta'' \to \eta'}^{-1} \right](0, \cdot) \circ \left[ T_{y, x} \varphi_{\eta' \to \eta}^{-1} \right](0, \cdot) = p'' \circ \left[ T_{z, x} \varphi_{\eta'' \to \eta}^{-1} \right](0, \cdot),$

thus, we get:

$$t_{\eta'' \to \eta' \to \eta} \circ \Psi = s_{\eta'' \to \eta},$$

and $(0, 0, \cdot) = [T\Psi](0, \cdot),$

where $\Psi := \left( \mathrm{id}_{\mathcal{C}_{\eta'' \to \eta'}} \times \varphi_{\eta' \to \eta} \right) \circ \varphi_{\eta'' \to \eta'} \circ \varphi_{\eta'' \to \eta}^{-1}$ and $t_{\eta'' \to \eta' \to \eta} : \mathcal{C}_{\eta'' \to \eta'} \times \mathcal{C}_{\eta' \to \eta} \times \mathcal{C}_{\eta} \to \mathcal{C}_{\eta}$ is the projection on the third Cartesian factor.

Finally, since $\mathcal{C}_{\eta}$ is connected, there exists a diffeomorphism $\varphi_{\eta'' \to \eta' \to \eta} : \mathcal{C}_{\eta'' \to \eta} \to \mathcal{C}_{\eta'' \to \eta'} \times \mathcal{C}_{\eta' \to \eta}$ such that $\Psi = \varphi_{\eta'' \to \eta' \to \eta} \times \mathrm{id}_{\mathcal{C}_{\eta}}$. $\square$

# 3 Constraints and regularization

When we try to incorporate the dynamics in the formalism described in the previous section, we quickly realize that the intuitive picture we were relying on was quite oversimplified. For, although it should be true that we only need a finite dimensional truncation of the kinematical theory to hold the elementary kinematical observables associated to any given real experiment, in general we cannot write the dynamics in a closed form within such a truncation.

As developed in appendix A, we take the point of view that from each kinematical observable arises a corresponding dynamical observable and, considering a family of functionally independent kinematical observables, it might be possible to write functional relations connecting the associated dynamical observables: here lies the predictive contents of the theory. However, such a functional relation can involve an infinite number of observables, and thus get silently dropped, if we never look at more than a finite number of observables at a time. When looking at a typical field theory, the interesting content of the dynamics lies precisely in those functional relations that can only be written over an infinite number of observables, and do not emerge from simpler relations within finite set of observables (a partial differential equation is mostly useless if we only dispose of a discrete, finite set of initial values).

On the other hand, if the the theory is to have any physically relevant predictivity, namely if it is to be usable to formulate predictions for the output of some real experiments, it should at least be possible to *approximate* the dynamics with relations over finite sets of elementary dynamical observables (we do nothing else when elaborating numerical techniques to deal with partial differential equations). In other words, although we may not be able to state *exact* predictions for any specific realistic experiment, we can restore predictivity in a weaker sense, by describing how to refine an experiment and the associated approximate predictions to make them better and better.

This concept of convergence is physically useful, notwithstanding the fact that we will *not* perform the infinite chain of experiments (that would again be a case of measuring an infinite number of observables, and we already mentioned that this is excluded in practice), because we can convert it into a notion of *plausibility*, by stating how to design an experimental protocol such that it will be highly unlikely that the output lies outside some confidence domain.

The object of this section is to formulate this raw idea more precisely, in order to develop a procedure to solve constraints in a projective system of phase spaces.

## 3.1 Elementary reductions

We begin by studying in detail under which conditions the dynamics actually *can* be formulated straightforwardly within a projective system of phase spaces, for this will be our building block when addressing the generic case.

Our aim here is the following: we want to write in each partial kinematical theory $\mathcal{M}_\eta^{\text{KIN}}$ a constraint surface $\mathcal{M}_\eta^{\text{SHELL}}$, and to reassemble the resulting reduced phase spaces $\mathcal{M}_\eta^{\text{DYN}}$ (see appendix A) into a new projective system of phase spaces. And we want to accomplish this in such a way that we can glue together the maps that, for each $\eta$, associate to the kinematical observables on $\mathcal{M}_\eta^{\text{KIN}}$ the corresponding dynamical observables on $\mathcal{M}_\eta^{\text{DYN}}$, thus building a map from the set of all observables on the kinematical projective system into the set of all observables on the dynamical projective system. For this map to accurately reproduce a given dynamics, it should give rise to functional relations between the dynamical observables that catch the full predictive power of the theory and it should account for the correct dynamical Poisson commutation relations.

We start by looking at a symplectic manifold $\mathcal{N}^{\text{KIN}}$, that extracts, via a projection $\pi^{\text{KIN}}$, specific

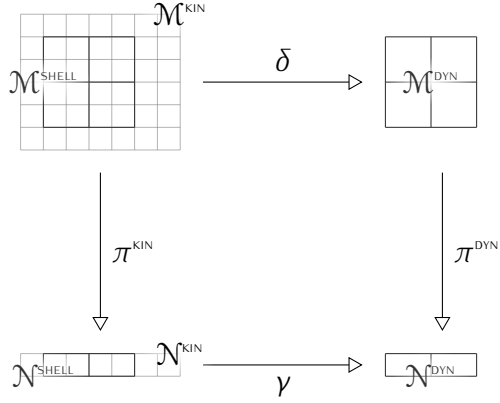

Here, we sketch a symplectic manifold by a grid, each square of which is to be thought as a point in the manifold (and is emblematic for infinitely many other points).

Figure 3.1 – Phase space reductions on $\mathcal{M}^{\text{KIN}}$ and $\mathcal{N}^{\text{KIN}}$, related by a projection $\pi^{\text{KIN}}$

degrees of freedom out of a bigger symplectic manifold $\mathcal{M}^{\text{KIN}}$ (as were introduced in def. 2.1). Given a phase space reduction on $\mathcal{M}^{\text{KIN}}$, with reduced phase space $\mathcal{M}^{\text{DYN}}$, we ask whether it is possible to write closed equations, involving only the degrees of freedom retained in $\mathcal{N}^{\text{KIN}}$, and capturing all what the dynamics on $\mathcal{M}^{\text{KIN}}$ has to say concerning these degrees of freedom.

More precisely, we are looking for a phase space reduction on $\mathcal{N}^{\text{KIN}}$, but also for a projection $\pi^{\text{DYN}}$ allowing to understand the reduced phase space $\mathcal{N}^{\text{DYN}}$ as a selection of dynamical degrees of freedom out of $\mathcal{M}^{\text{DYN}}$ (fig. 3.1). Indeed, if we consider an observable $O^{\text{KIN}}$ on $\mathcal{N}^{\text{KIN}}$, we can pull it back by $\pi^{\text{KIN}}$ into an observable $O^{\text{KIN}\prime}$ on $\mathcal{M}^{\text{KIN}}$. So using the dynamics on $\mathcal{M}^{\text{KIN}}$, we can obtain a corresponding dynamical observable $O^{\text{DYN}\prime}$ on $\mathcal{M}^{\text{DYN}}$. Now, if we can write the dynamics in closed form on $\mathcal{N}^{\text{KIN}}$, we can also map directly $O^{\text{KIN}}$ to a dynamical observable $O^{\text{DYN}}$ on $\mathcal{N}^{\text{DYN}}$. The role of the projection $\pi^{\text{DYN}}$ is then to ensure that the dynamics we have on $\mathcal{N}$ is actually consistent with the one on $\mathcal{M}$, by requiring $O^{\text{DYN}\prime}$ to be precisely the pullback of $O^{\text{DYN}}$ by $\pi^{\text{DYN}}$.

If this is at all possible, both the reduction on $\mathcal{N}^{\text{KIN}}$ and the projection $\pi^{\text{DYN}}$ are uniquely determined by the dynamics we choose on $\mathcal{M}^{\text{KIN}}$. Indeed, the constraint surface in $\mathcal{N}^{\text{KIN}}$ has to be the projection by $\pi^{\text{KIN}}$ of the one in $\mathcal{M}^{\text{KIN}}$ (for the constraint surface can be reconstructed if we know which kinematical observables are mapped to a vanishing dynamical observables), and we have from prop. A.6 (at least in the finite dimensional case) that a reduction is completely determined by its constraint surface. Then, the uniqueness of $\pi^{\text{DYN}}$ is enforced by requiring that it correctly makes the connection between the dynamics on $\mathcal{N}^{\text{KIN}}$ and the aforementioned map $O^{\text{KIN}} \mapsto O^{\text{DYN}\prime}$ (that only depends of $\pi^{\text{KIN}}$ and of the reduction on $\mathcal{M}^{\text{KIN}}$).

**Definition 3.1** Let $\mathcal{M}^{\text{KIN}}$ and $\mathcal{N}^{\text{KIN}}$ be two symplectic manifolds and $\pi^{\text{KIN}} : \mathcal{M}^{\text{KIN}} \to \mathcal{N}^{\text{KIN}}$ a surjective map compatible with the symplectic structures (def. 2.1). Let $(\mathcal{M}^{\text{DYN}}, \mathcal{M}^{\text{SHELL}}, \delta)$, resp. $(\mathcal{N}^{\text{DYN}}, \mathcal{N}^{\text{SHELL}}, \gamma)$, be phase space reductions of $\mathcal{M}^{\text{KIN}}$, resp. $\mathcal{N}^{\text{KIN}}$ (def. A.1). We say that these reductions are related by $\pi^{\text{KIN}}$ if:

1. $\pi^{\text{KIN}} \langle \mathcal{M}^{\text{SHELL}} \rangle = \mathcal{N}^{\text{SHELL}}$;

2. there exists a surjective map $\pi^{\text{DYN}} : \mathcal{M}^{\text{DYN}} \to \mathcal{N}^{\text{DYN}}$, compatible with the symplectic structures, such that:

$$\forall x \in \mathcal{N}^{\text{SHELL}}, \forall y' \in \mathcal{M}^{\text{DYN}}, \; \left(\exists x' \in \mathcal{M}^{\text{SHELL}} / \delta(x') = y' \; \& \; \pi^{\text{KIN}}(x') = x\right) \Leftrightarrow \left(\gamma(x) = \pi^{\text{DYN}}(y')\right). \quad (3.1.1)$$

**Proposition 3.2** With the notations of def. 3.1, if $\pi^{\text{DYN},1}$ and $\pi^{\text{DYN},2}$ are two surjective maps satisfying eq. (3.1.1), then $\pi^{\text{DYN},1} = \pi^{\text{DYN},2}$.

**Proof** Let $y' \in \mathcal{M}^{\text{DYN}}$. Since $\delta$ is surjective, there exists $x' \in \mathcal{M}^{\text{SHELL}}$ such that $\delta(x') = y'$. Hence, $\pi^{\text{DYN},1}(y') = \gamma \circ \pi^{\text{KIN}}(x') = \pi^{\text{DYN},2}(y')$. □

**Proposition 3.3** We consider the same objects as in def. 3.1 and use the notations introduced in def. A.2. For $f \in B(\mathcal{N}^{\text{KIN}})$, we have $f \circ \pi^{\text{KIN}} \in B(\mathcal{M}^{\text{KIN}})$ and:

$$(f \circ \pi^{\text{KIN}})^{\text{DYN}} = f^{\text{DYN}} \circ \pi^{\text{DYN}}.$$

**Proof** Let $y' \in \mathcal{M}^{\text{DYN}}$. Using eq. (3.1.1) into eq. (A.2.1), we have:

$$(f \circ \pi^{\text{KIN}})^{\text{DYN}}(y') = \sup \left\{ f \circ \pi^{\text{KIN}}(x') \mid x' \in \delta^{-1}\langle y' \rangle \right\}$$

$$= \sup \left\{ f(x) \mid x \in \gamma^{-1}\langle \pi^{\text{DYN}}(y') \rangle \right\} = f^{\text{DYN}} \circ \pi^{\text{DYN}}(y').$$

□

Why do we need to require eq. (3.1.1) for $\pi^{\text{DYN}}$ instead of the seemingly more natural condition $\gamma \circ \pi^{\text{KIN}} = \pi^{\text{DYN}} \circ \delta$? The physical reason behind eq. (3.1.1) is that we shall not look at the map $\delta$ but rather at $\delta^{-1}\langle \cdot \rangle$, that sends a point in $\mathcal{M}^{\text{DYN}}$ to an orbit in $\mathcal{M}^{\text{SHELL}}$ (and similarly at $\gamma^{-1}\langle \cdot \rangle$ instead of $\gamma$), for this is the map that is dual to the application associating a kinematical observable to a dynamical one (in a way similar to $\pi^{\text{KIN}}$ being dual to the application that sends an observable on $\mathcal{N}^{\text{KIN}}$ into an observable on $\mathcal{M}^{\text{KIN}}$). And, indeed, we can rewrite eq. (3.1.1) as $\pi^{\text{KIN}}\langle \cdot \rangle \circ \delta^{-1}\langle \cdot \rangle = \gamma^{-1}\langle \cdot \rangle \circ \pi^{\text{DYN}}$.

That eq. (3.1.1) could fail in situations where $\gamma \circ \pi^{\text{KIN}} = \pi^{\text{DYN}} \circ \delta$ does hold, can have local as well as global causes, as illustrated by the examples below. It happens when the projection of an orbit in $\mathcal{M}^{\text{SHELL}}$, though included in an orbit of $\mathcal{N}^{\text{SHELL}}$, does not *fill* it.

**Proposition 3.4** If we replace in def. 3.1 the condition given by eq. (3.1.1) by the weaker assumption:

$$\gamma \circ \pi^{\text{KIN}} = \pi^{\text{DYN}} \circ \delta, \tag{3.4.1}$$

then the previous result (prop. 3.3) does not hold.

**Proof** As a counter example, we consider the following situation:

1. $\mathcal{M}^{\text{KIN}} = (\mathbb{R}^2)^3$, $\mathcal{M}^{\text{DYN}} = (\mathbb{R}^2)^2$, $\mathcal{N}^{\text{KIN}} = (\mathbb{R}^2)^2$, $\mathcal{N}^{\text{DYN}} = \mathbb{R}^2$ (with the standard symplectic structure on $\mathbb{R}^2$: $\Omega_{\mathbb{R}^2}(x, p; x', p') = x p' - x' p$);

2. $\forall (x_i, p_i)_{i \in \{0,\ldots,2\}} \in \mathcal{M}^{\text{KIN}}$, $\pi^{\text{KIN}}\left((x_i, p_i)_{i \in \{0,\ldots,2\}}\right) = (x_i, p_i)_{i \in \{0,1\}}$;

3. $\mathcal{M}^{\text{SHELL}} = \left\{ (x_i, p_i)_{i \in \{0,\ldots,2\}} \mid p_1 = 0 \ \& \ x_1 = x_2 \right\}$ and $\forall (x_i, p_i)_{i \in \{0,\ldots,2\}} \in \mathcal{M}^{\text{SHELL}}$, $\delta\left((x_i, p_i)_{i \in \{0,\ldots,2\}}\right) = (x_i, p_i)_{i \in \{0,2\}}$;

4. $\mathcal{N}^{\text{SHELL}} = \left\{ (x_i, p_i)_{i \in \{0,1\}} \mid p_1 = 0 \right\}$ and $\forall (x_i, p_i)_{i \in \{0,1\}} \in \mathcal{N}^{\text{SHELL}}$, $\gamma\left((x_i, p_i)_{i \in \{0,1\}}\right) = (x_0, p_0)$;

5. $\forall (x_i, p_i)_{i \in \{0,2\}} \in \mathcal{M}^{\text{DYN}}$, $\pi^{\text{DYN}}\left((x_i, p_i)_{i \in \{0,2\}}\right) = (x_0, p_0)$.

We can check that $(\mathcal{M}^{\text{DYN}}, \mathcal{M}^{\text{SHELL}}, \delta)$ is a phase space reduction of $\mathcal{M}^{\text{KIN}}$ and $(\mathcal{N}^{\text{DYN}}, \mathcal{N}^{\text{SHELL}}, \gamma)$ is a phase space reduction of $\mathcal{N}^{\text{KIN}}$. $\pi^{\text{KIN}}$ and $\pi^{\text{DYN}}$ are surjective maps compatible with the symplectic structures, satisfying $\pi^{\text{KIN}} \langle \mathcal{M}^{\text{SHELL}} \rangle = \mathcal{N}^{\text{SHELL}}$ and $\gamma \circ \pi^{\text{KIN}} = \pi^{\text{DYN}} \circ \delta$.

However, if we consider $f \in B(\mathcal{N}^{\text{KIN}})$ defined by:

$$\forall (x_i, p_i)_{i \in \{0,1\}} \in \mathcal{N}^{\text{KIN}}, \ f\left((x_i, p_i)_{i \in \{0,1\}}\right) = \begin{cases} 1 \text{ if } x_1 \leqslant 0 \\ 0 \text{ else} \end{cases},$$

we have $f^{\text{DYN}} \circ \pi^{\text{DYN}} \equiv 1$, but:

$$\forall (x_i, p_i)_{i \in \{0,2\}} \in \mathcal{M}^{\text{DYN}}, \ (f \circ \pi^{\text{KIN}})^{\text{DYN}} \left((x_i, p_i)_{i \in \{0,2\}}\right) = \begin{cases} 1 \text{ if } x_2 \leqslant 0 \\ 0 \text{ else} \end{cases}.$$

Requiring local conditions in addition to eq. (3.4.1) would not help either, since even if everything works well locally, it may still goes wrong globally, as the following example shows:

6. $\mathcal{M}^{\text{KIN}} = \left(\mathbb{R}^2\right)^4$, $\mathcal{M}^{\text{DYN}} = \left(\mathbb{R}^2\right)^2$, $\mathcal{N}^{\text{KIN}} = \left(\mathbb{R}^2\right)^2$, $\mathcal{N}^{\text{DYN}} = \mathbb{R}^2$;

7. $\forall (x_i, p_i)_{i \in \{0,\ldots,3\}} \in \mathcal{M}^{\text{KIN}}$, $\pi^{\text{KIN}}\left((x_i, p_i)_{i \in \{0,\ldots,3\}}\right) = (x_i, p_i)_{i \in \{0,1\}}$;

8. $\mathcal{M}^{\text{SHELL}} = \left\{(x_i, p_i)_{i \in \{0,\ldots,3\}} \mid p_1 = 0, \ p_2 = 0 \ \& \ x_1 = x_3 + \exp(x_2)\right\}$ and $\forall (x_i, p_i)_{i \in \{0,\ldots,3\}} \in \mathcal{M}^{\text{SHELL}}$, $\delta\left((x_i, p_i)_{i \in \{0,\ldots,3\}}\right) = (x_i, p_i)_{i \in \{0,3\}}$;

9. $\mathcal{N}^{\text{SHELL}} = \left\{(x_i, p_i)_{i \in \{0,1\}} \mid p_1 = 0\right\}$ and $\forall (x_i, p_i)_{i \in \{0,1\}} \in \mathcal{N}^{\text{SHELL}}$, $\gamma\left((x_i, p_i)_{i \in \{0,1\}}\right) = (x_0, p_0)$;

10. $\forall (x_i, p_i)_{i \in \{0,3\}} \in \mathcal{M}^{\text{DYN}}$, $\pi^{\text{DYN}}\left((x_i, p_i)_{i \in \{0,3\}}\right) = (x_0, p_0)$.

We can check that $(\mathcal{M}^{\text{DYN}}, \mathcal{M}^{\text{SHELL}}, \delta)$ is a phase space reduction of $\mathcal{M}^{\text{KIN}}$ and $(\mathcal{N}^{\text{DYN}}, \mathcal{N}^{\text{SHELL}}, \gamma)$ is a phase space reduction of $\mathcal{N}^{\text{KIN}}$. $\pi^{\text{KIN}}$ and $\pi^{\text{DYN}}$ are surjective maps compatible with the symplectic structures, satisfying $\pi^{\text{KIN}} \langle \mathcal{M}^{\text{SHELL}} \rangle = \mathcal{N}^{\text{SHELL}}$ and $\gamma \circ \pi^{\text{KIN}} = \pi^{\text{DYN}} \circ \delta$.

Moreover, eq. (3.1.1) holds at the linear level, namely:

$\forall x' \in \mathcal{M}^{\text{SHELL}}, \quad \forall v \in T_{\pi^{\text{KIN}}(x')}(\mathcal{N}^{\text{SHELL}}), \forall w' \in T_{\delta(x')}(\mathcal{M}^{\text{DYN}}),$

$$\left(\exists v' \in T_{x'}(\mathcal{M}^{\text{SHELL}}) \ / \ T_{x'}\delta(v') = w' \ \& \ T_{x'}\pi^{\text{KIN}}(v') = v\right) \Leftrightarrow \left(T_{\pi^{\text{KIN}}(x')}\gamma(v) = T_{\delta(x')}\pi^{\text{DYN}}(w')\right),$$

for this reduces in the present example to:

$$\forall x_2 \in \mathbb{R}, \ \forall x_1^v \in \mathbb{R}, \forall x_3^{w'} \in \mathbb{R}, \ \left(\exists x_2^{v'}, x_3^{v'} \in \mathbb{R}^2 \ / \ x_3^{v'} = x_3^{w'} \ \& \ x_3^{v'} + \exp(x_2) x_2^{v'} = x_1^v\right).$$

However, if we consider the same $f \in B(\mathcal{N}^{\text{KIN}})$ as before, we have $f^{\text{DYN}} \circ \pi^{\text{DYN}} \equiv 1$, but:

$$\forall (x_i, p_i)_{i \in \{0,3\}} \in \mathcal{M}^{\text{DYN}}, \ (f \circ \pi^{\text{KIN}})^{\text{DYN}} \left((x_i, p_i)_{i \in \{0,3\}}\right) = \begin{cases} 1 \text{ if } x_3 \leqslant 0 \\ 0 \text{ else} \end{cases}.$$

$\square$

Asking for the dynamics on $\mathcal{M}^{\text{KIN}}$ to define a dynamics on $\mathcal{N}^{\text{KIN}}$ in the sense above actually puts strong restrictions (local as well as global ones) on what the constraint surface in $\mathcal{M}^{\text{KIN}}$ can be.

If we consider the special case where $\mathcal{M}^{\text{KIN}}$ and $\mathcal{N}^{\text{KIN}}$ are symplectic vector spaces, and $\pi^{\text{KIN}}$ is a linear map, the symplectic structure provides a natural decomposition of $\mathcal{M}^{\text{KIN}}$ as $\mathcal{P}^{\text{KIN}} \oplus (\mathcal{P}^{\text{KIN}})^\perp$, with $\mathcal{P}^{\text{KIN}} = \operatorname{Ker} \pi^{\text{KIN}}$ and $(\mathcal{P}^{\text{KIN}})^\perp \approx \mathcal{N}^{\text{KIN}}$ (where the orthogonal subspace is defined with respect to the symplectic structure; this is the linear version of prop. 2.10). What are the conditions for a vector subspace $\mathcal{M}^{\text{SHELL}}$ of $\mathcal{M}^{\text{KIN}}$ to define a (linear) dynamics that will descend well through $\pi^{\text{KIN}}$? An obvious way of fulfilling this wish is to have a constraint surface $\mathcal{M}^{\text{SHELL}}$ that decomposes as $\mathcal{M}^{\text{SHELL}} = W \oplus V$ where $W$ and $V$ are vector subsets of $\mathcal{P}^{\text{KIN}}$ and $(\mathcal{P}^{\text{KIN}})^\perp$ respectively: this would be a dynamics with no interaction between the degrees of freedom in $\mathcal{N}^{\text{KIN}}$ and the ones in $\mathcal{P}^{\text{KIN}}$, so clearly we can write separately the dynamics on $\mathcal{N}^{\text{KIN}}$. However, a closer study of what is really needed shows that we have an additional freedom to construct admissible constraint surfaces $\mathcal{M}^{\text{SHELL}}$: instead of choosing $V$ as a vector subset of $(\mathcal{P}^{\text{KIN}})^\perp$, it is enough for $V$ to be included in $W^\perp$, provided $\pi^{\text{KIN}}$ identifies the restriction to $V$ of the symplectic structure $\Omega_{\mathcal{M}^{\text{KIN}}}$ with the restriction to $\mathcal{N}^{\text{SHELL}} = \pi^{\text{KIN}} \langle V \rangle$ of $\Omega_{\mathcal{N}^{\text{KIN}}}$.

This study of the linear case essentially translates to local necessary conditions in the generic case. However, this holds only at the points in $\mathcal{M}^{\text{SHELL}}$ where the derivative of $\pi^{\text{KIN}}$ maps the tangent space of the orbit of $\mathcal{M}^{\text{SHELL}}$ going through that point into the tangent space of an orbit of $\mathcal{N}^{\text{SHELL}}$: for, although eq. (3.1.1) implies that $\pi^{\text{KIN}}$ should map an orbit into an orbit, this does not need to hold at the linear level (the derivative of a surjective map does not need to be surjective; nonetheless Sard's theorem [16] tells us, in a specific sense, that this 'rarely' fails).

**Proposition 3.5** Let $\mathcal{M}^{\text{KIN}}$ and $\mathcal{N}^{\text{KIN}}$ be two finite dimensional symplectic manifolds and $\pi^{\text{KIN}} : \mathcal{M}^{\text{KIN}} \to \mathcal{N}^{\text{KIN}}$ a surjective map compatible with the symplectic structures. Let $(\mathcal{M}^{\text{DYN}}, \mathcal{M}^{\text{SHELL}}, \delta)$, resp. $(\mathcal{N}^{\text{DYN}}, \mathcal{N}^{\text{SHELL}}, \gamma)$, be phase space reductions of $\mathcal{M}^{\text{KIN}}$, resp. $\mathcal{N}^{\text{KIN}}$. Assume these reductions are related by $\pi^{\text{KIN}}$, and let $x' \in \mathcal{M}^{\text{SHELL}}$, $y' := \delta(x') \in \mathcal{M}^{\text{DYN}}$, $x := \pi^{\text{KIN}}(x') \in \mathcal{N}^{\text{SHELL}}$ and $y := \gamma(x) \in \mathcal{N}^{\text{DYN}}$. Then, $\pi^{\text{KIN}}$ induces a surjective map $\delta^{-1}\langle y' \rangle \to \gamma^{-1}\langle y \rangle$.

If moreover $T_{x'}\pi^{\text{KIN}} \langle T_{x'}(\delta^{-1}\langle y' \rangle) \rangle = T_x(\gamma^{-1}\langle y \rangle)$, then there exist $V_{x'}$, $W_{x'}$ vector subspaces of $T_{x'}(\mathcal{M}^{\text{SHELL}})$ such that:

1. $T_{x'}(\mathcal{M}^{\text{SHELL}}) = V_{x'} \oplus W_{x'}$ & $\Omega_{\mathcal{M}^{\text{KIN}},x'}(V_{x'}, W_{x'}) = \{0\}$;

2. $V_{x'} \cap \operatorname{Ker} T_{x'}\pi^{\text{KIN}} = \{0\}$ & $W_{x'} \subset \operatorname{Ker} T_{x'}\pi^{\text{KIN}}$;

3. $\pi_{x'}^{\text{KIN},*} \Omega_{\mathcal{N}^{\text{KIN}},x}\big|_{V_{x'}} = \Omega_{\mathcal{M}^{\text{KIN}},x'}\big|_{V_{x'}}$.

**Proof** Let $\pi^{\text{DYN}}$ be as in def. 3.1.2. From eq. (3.1.1), we have $\gamma \circ \pi^{\text{KIN}} = \pi^{\text{DYN}} \circ \delta$, hence $y = \pi^{\text{DYN}}(y')$, and:

$$\forall z \in \mathcal{N}^{\text{SHELL}}, \ \left(z \in \gamma^{-1}\langle y \rangle\right) \Leftrightarrow \left(\gamma(z) = \pi^{\text{DYN}}(y')\right) \Leftrightarrow \left(\exists z' \in \delta^{-1}\langle y' \rangle \ / \ \pi^{\text{KIN}}(z') = z\right)$$

$$\Leftrightarrow \left(z \in \pi^{\text{KIN}}\langle \delta^{-1}\langle y' \rangle \rangle\right),$$

therefore $\pi^{\text{KIN}}\langle \delta^{-1}\langle y' \rangle \rangle = \gamma^{-1}\langle y \rangle$.

We now moreover assume that $T_{x'}\pi^{\text{KIN}}$ induces a surjective linear map from $\operatorname{Ker} T_{x'}\delta = T_{x'}(\delta^{-1}\langle y' \rangle)$ into $\operatorname{Ker} T_x\gamma = T_x(\gamma^{-1}\langle y \rangle)$. Then, there exist vector subspaces $V^o_{x'}$ and $W^o_{x'}$ of $\operatorname{Ker} T_{x'}\delta$ such that:

$$W^o_{x'} = \operatorname{Ker} T_{x'}\pi^{\text{KIN}} \cap \operatorname{Ker} T_{x'}\delta \quad \& \quad \operatorname{Ker} T_{x'}\delta = V^o_{x'} \oplus W^o_{x'},$$

and $T_{x'}\pi^{\text{KIN}}$ induces a bijection $V^o_{x'} \to \operatorname{Ker} T_x\gamma$.

Next, we define the vector subspaces $V^1_{y'}$ and $W^1_{y'}$ of $T_{y'}(\mathcal{M}^{\text{DYN}})$ by:

$$W^1_{y'} := \operatorname{Ker} T_{y'}\pi^{\text{DYN}} \quad \& \quad V^1_{y'} := \left(W^1_{y'}\right)^\perp = \left\{v \in T_{y'}(\mathcal{M}^{\text{DYN}}) \mid \Omega_{\text{DYN},y'}\left(v, W^1_{y'}\right) = \{0\}\right\},$$

and since $\pi^{\text{DYN}}$ is compatible with the symplectic structures, we have $T_{y'}(\mathcal{M}^{\text{DYN}}) = V^1_{y'} \oplus W^1_{y'}$ and $T_{y'}\pi^{\text{DYN}}$ being surjective, it induces a bijection $V^1_{y'} \to T_y(\mathcal{N}^{\text{DYN}})$, such that:

$$\pi^{\text{DYN},*}_{y'} \Omega_{\mathcal{N}^{\text{DYN}},y} \big|_{V^1_{y'}} = \Omega_{\mathcal{M}^{\text{DYN}},y'} \big|_{V^1_{y'}}.$$

Let $\widetilde{W}^2_{x'} := [T_{x'}\delta]^{-1} \langle W^1_{y'} \rangle$. We have $\operatorname{Ker} T_{x'}\delta \subset \widetilde{W}^2_{x'}$, and from $T_x\gamma \circ T_{x'}\pi^{\text{KIN}} \langle \widetilde{W}^2_{x'} \rangle = T_{y'}\pi^{\text{DYN}} \circ T_{x'}\delta \langle \widetilde{W}^2_{x'} \rangle = \{0\}$ and $\operatorname{Ker} T_x\gamma = T_{x'}\pi^{\text{KIN}} \langle \operatorname{Ker} T_{x'}\delta \rangle \subset T_{x'}\pi^{\text{KIN}} \langle \widetilde{W}^2_{x'} \rangle$, we also have $T_{x'}\pi^{\text{KIN}} \langle \widetilde{W}^2_{x'} \rangle = \operatorname{Ker} T_x\gamma$, hence there exists a vector subspace $W^2_{x'}$ of $\widetilde{W}^2_{x'}$ such that:

$$W^o_{x'} \oplus W^2_{x'} = \operatorname{Ker} T_{x'}\pi^{\text{KIN}} \cap \widetilde{W}^2_{x'} \quad \& \quad \widetilde{W}^2_{x'} = V^o_{x'} \oplus W^o_{x'} \oplus W^2_{x'},$$

and $T_{x'}\delta \langle W^2_{x'} \rangle = T_{x'}\delta \langle \widetilde{W}^2_{x'} \rangle = W^1_{y'}$ for $T_{x'}\delta$ is surjective. Additionally, since $T_{x'}\delta$ is surjective, there exists a vector subspace $V^2_{x'}$ of $T_{x'}(\mathcal{M}^{\text{SHELL}})$ such that:

$$T_{x'}(\mathcal{M}^{\text{SHELL}}) = V^o_{x'} \oplus W^o_{x'} \oplus V^2_{x'} \oplus W^2_{x'},$$

with $T_{x'}\delta$ inducing a bijective map $V^2_{x'} \to V^1_{y'}$. So $T_x\gamma \circ T_{x'}\pi^{\text{KIN}} = T_{y'}\pi^{\text{DYN}} \circ T_{x'}\delta$ induce a bijective map $V^2_{x'} \to T_y(\mathcal{N}^{\text{DYN}}) = T_x\gamma \langle T_x(\mathcal{N}^{\text{SHELL}}) \rangle$, therefore $T_{x'}\pi^{\text{KIN}}$ induce a bijective map $V^o_{x'} \oplus V^2_{x'} \to T_x(\mathcal{N}^{\text{SHELL}})$, such that, for all $u, v \in V^o_{x'} \oplus V^2_{x'}$:

$$\Omega_{\mathcal{N}^{\text{KIN}},x}\left(T_{x'}\pi^{\text{KIN}}(u), T_{x'}\pi^{\text{KIN}}(v)\right) = \Omega_{\mathcal{N}^{\text{DYN}},y}\left(T_x\gamma \circ T_{x'}\pi^{\text{KIN}}(u), T_x\gamma \circ T_{x'}\pi^{\text{KIN}}(v)\right)$$

$$= \Omega_{\mathcal{N}^{\text{DYN}},y}\left(T_{y'}\pi^{\text{DYN}} \circ T_{x'}\delta(u), T_{y'}\pi^{\text{DYN}} \circ T_{x'}\delta(v)\right)$$

$$= \Omega_{\mathcal{M}^{\text{DYN}},y'}\left(T_{x'}\delta(u), T_{x'}\delta(v)\right)$$

$$= \Omega_{\mathcal{M}^{\text{KIN}},x'}(u, v).$$

Finally, defining $V_{x'} := V^o_{x'} \oplus V^2_{x'}$ and $W_{x'} := W^o_{x'} \oplus W^2_{x'}$, we have:

$$\Omega_{\text{KIN},x'}(V_{x'}, W_{x'}) = \Omega_{\text{KIN},x'}\left(V^2_{x'}, W^2_{x'}\right) \text{ (for } \Omega_{\text{KIN},x'}(T_{x'}(\mathcal{M}^{\text{SHELL}}), \operatorname{Ker} T_{x'}\delta) = \{0\})$$

$$= \Omega_{\text{DYN},y'}\left(V^1_{y'}, W^1_{y'}\right) \text{ (for } \Omega_{\text{KIN},x'}\big|_{T_{x'}(\mathcal{M}^{\text{SHELL}})} = \delta^*_{x'}\Omega_{\text{DYN},y'})$$

$$= \{0\} \text{ (for } V^1_{y'} = \left(W^1_{y'}\right)^\perp),$$

and $W_{x'} \subset \operatorname{Ker} T_{x'}\pi^{\text{KIN}}$, while $\operatorname{Ker} T_{x'}\pi^{\text{KIN}} \cap V_{x'} = \{0\}$. $\square$

Returning to the linear case previously mentioned, we can reformulate in terms of constraints the condition we had for $\mathcal{M}^{\text{SHELL}}$ to define a closed dynamics on $\mathcal{N}^{\text{KIN}}$ (through the straightforward duality between the description of $\mathcal{M}^{\text{SHELL}}$ as a vector subspace and its description by linear constraints). This provides a specification of $\mathcal{M}^{\text{SHELL}}$ as characterized by three sets of constraints $C^{\mathcal{P}}_i$, $C^{\mathcal{N}}_j$, and $C^{\text{mix}}_k$, where the $C^{\mathcal{P}}_i$ only depend on the variables from $\mathcal{P}^{\text{KIN}}$ and characterize in $\mathcal{P}^{\text{KIN}}$ the projection $\mathcal{P}^{\text{SHELL}}$ of $\mathcal{M}^{\text{SHELL}}$, similarly the $C^{\mathcal{N}}_j$ only depend on the variables from $\mathcal{N}^{\text{KIN}}$ and characterize $\mathcal{N}^{\text{SHELL}}$ in $\mathcal{N}^{\text{KIN}}$, while the $C^{\text{mix}}_k$ account for possible interactions. These interactions cannot be arbitrary: the requirements on $V$ discussed above prescribe that the constraints $C_k^{\text{mix},x}$, obtained on $\mathcal{P}^{\text{KIN}}$ from the $C_k^{\text{mix}}$ by fixing some $x \in \mathcal{N}^{\text{SHELL}}$, should perform a partial gauge fixing of $\mathcal{P}^{\text{SHELL}}$ (prop. A.8).

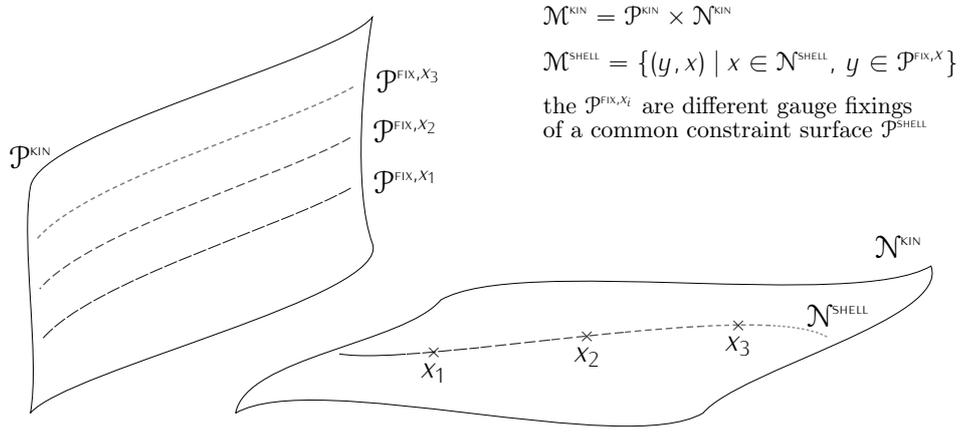

Figure 3.2 – A (rather broad but not exhaustive) way to construct an admissible dynamics on $\mathcal{M}^{\text{KIN}}$ in the factorizing case

In the generic case of a symplectic manifold $\mathcal{M}^{\text{KIN}}$ factorizing as $\mathcal{M}^{\text{KIN}} = \mathcal{P}^{\text{KIN}} \times \mathcal{N}^{\text{KIN}}$ (such as considered in subsection 2.3), the insight we gain from the linear case suggests a possibility, depicted in fig. 3.2, to design dynamics on $\mathcal{M}^{\text{SHELL}}$ that will project well on $\mathcal{N}^{\text{KIN}}$. This provides a much broader class of admissible dynamics than the trivial ones splitting into independent dynamics on $\mathcal{P}^{\text{KIN}}$ and $\mathcal{N}^{\text{KIN}}$.

Nevertheless, this procedure only corresponds to a *sufficient* condition for def. 3.1 to be fulfilled. Note that the gap between the necessary condition at the linear level supplied by prop. 3.5 and the characterization of $\mathcal{M}^{\text{SHELL}}$ considered here does not solely arise from global considerations: for $\mathcal{M}^{\text{SHELL}}$ to be of this form, some additional integrability conditions (ie. requirements at the second order) need to hold, so that we can combine the prescriptions in the tangent space of each point into prescriptions in small open patches.

**Proposition 3.6** Let $\mathcal{M}^{\text{KIN}} = \mathcal{P}^{\text{KIN}} \times \mathcal{N}^{\text{KIN}}$, where $\mathcal{M}^{\text{KIN}}$, $\mathcal{P}^{\text{KIN}}$ and $\mathcal{N}^{\text{KIN}}$ are finite dimensional symplectic manifolds, and define:

$$\pi^{\text{KIN}} : \mathcal{M}^{\text{KIN}} \to \mathcal{N}^{\text{KIN}}$$
$$y, x \mapsto x \qquad .$$

Let $(\mathcal{M}^{\text{DYN}}, \mathcal{M}^{\text{SHELL}}, \delta)$, $(\mathcal{P}^{\text{DYN}}, \mathcal{P}^{\text{SHELL}}, \theta)$, resp. $(\mathcal{N}^{\text{DYN}}, \mathcal{N}^{\text{SHELL}}, \gamma)$, be phase space reductions of $\mathcal{M}^{\text{KIN}}$, $\mathcal{P}^{\text{KIN}}$, resp. $\mathcal{N}^{\text{KIN}}$. Assume that there exist a submanifold $\mathcal{P}^{\text{FIX}}$ of $\mathcal{P}^{\text{SHELL}}$ and a smooth map:

$$\Psi : \mathcal{P}^{\text{FIX}} \times \mathcal{N}^{\text{SHELL}} \to \mathcal{P}^{\text{SHELL}} \times \mathcal{N}^{\text{SHELL}}$$
$$y, x \mapsto \psi(y, x), x \qquad ,$$

such that:

1. $\text{Im}\,\Psi = \mathcal{M}^{\text{SHELL}}$ and $\Psi|_{\mathcal{P}^{\text{FIX}} \times \mathcal{N}^{\text{SHELL}} \to \mathcal{M}^{\text{SHELL}}}$ is a diffeomorphism;

2. $\Psi^* \left( \Omega_{\mathcal{M}^{\text{KIN}}}|_{T(\mathcal{M}^{\text{SHELL}})} \right) = \Omega_{\mathcal{P}^{\text{KIN}}}|_{T(\mathcal{P}^{\text{FIX}})} \times \Omega_{\mathcal{N}^{\text{KIN}}}|_{T(\mathcal{N}^{\text{SHELL}})}$;

3. $\forall x \in \mathcal{N}^{\text{SHELL}}$, $\mathcal{P}^{\text{FIX},x} := \psi \langle \mathcal{P}^{\text{FIX}} \times \{x\} \rangle$ defines a partial gauge fixing of $(\mathcal{P}^{\text{DYN}}, \mathcal{P}^{\text{SHELL}}, \theta)$ (prop. A.8).

Then, $(\mathcal{M}^{\text{DYN}}, \mathcal{M}^{\text{SHELL}}, \delta)$ and $(\mathcal{N}^{\text{DYN}}, \mathcal{N}^{\text{SHELL}}, \gamma)$ are related by $\pi^{\text{KIN}}$.

**Proof** From the definition of $\Psi$, we have $\pi^{\text{KIN}} \langle \mathcal{M}^{\text{SHELL}} \rangle = \text{Im}\, \pi^{\text{KIN}} \circ \Psi = \mathcal{N}^{\text{SHELL}}$.

Let $x \in \mathcal{N}^{\text{SHELL}}$. Using assumption 3.6.1 together with the definition of $\mathcal{P}^{\text{FIX},x}$, the map:

$$\psi^x \,:\, \mathcal{P}^{\text{FIX}} \to \mathcal{P}^{\text{FIX},x}$$
$$y \mapsto \psi(y, x)$$

is a diffeomorphism and, by 3.6.2, it satisfies $\psi^{x,*}\left(\Omega_{\mathcal{P}^{\text{KIN}}}|_{T(\mathcal{P}^{\text{FIX},x})}\right) = \Omega_{\mathcal{P}^{\text{KIN}}}|_{T(\mathcal{P}^{\text{FIX}})}$.

Now, from prop. A.8, $\left(\mathcal{P}^{\text{DYN}}, \mathcal{P}^{\text{FIX},x}, \theta|_{\mathcal{P}^{\text{FIX},x}}\right)$ is a phase space reduction of $\mathcal{P}^{\text{KIN}}$. Hence, defining $\theta^{\text{FIX},x} := \theta|_{\mathcal{P}^{\text{FIX},x}} \circ \psi^x$, $(\mathcal{P}^{\text{DYN}}, \mathcal{P}^{\text{FIX}}, \theta^{\text{FIX},x})$ is a phase space reduction of $\mathcal{P}^{\text{KIN}}$.

Using 3.6.2, we have for all $x \in \mathcal{N}^{\text{SHELL}}$, $y \in \mathcal{P}^{\text{FIX}}$ and for all $v \in T_x(\mathcal{N}^{\text{SHELL}})$, $w \in T_y(\mathcal{P}^{\text{FIX}})$:

$$0 = \Omega_{\mathcal{M}^{\text{KIN}}}\left(\left(\left[T_{(y,x)}\psi\right](0,v), v\right), \left(\left[T_{(y,x)}\psi\right](w,0), 0\right)\right) = \Omega_{\mathcal{P}^{\text{KIN}}}\left(\left[T_{(y,x)}\psi\right](0,v), \left[T_{(y,x)}\psi\right](w,0)\right).$$

However, since $\psi^x$ is a diffeomorphism, $\left[T_{(y,x)}\psi\right](w,0)$ runs through $T_{\psi(y,x)}(\mathcal{P}^{\text{FIX},x})$ when $w$ runs through $T_y(\mathcal{P}^{\text{FIX}})$, so $\left[T_{(y,x)}\psi\right](0,v) \in \left(T_{\psi(y,x)}(\mathcal{P}^{\text{FIX},x})\right)^{\perp} \cap T_{\psi(y,x)}(\mathcal{P}^{\text{SHELL}})$. As $\mathcal{P}^{\text{FIX},x}$ defines a partial gauge fixing of $(\mathcal{P}^{\text{DYN}}, \mathcal{P}^{\text{SHELL}}, \theta)$, we have $T_{\psi(y,x)}(\mathcal{P}^{\text{SHELL}}) = T_{\psi(y,x)}(\mathcal{P}^{\text{FIX},x}) + K_{\psi(y,x)}(\mathcal{P}^{\text{SHELL}})$, hence $\left[T_{(y,x)}\psi\right](0,v) \in K_{\psi(y,x)}(\mathcal{P}^{\text{SHELL}})$. Therefore, $\partial_x \theta^{\text{FIX},x} = 0$.

Without loss of generality, we can assume that $\mathcal{N}^{\text{SHELL}}$ is connected (otherwise $\mathcal{M}^{\text{SHELL}}$ is not connected either and we can consider each connected part of $\mathcal{N}^{\text{SHELL}}$ separately). Then, we can define $\theta^{\text{FIX}} := \theta^{\text{FIX},x}$.

Using $\theta^{\text{FIX}}$, we define:

$$\widetilde{\delta} \,:\, \mathcal{M}^{\text{SHELL}} \to \mathcal{P}^{\text{DYN}} \times \mathcal{N}^{\text{DYN}}$$
$$y, x \mapsto (\theta^{\text{FIX}} \times \gamma) \circ \left(\Psi|_{\mathcal{P}^{\text{FIX}} \times \mathcal{N}^{\text{SHELL}} \to \mathcal{M}^{\text{SHELL}}}\right)^{-1}(y,x) = \left(\theta(y), \gamma(x)\right).$$

We want to prove that $\left(\mathcal{P}^{\text{DYN}} \times \mathcal{N}^{\text{DYN}}, \mathcal{M}^{\text{SHELL}}, \widetilde{\delta}\right)$ is a phase space reduction of $\mathcal{M}^{\text{KIN}}$.

First, we need to show that $\widetilde{\delta}$ is surjective, that its derivative is surjective at each point, and transports correctly the restriction to $\mathcal{M}^{\text{SHELL}}$ of the symplectic structure. Since $\left(\Psi|_{\mathcal{P}^{\text{SHELL}} \times \mathcal{N}^{\text{FIX}} \to \mathcal{M}^{\text{SHELL}}}\right)^{-1}$ is a diffeomorphism and transports the symplectic structure, we need only to check the corresponding properties of $\theta^{\text{FIX}} \times \gamma$. Now, since $\theta^{\text{FIX}}$ and $\gamma$ corresponds to phase space reductions, they indeed have the required properties, and so does $\theta^{\text{FIX}} \times \gamma$.

Let $(y,x) \in \mathcal{P}^{\text{FIX}} \times \mathcal{N}^{\text{SHELL}}$. We choose a basis $(e_i)_{i \leq k}$ of $K_x(\mathcal{N}^{\text{SHELL}})$ (with $k := \dim K_x(\mathcal{N}^{\text{SHELL}})$) and we complete it into a basis $(e_i)_{i \leq n}$ of $T_x(\mathcal{N}^{\text{SHELL}})$ (with $n := \dim T_x(\mathcal{N}^{\text{SHELL}})$). We also choose a basis $(f_j)_{j \leq l}$ of $K_y(\mathcal{P}^{\text{FIX}})$ (with $l := \dim K_y(\mathcal{P}^{\text{FIX}})$) and complete it into a basis $(f_j)_{j \leq p}$ of $T_y(\mathcal{P}^{\text{FIX}})$ (with $p := \dim T_y(\mathcal{P}^{\text{FIX}})$). Then, we have:

$$T_{\psi(y,x)}(\mathcal{M}^{\text{SHELL}}) = \text{Vect}\left\{\left(\left[T_{(y,x)}\psi\right](0,e_i), e_i\right) \,\Big|\, i \leq n\right\} + \text{Vect}\left\{\left(\left[T_{(y,x)}\psi\right](f_j,0), 0\right) \,\Big|\, j \leq p\right\}.$$

As proved above, we have $\forall v \in T_x(\mathcal{N}^{\text{SHELL}})$, $\left[T_{(y,x)}\psi\right](0,v) \in K_{\psi(y,x)}(\mathcal{P}^{\text{SHELL}})$. Since $\psi^x$ is a diffeomorphism $\mathcal{P}^{\text{FIX}} \to \mathcal{P}^{\text{FIX},x}$, the $\left[T_{(y,x)}\psi\right](f_j,0)$ for $j \leq p$ span $T_{\psi(y,x)}(\mathcal{P}^{\text{FIX},x})$. And since $\psi^x$ transports the symplectic structure, we also have $\forall w \in T_y(\mathcal{P}^{\text{FIX}})$, $\left[T_{(y,x)}\psi\right](w,0) \in K_{\psi(y,x)}(\mathcal{P}^{\text{FIX},x}) \Leftrightarrow w \in K_y(\mathcal{P}^{\text{FIX}})$.

Therefore, we have:

$$K_{\psi(y,x)}(\mathcal{M}^{\text{SHELL}}) = \text{Vect}\left\{\left([T_{(y,x)}\psi](0, e_i), e_i\right) \mid i \leqslant k\right\} + \text{Vect}\left\{\left([T_{(y,x)}\psi](f_j, 0), 0\right) \mid j \leqslant l\right\}.$$

Using $K_{\psi(y,x)}(\mathcal{P}^{\text{FIX},x}) \subset K_{\psi(y,x)}(\mathcal{P}^{\text{SHELL}})$, we can now check that $T_{\psi(y,x)}\widetilde{\delta}\left\langle K_{\psi(y,x)}(\mathcal{M}^{\text{SHELL}})\right\rangle = \{0\}$. Therefore, the leaf of the foliation $K(\mathcal{M}^{\text{SHELL}})$ that goes through $\Psi(y,x)$ is included in $\widetilde{\delta}^{-1}\left\langle\{\theta^{\text{FIX}}(y), \gamma(x)\}\right\rangle$ (as leaves of foliation are by definition connected).

On the other hand, $\widetilde{\delta}^{-1}\left\langle\{\theta^{\text{FIX}}(y), \gamma(x)\}\right\rangle = \Psi\left\langle \theta^{\text{FIX},-1}\left\langle\{\theta^{\text{FIX}}(y)\}\right\rangle \times \gamma^{-1}\left\langle\{\gamma(x)\}\right\rangle\right\rangle$ is connected as image by a continuous map of a Cartesian product of connected spaces. And its tangent space at $\Psi(y,x)$ is given by:

$$T_{(y,x)}\Psi\left\langle K_y(\mathcal{P}^{\text{FIX}}) \times K_x(\mathcal{N}^{\text{SHELL}})\right\rangle = K_{\psi(y,x)}(\mathcal{M}^{\text{SHELL}}) \text{ (using 3.6.2).}$$

Therefore, $\widetilde{\delta}^{-1}\left\langle\{\theta^{\text{FIX}}(y), \gamma(x)\}\right\rangle$ is included in the leaf of the foliation $K(\mathcal{M}^{\text{SHELL}})$ that goes through $\Psi(y,x)$.

This concludes the proof that $\left(\mathcal{P}^{\text{DYN}} \times \mathcal{N}^{\text{DYN}}, \mathcal{M}^{\text{SHELL}}, \widetilde{\delta}\right)$ is a phase space reduction of $\mathcal{M}^{\text{KIN}}$. Now, using prop. A.6, there exists a symplectomorphism $\Phi : \mathcal{M}^{\text{DYN}} \to \mathcal{P}^{\text{DYN}} \times \mathcal{N}^{\text{DYN}}$ such that $\Phi \circ \delta = \widetilde{\delta}$. $\pi^{\text{DYN}}$ is then the projection corresponding to this factorization of $\mathcal{M}^{\text{DYN}}$. $\square$

We are now ready to consider a projective system of phase spaces $\mathcal{M}^{\text{KIN}}_\eta$, with a phase space reduction of $\mathcal{M}^{\text{KIN}}_\eta$ for each $\eta$. As announced at the beginning of this subsection, we want to examine the situation where the reduced phase spaces $\mathcal{M}^{\text{DYN}}_\eta$ can be arranged into a new projective system of phase spaces, in such a way that the maps, that translate the kinematical observables into dynamical ones for each $\eta$, are intertwined by the projections on both sides. Thus, we can associate to an observable on the projective limit of the $\mathcal{M}^{\text{KIN}}_\eta$ an observable on the projective limit of the $\mathcal{M}^{\text{DYN}}_\eta$. In a dual way, to each state on this dynamical projective system of phase spaces corresponds a projective family of orbits in the constraint surfaces $\mathcal{M}^{\text{SHELL}}_\eta$ (another option here would be to consider projective family of probability measures, aka. statistical states, in which case we would map dynamical statistical states to on-shell supported, gauge invariant, kinematical statistical states).

The previous study, examining a projection that relates the phase space reductions on two symplectic manifolds, is the key element for this construction. Indeed the requirement that the dynamical phase spaces should readily assemble into a new projective system can actually be enforced by asking, for each pair of index $\eta \preccurlyeq \eta'$, that the reductions on $\mathcal{M}^{\text{KIN}}_\eta$ and $\mathcal{M}^{\text{KIN}}_{\eta'}$ should be related by $\pi^{\text{KIN}}_{\eta' \to \eta}$.

**Definition 3.7** Let $(\mathcal{L}, \mathcal{M}^{\text{KIN}}, \pi^{\text{KIN}})^{\downarrow}$ be a projective system of phase spaces. An elementary reduction of $(\mathcal{L}, \mathcal{M}^{\text{KIN}}, \pi^{\text{KIN}})^{\downarrow}$ is a quadruple $\left(\left(\mathcal{M}^{\text{DYN}}_\eta\right)_{\eta \in \mathcal{L}}, \left(\mathcal{M}^{\text{SHELL}}_\eta\right)_{\eta \in \mathcal{L}}, \left(\pi^{\text{DYN}}_{\eta' \to \eta}\right)_{\eta \preccurlyeq \eta'}, \left(\delta_\eta\right)_{\eta \in \mathcal{L}}\right)$ such that:

1. $(\mathcal{L}, \mathcal{M}^{\text{DYN}}, \pi^{\text{DYN}})^{\downarrow}$ is a projective system of phase spaces;
2. $\forall \eta \in \mathcal{L}, (\mathcal{M}^{\text{DYN}}_\eta, \mathcal{M}^{\text{SHELL}}_\eta, \delta_\eta)$ is a phase space reduction of $\mathcal{M}^{\text{KIN}}_\eta$;

3. $\forall \eta \preccurlyeq \eta' \in \mathcal{L}$, $\pi^{\text{KIN}}_{\eta' \to \eta} \langle \mathcal{M}^{\text{SHELL}}_{\eta'} \rangle = \mathcal{M}^{\text{SHELL}}_{\eta}$ and:

$$\forall x_\eta \in \mathcal{M}^{\text{SHELL}}_{\eta}, \forall y_{\eta'} \in \mathcal{M}^{\text{DYN}}_{\eta'},$$

$$\left(\exists x_{\eta'} \in \mathcal{M}^{\text{SHELL}}_{\eta'} \;/\; \delta_{\eta'}(x_{\eta'}) = y_{\eta'} \quad \& \quad \pi^{\text{KIN}}_{\eta' \to \eta}(x_{\eta'}) = x_\eta\right) \Leftrightarrow \left(\delta_\eta(x_\eta) = \pi^{\text{DYN}}_{\eta' \to \eta}(y_{\eta'})\right).$$

Whenever possible, we will use the shortened notation $(\mathcal{L}, \mathcal{M}, \pi, \delta)^{\text{DYN}}$ instead of $\left(\left(\mathcal{M}^{\text{DYN}}_\eta\right)_{\eta \in \mathcal{L}}, \left(\mathcal{M}^{\text{SHELL}}_\eta\right)_{\eta \in \mathcal{L}}, \left(\pi^{\text{DYN}}_{\eta' \to \eta}\right)_{\eta \preccurlyeq \eta'}, \left(\delta_\eta\right)_{\eta \in \mathcal{L}}\right)$.

**Definition 3.8** We consider the same objects as in def. 3.7 and we define (in analogy to the definition of $\mathcal{S}^{\downarrow}_{(\mathcal{L}, \mathcal{M}, \pi)}$ in def. 2.3) $\widehat{\mathcal{S}}^{\downarrow}_{(\mathcal{L}, \mathcal{M}^{\text{KIN}}, \pi^{\text{KIN}})}$ as:

$$\widehat{\mathcal{S}}^{\downarrow}_{(\mathcal{L}, \mathcal{M}^{\text{KIN}}, \pi^{\text{KIN}})} := \left\{ (D_\eta)_{\eta \in \mathcal{L}} \in \prod_{\eta \in \mathcal{L}} \mathcal{P}(\mathcal{M}^{\text{KIN}}_\eta) \;\middle|\; \forall \eta \preccurlyeq \eta', \pi^{\text{KIN}}_{\eta' \to \eta}\langle D_{\eta'} \rangle = D_\eta \right\},$$

where, for $\eta \in \mathcal{L}$, $\mathcal{P}(\mathcal{M}^{\text{KIN}}_\eta)$ is the set of subsets of $\mathcal{M}^{\text{KIN}}_\eta$.

Then, we define:

$$\Delta \;:\; \mathcal{S}^{\downarrow}_{(\mathcal{L}, \mathcal{M}^{\text{DYN}}, \pi^{\text{DYN}})} \;\to\; \widehat{\mathcal{S}}^{\downarrow}_{(\mathcal{L}, \mathcal{M}^{\text{KIN}}, \pi^{\text{KIN}})}$$
$$(y_\eta)_{\eta \in \mathcal{L}} \;\mapsto\; \left(\delta_\eta^{-1}\langle \{y_\eta\} \rangle\right)_{\eta \in \mathcal{L}},$$

which is well-defined as a map $\mathcal{S}^{\downarrow}_{(\mathcal{L}, \mathcal{M}^{\text{DYN}}, \pi^{\text{DYN}})} \to \widehat{\mathcal{S}}^{\downarrow}_{(\mathcal{L}, \mathcal{M}^{\text{KIN}}, \pi^{\text{KIN}})}$, for we have $\forall \eta \preccurlyeq \eta' \in \mathcal{L}, \forall y_{\eta'} \in \mathcal{M}^{\text{DYN}}_{\eta'}, \delta_\eta^{-1}\langle \{\pi^{\text{DYN}}_{\eta' \to \eta}(y_{\eta'})\} \rangle = \pi^{\text{KIN}}_{\eta' \to \eta}\langle \delta_{\eta'}^{-1}\langle y_{\eta'} \rangle \rangle$.

**Proposition 3.9** Let $(\mathcal{L}, \mathcal{M}^{\text{KIN}}, \pi^{\text{KIN}})^{\downarrow}$ be a projective system of phase spaces and let $(\mathcal{L}, \mathcal{M}, \pi, \delta)^{\text{DYN}}$ be an elementary reduction of $(\mathcal{L}, \mathcal{M}^{\text{KIN}}, \pi^{\text{KIN}})^{\downarrow}$. We define (in analogy to def. 2.4) $\mathcal{A}^{\downarrow}_{(\mathcal{L}, \mathcal{M}^{\text{KIN}}, \pi^{\text{KIN}})}$ as the set of equivalence classes in $\bigcup_{\eta \in \mathcal{L}} B(\mathcal{M}_\eta)$ for the equivalence relation defined by:

$$\forall \eta, \eta' \in \mathcal{L}, \forall f_\eta \in B(\mathcal{M}^{\text{KIN}}_\eta), \forall f_{\eta'} \in B(\mathcal{M}^{\text{KIN}}_{\eta'}),$$

$$f_\eta \sim^{\text{KIN}} f_{\eta'} \Leftrightarrow \left(\exists \eta'' \in \mathcal{L} \;/\; \eta \preccurlyeq \eta'', \eta' \preccurlyeq \eta'' \quad \& \quad f_\eta \circ \pi^{\text{KIN}}_{\eta'' \to \eta} = f_{\eta'} \circ \pi^{\text{KIN}}_{\eta'' \to \eta'}\right),$$

and similarly $\mathcal{A}^{\downarrow}_{(\mathcal{L}, \mathcal{M}^{\text{DYN}}, \pi^{\text{DYN}})}$ with the equivalence relation $\sim^{\text{DYN}}$.

Then, the map:

$$(\cdot)^{\text{DYN}} \;:\; \mathcal{A}^{\downarrow}_{(\mathcal{L}, \mathcal{M}^{\text{KIN}}, \pi^{\text{KIN}})} \;\to\; \mathcal{A}^{\downarrow}_{(\mathcal{L}, \mathcal{M}^{\text{DYN}}, \pi^{\text{DYN}})}$$
$$[f_\eta]_{\sim^{\text{KIN}}} \;\mapsto\; [f^{\text{DYN}}_\eta]_{\sim^{\text{DYN}}}$$

is well-defined.

For $(D_\eta)_{\eta \in \mathcal{L}} \in \widehat{\mathcal{S}}^{\downarrow}_{(\mathcal{L}, \mathcal{M}^{\text{KIN}}, \pi^{\text{KIN}})}$ and $f = [f_\eta]_{\sim^{\text{KIN}}} \in \mathcal{A}^{\downarrow}_{(\mathcal{L}, \mathcal{M}^{\text{KIN}}, \pi^{\text{KIN}})}$, we define:

$$[f_\eta]_{\sim^{\text{KIN}}}\left((D_\eta)_{\eta \in \mathcal{L}}\right) := \sup\{f_\eta(x) \mid x \in D_\eta\},$$

(the definition of the equivalence relation $\sim^{\text{KIN}}$ ensures that this is well-defined)

Then, we have for all $y \in \mathcal{S}^{\downarrow}_{(\mathcal{L}, \mathcal{M}^{\text{DYN}}, \pi^{\text{DYN}})}$ and all $f \in \mathcal{A}^{\downarrow}_{(\mathcal{L}, \mathcal{M}^{\text{KIN}}, \pi^{\text{KIN}})}$:

$$f^{\text{DYN}}(y) = f(\Delta(y)). \tag{3.9.1}$$

**Proof** What we need to show is that for $\eta, \eta' \in \mathcal{L}$, $f_\eta \in B(\mathcal{M}_\eta)$ and $f_{\eta'} \in B(\mathcal{M}_{\eta'})$, $\left(f_\eta \sim^{\text{KIN}} f_{\eta'}\right) \Rightarrow \left(f_\eta^{\text{DYN}} \sim^{\text{DYN}} f_{\eta'}^{\text{DYN}}\right)$. Indeed if there exist $\eta'' \in \mathcal{L}$, with $\eta'' \succcurlyeq \eta$, $\eta'' \succcurlyeq \eta'$, and $f_{\eta''} \in B(\mathcal{M}_{\eta''})$ such that $f_\eta \circ \pi^{\text{KIN}}_{\eta'' \to \eta} = f_{\eta'} \circ \pi^{\text{KIN}}_{\eta'' \to \eta'}$, then, from prop. 3.3:

$$f_\eta^{\text{DYN}} \circ \pi^{\text{DYN}}_{\eta'' \to \eta} = \left(f_\eta \circ \pi^{\text{KIN}}_{\eta'' \to \eta}\right)^{\text{DYN}} = \left(f_{\eta'} \circ \pi^{\text{KIN}}_{\eta'' \to \eta'}\right)^{\text{DYN}} = f_{\eta'}^{\text{DYN}} \circ \pi^{\text{DYN}}_{\eta'' \to \eta'}.$$

Then, we only need to check eq. (3.9.1) for a particular representative $f_\eta$ of $f$:

$$f^{\text{DYN}}(y) = f_\eta^{\text{DYN}}(y_\eta) = \sup_{x \in \delta_\eta^{-1}\langle\{y_\eta\}\rangle} f_\eta(x) = \sup_{x \in (\delta^{\text{DYN}}(y))_\eta} f_\eta(x) = f(\Delta(y)).$$

$\square$

**Proposition 3.10** Let $(\mathcal{L}, \mathcal{M}^{\text{KIN}}, \pi^{\text{KIN}})^\downarrow$ be a projective system of phase spaces. For all $\eta \in \mathcal{L}$, we give ourselves a phase space reduction $(\mathcal{M}_\eta^{\text{DYN}}, \mathcal{M}_\eta^{\text{SHELL}}, \delta_\eta)$ of $\mathcal{M}_\eta$. The following statements are equivalent:

1. there exists a family of surjective maps $\left(\pi^{\text{DYN}}_{\eta' \to \eta}\right)_{\eta \preccurlyeq \eta'}$ such that $(\mathcal{L}, \mathcal{M}, \pi, \delta)^{\text{DYN}}$ is an elementary reduction of $(\mathcal{L}, \mathcal{M}^{\text{KIN}}, \pi^{\text{KIN}})^\downarrow$;

2. $\forall \eta \preccurlyeq \eta'$, $(\mathcal{M}_{\eta'}^{\text{DYN}}, \mathcal{M}_{\eta'}^{\text{SHELL}}, \delta_{\eta'})$ and $(\mathcal{M}_\eta^{\text{DYN}}, \mathcal{M}_\eta^{\text{SHELL}}, \delta_\eta)$ are related by $\pi^{\text{KIN}}_{\eta' \to \eta}$.

**Proof** By definition of an elementary reduction of $(\mathcal{L}, \mathcal{M}^{\text{KIN}}, \pi^{\text{KIN}})^\downarrow$, we have 3.10.1 $\Rightarrow$ 3.10.2.

To prove the other direction, we need to show that the $\pi^{\text{DYN}}_{\eta' \to \eta}$ induced by the $\pi^{\text{KIN}}_{\eta' \to \eta}$ satisfy the three-spaces consistency condition:

$$\forall \eta \preccurlyeq \eta' \preccurlyeq \eta'' \in \mathcal{L}, \; \pi^{\text{DYN}}_{\eta'' \to \eta} = \pi^{\text{DYN}}_{\eta' \to \eta} \circ \pi^{\text{DYN}}_{\eta'' \to \eta'}.$$

For $x_\eta \in \mathcal{M}_\eta^{\text{SHELL}}$, $y_{\eta''} \in \mathcal{M}_{\eta''}^{\text{DYN}}$, we have (using def. 3.7.3 for $\eta \preccurlyeq \eta'$ and $\eta' \preccurlyeq \eta''$):

$$\left(\delta_\eta(x_\eta) = \pi^{\text{DYN}}_{\eta' \to \eta}\left(\pi^{\text{DYN}}_{\eta'' \to \eta'}(y_{\eta''})\right)\right) \Leftrightarrow \left(\exists x_{\eta'} \in \mathcal{M}_{\eta'}^{\text{SHELL}} / \delta_{\eta'}(x_{\eta'}) = \pi^{\text{DYN}}_{\eta'' \to \eta'}(y_{\eta''}) \;\&\; \pi^{\text{KIN}}_{\eta' \to \eta}(x_{\eta'}) = x_\eta\right)$$

$$\Leftrightarrow \left(\exists x_{\eta'} \in \mathcal{M}_{\eta'}^{\text{SHELL}}, \exists x_{\eta''} \in \mathcal{M}_{\eta''}^{\text{SHELL}} / \delta_{\eta''}(x_{\eta''}) = y_{\eta''} \;\&\; \pi^{\text{KIN}}_{\eta'' \to \eta'}(x_{\eta''}) = x_{\eta'} \;\&\; \pi^{\text{KIN}}_{\eta' \to \eta}(x_{\eta'}) = x_\eta\right)$$

$$\Leftrightarrow \left(\exists x_{\eta''} \in \mathcal{M}_{\eta''}^{\text{SHELL}} / \delta_{\eta''}(x_{\eta''}) = y_{\eta''} \;\&\; \pi^{\text{KIN}}_{\eta' \to \eta}\left(\pi^{\text{KIN}}_{\eta'' \to \eta'}(x_{\eta''})\right) = x_\eta\right).$$

Hence, using $\pi^{\text{KIN}}_{\eta' \to \eta} \circ \pi^{\text{KIN}}_{\eta'' \to \eta'} = \pi^{\text{KIN}}_{\eta'' \to \eta}$, and applying prop. 3.2 with $\pi^{\text{DYN}}_{\eta'' \to \eta}$ and $\pi^{\text{DYN}}_{\eta' \to \eta} \circ \pi^{\text{DYN}}_{\eta'' \to \eta'}$, we have $\pi^{\text{DYN}}_{\eta'' \to \eta} = \pi^{\text{DYN}}_{\eta' \to \eta} \circ \pi^{\text{DYN}}_{\eta'' \to \eta'}$. $\square$

Recalling the discussion of subsection 2.2, regarding restrictions and extensions of the label set, we would like to understand how elementary reductions pass through these operations. It is quite straightforward that everything will go smoothly if we *restrict* the label set.

The interesting question occurs when we have an elementary reduction on a subset $\mathcal{L}'$ of $\mathcal{L}$. In particular, if $\mathcal{L}'$ is cofinal in $\mathcal{L}$, we can identify the kinematical spaces of states and observables over the projective system restricted to $\mathcal{L}'$ with the ones over the original projective system on $\mathcal{L}$ (prop. 2.5), thus the transport of observables (from the kinematical to the dynamical theory) and states (from the dynamical to the kinematical theory) arising from an elementary reduction on $\mathcal{L}'$ immediately defines corresponding transport maps between the kinematical projective structure on $\mathcal{L}$ and the dynamical projective structure (which is then only defined for the label set $\mathcal{L}'$). In other words, we are still able to glue together the dynamical phase spaces $\mathcal{M}_\eta^{\text{DYN}}$ ($\eta \in \mathcal{L}'$) into a dynamical projective structure, to inherit observables on this structure and to project back its states. However, in general, there will *not* exist an elementary reduction on $\mathcal{L}$, that would reproduce the same transport maps (modulo the identification of the thus obtained dynamical structure on $\mathcal{L}$ with its restriction to $\mathcal{L}'$). This point will play a key role when moving to the regularization of a dynamics that does not break down well on the projective structure (subsection 3.2).

What is lacking, when trying to extend to $\mathcal{L}$ an elementary reduction on $\mathcal{L}'$, is the assurance that there will exist phase space reductions of the $\mathcal{M}_\eta^{\text{KIN}}$ for $\eta \in \mathcal{L} \setminus \mathcal{L}'$, and that these reductions will be compatible with each other, as well as with the given reductions on $\mathcal{L}'$: specifically, for any pair of labels $\eta \preccurlyeq \eta'$ (one or both being in $\mathcal{L} \setminus \mathcal{L}'$) the reductions should be related by $\pi_{\eta' \to \eta}^{\text{KIN}}$ (the elementary reduction on $\mathcal{L}'$ already accounts for the compatibility when both labels are in $\mathcal{L}'$). Prop. 3.12 shows slightly weaker hypotheses under which the extension *is* possible, provided the $\mathcal{M}_\eta^{\text{KIN}}$ for $\eta \in \mathcal{L} \setminus \mathcal{L}'$ are finite dimensional.

**Proposition 3.11** Let $(\mathcal{L}, \mathcal{M}^{\text{KIN}}, \pi^{\text{KIN}})^{\downarrow}$ be a projective system of phase spaces and $(\mathcal{L}, \mathcal{M}, \pi, \delta)^{\text{DYN}}$ be an elementary reduction of $(\mathcal{L}, \mathcal{M}^{\text{KIN}}, \pi^{\text{KIN}})^{\downarrow}$. If $\mathcal{L}'$ is a directed subset of $\mathcal{L}$, $(\mathcal{L}', \mathcal{M}, \pi, \delta)^{\text{DYN}}$ is an elementary reduction of $(\mathcal{L}', \mathcal{M}^{\text{KIN}}, \pi^{\text{KIN}})^{\downarrow}$ and we have:

$$\widehat{\sigma}_{\mathcal{L} \to \mathcal{L}'}^{\text{KIN}} \circ \Delta = \Delta' \circ \sigma_{\mathcal{L} \to \mathcal{L}'}^{\text{DYN}},$$

where $\widehat{\sigma}_{\mathcal{L} \to \mathcal{L}'}^{\text{KIN}} : \widehat{\mathcal{S}}_{(\mathcal{L}, \mathcal{M}^{\text{KIN}}, \pi^{\text{KIN}})}^{\downarrow} \to \widehat{\mathcal{S}}_{(\mathcal{L}', \mathcal{M}^{\text{KIN}}, \pi^{\text{KIN}})}^{\downarrow}$, $\sigma_{\mathcal{L} \to \mathcal{L}'}^{\text{DYN}} : \mathcal{S}_{(\mathcal{L}, \mathcal{M}^{\text{DYN}}, \pi^{\text{DYN}})}^{\downarrow} \to \mathcal{S}_{(\mathcal{L}', \mathcal{M}^{\text{DYN}}, \pi^{\text{DYN}})}^{\downarrow}$ are defined in analogy to prop. 2.5, while $\Delta : \mathcal{S}_{(\mathcal{L}, \mathcal{M}^{\text{DYN}}, \pi^{\text{DYN}})}^{\downarrow} \to \widehat{\mathcal{S}}_{(\mathcal{L}, \mathcal{M}^{\text{KIN}}, \pi^{\text{KIN}})}^{\downarrow}$ and $\Delta' : \mathcal{S}_{(\mathcal{L}', \mathcal{M}^{\text{DYN}}, \pi^{\text{DYN}})}^{\downarrow} \to \widehat{\mathcal{S}}_{(\mathcal{L}', \mathcal{M}^{\text{KIN}}, \pi^{\text{KIN}})}^{\downarrow}$ are defined as in def. 3.8.

In addition, for any $f \in \mathcal{A}_{(\mathcal{L}', \mathcal{M}^{\text{KIN}}, \pi^{\text{KIN}})}^{\downarrow}$, we have:

$$\left(\beta_{\mathcal{L} \leftarrow \mathcal{L}'}^{\text{KIN}}(f)\right)^{\text{DYN}} = \beta_{\mathcal{L} \leftarrow \mathcal{L}'}^{\text{DYN}}\left(f^{\text{DYN}}\right),$$

where $\beta_{\mathcal{L} \leftarrow \mathcal{L}'}^{\text{KIN/DYN}} : \mathcal{A}_{(\mathcal{L}', \mathcal{M}^{\text{KIN/DYN}}, \pi^{\text{KIN/DYN}})}^{\downarrow} \to \mathcal{A}_{(\mathcal{L}, \mathcal{M}^{\text{KIN/DYN}}, \pi^{\text{KIN/DYN}})}^{\downarrow}$ are defined in analogy to prop. 2.5, while $(\,\cdot\,)^{\text{DYN}} : \mathcal{A}_{(\mathcal{L}, \mathcal{M}^{\text{KIN}}, \pi^{\text{KIN}})}^{\downarrow} \to \mathcal{A}_{(\mathcal{L}, \mathcal{M}^{\text{DYN}}, \pi^{\text{DYN}})}^{\downarrow}$ and $(\,\cdot\,)^{\text{DYN}} : \mathcal{A}_{(\mathcal{L}', \mathcal{M}^{\text{KIN}}, \pi^{\text{KIN}})}^{\downarrow} \to \mathcal{A}_{(\mathcal{L}', \mathcal{M}^{\text{DYN}}, \pi^{\text{DYN}})}^{\downarrow}$ are defined as in prop. 3.9.

**Proof** That $(\mathcal{L}', \mathcal{M}, \pi, \delta)^{\text{DYN}}$ is an elementary reduction of $(\mathcal{L}', \mathcal{M}^{\text{KIN}}, \pi^{\text{KIN}})^{\downarrow}$ can be immediately checked from def. 3.7.

Let $(y_\eta)_{\eta \in \mathcal{L}} \in \mathcal{S}_{(\mathcal{L}, \mathcal{M}^{\text{DYN}}, \pi^{\text{DYN}})}^{\downarrow}$. We have:

$$\widehat{\sigma}_{\mathcal{L} \to \mathcal{L}'}^{\text{KIN}} \circ \Delta\left((y_\eta)_{\eta \in \mathcal{L}}\right) = \widehat{\sigma}_{\mathcal{L} \to \mathcal{L}'}^{\text{KIN}}\left(\left(\delta_\eta^{-1} \langle y_\eta \rangle\right)_{\eta \in \mathcal{L}}\right) = \left(\delta_\eta^{-1} \langle y_\eta \rangle\right)_{\eta \in \mathcal{L}'}$$

$$= \Delta'\left((y_\eta)_{\eta \in \mathcal{L}'}\right) = \Delta' \circ \sigma_{\mathcal{L} \to \mathcal{L}'}^{\text{DYN}}\left((y_\eta)_{\eta \in \mathcal{L}}\right).$$

Let $f = [f_\eta]_{\sim_{\text{KIN}}} \in \mathcal{A}_{(\mathcal{L}', \mathcal{M}^{\text{KIN}}, \pi^{\text{KIN}})}^{\downarrow}$. We have:

$$\left(\beta_{\mathcal{L} \leftarrow \mathcal{L}'}^{\text{KIN}}(f)\right)^{\text{DYN}} = \left[f_\eta^{\text{DYN}}\right]_{\sim_{\text{DYN}}} = \beta_{\mathcal{L} \leftarrow \mathcal{L}'}^{\text{DYN}}\left(f^{\text{DYN}}\right).$$

**Proposition 3.12** Let $(\mathcal{L}, \mathcal{M}^{\text{KIN}}, \pi^{\text{KIN}})^{\downarrow}$ be a projective system of phase spaces and let $\mathcal{L}'$ be a cofinal subset of $\mathcal{L}$. We assume:

1. that we are given an elementary reduction $(\mathcal{L}', \mathcal{M}, \pi, \delta)^{\text{DYN}}$ of $(\mathcal{L}', \mathcal{M}^{\text{KIN}}, \pi^{\text{KIN}})^{\downarrow}$;

2. that for any $\eta \in \mathcal{L} \setminus \mathcal{L}'$, $\mathcal{M}_\eta^{\text{KIN}}$ is finite dimensional and we are given a phase space reduction $(\mathcal{M}_\eta^{\text{DYN}}, \mathcal{M}_\eta^{\text{SHELL}}, \delta_\eta)$ of $\mathcal{M}_\eta^{\text{KIN}}$;

3. that for any $\eta \in \mathcal{L}$, and for any $\eta' \in \mathcal{L}'$ with $\eta' \succcurlyeq \eta$, the reductions on $\mathcal{M}_{\eta'}^{\text{KIN}}$ and $\mathcal{M}_\eta^{\text{KIN}}$ are related by $\pi_{\eta' \to \eta}^{\text{KIN}}$.

Then, $(\mathcal{L}', \mathcal{M}, \pi, \delta)^{\text{DYN}}$ can be completed into an elementary reduction $(\mathcal{L}, \mathcal{M}, \pi, \delta)^{\text{DYN}}$ of $(\mathcal{L}, \mathcal{M}^{\text{KIN}}, \pi^{\text{KIN}})^{\downarrow}$.

**Lemma 3.13** Let $\mathcal{M}^{\text{KIN}}$, $\mathcal{N}^{\text{KIN}}$ and $\mathcal{P}^{\text{KIN}}$ be symplectic manifolds and assume that $\mathcal{N}^{\text{KIN}}$ and $\mathcal{P}^{\text{KIN}}$ are finite dimensional. Let $\pi_1^{\text{KIN}} : \mathcal{M}^{\text{KIN}} \to \mathcal{N}^{\text{KIN}}$, $\pi_2^{\text{KIN}} : \mathcal{N}^{\text{KIN}} \to \mathcal{P}^{\text{KIN}}$, and $\pi_3^{\text{KIN}} : \mathcal{M}^{\text{KIN}} \to \mathcal{P}^{\text{KIN}}$ be projections compatible with the symplectic structures, satisfying $\pi_3^{\text{KIN}} = \pi_2^{\text{KIN}} \circ \pi_1^{\text{KIN}}$. Let $(\mathcal{M}^{\text{DYN}}, \mathcal{M}^{\text{SHELL}}, \delta)$, $(\mathcal{N}^{\text{DYN}}, \mathcal{N}^{\text{SHELL}}, \gamma)$ and $(\mathcal{P}^{\text{DYN}}, \mathcal{P}^{\text{SHELL}}, \eta)$ be phase space reductions of $\mathcal{M}^{\text{KIN}}$, $\mathcal{N}^{\text{KIN}}$ and $\mathcal{P}^{\text{KIN}}$ respectively.

If the reductions on $\mathcal{M}^{\text{KIN}}$ and $\mathcal{N}^{\text{KIN}}$ are related by $\pi_1^{\text{KIN}}$ and the reductions on $\mathcal{M}^{\text{KIN}}$ and $\mathcal{P}^{\text{KIN}}$ are related by $\pi_3^{\text{KIN}}$, then the reductions on $\mathcal{N}^{\text{KIN}}$ and $\mathcal{P}^{\text{KIN}}$ are related by $\pi_2^{\text{KIN}}$.

**Proof** Applying def. 3.1.1 for $\pi_1^{\text{KIN}}$ and $\pi_3^{\text{KIN}}$, we have:

$$\pi_2^{\text{KIN}} \langle \mathcal{N}^{\text{SHELL}} \rangle = \pi_2^{\text{KIN}} \langle \pi_1^{\text{KIN}} \langle \mathcal{M}^{\text{SHELL}} \rangle \rangle = \pi_3^{\text{KIN}} \langle \mathcal{M}^{\text{SHELL}} \rangle = \mathcal{P}^{\text{SHELL}}.$$

Let $\pi_1^{\text{DYN}} : \mathcal{M}^{\text{DYN}} \to \mathcal{N}^{\text{DYN}}$ and $\pi_3^{\text{DYN}} : \mathcal{M}^{\text{DYN}} \to \mathcal{P}^{\text{DYN}}$ be as in def. 3.1.2. For any $y, y' \in \mathcal{M}^{\text{DYN}}$ such that $\pi_1^{\text{DYN}}(y) = \pi_1^{\text{DYN}}(y')$, there exists $z \in \gamma^{-1} \langle \pi_1^{\text{DYN}}(y) \rangle \subset \mathcal{N}^{\text{SHELL}}$ (for $\gamma$ is surjective from def. A.1.2) and, using eq. (3.1.1), there exist $x, x' \in \mathcal{M}^{\text{SHELL}}$ such that:

$$\delta(x) = y, \ \delta(x') = y' \text{ and } \pi_1^{\text{KIN}}(x) = z = \pi_1^{\text{KIN}}(x').$$

Therefore, $\pi_3^{\text{KIN}}(x) = \pi_2^{\text{KIN}}(z) = \pi_3^{\text{KIN}}(x')$, so using again eq. (3.1.1), we have $\pi_3^{\text{DYN}}(y) = \eta \circ \pi_2^{\text{KIN}}(z) = \pi_3^{\text{DYN}}(y')$.

Hence, $\pi_3^{\text{DYN}}$ is constant on the level sets of $\pi_1^{\text{DYN}}$, so there exists a map $\pi_2^{\text{DYN}} : \mathcal{N}^{\text{DYN}} \to \mathcal{P}^{\text{DYN}}$ such that $\pi_3^{\text{DYN}} = \pi_2^{\text{DYN}} \circ \pi_1^{\text{DYN}}$.

Now, for $z' \in \mathcal{P}^{\text{SHELL}}$ and $w \in \mathcal{N}^{\text{DYN}}$, there exists $y \in \mathcal{M}^{\text{DYN}}$ such that $\pi_1^{\text{DYN}}(y) = w$ (for $\pi_1^{\text{DYN}}$ is surjective) and we have:

$$\left(\eta(z') = \pi_2^{\text{DYN}}(w)\right) \Leftrightarrow \left(\eta(z') = \pi_3^{\text{DYN}}(y)\right) \Leftrightarrow \left(\exists x \in \mathcal{M}^{\text{SHELL}} \,/\, \delta(x) = y \ \& \ \pi_3^{\text{KIN}}(x) = z'\right)$$

$$\Leftrightarrow \left(\exists z \in \mathcal{N}^{\text{SHELL}} \,/\, \gamma(z) = w \ \& \ \pi_2^{\text{KIN}}(z) = z'\right),$$

where the last equivalence comes from setting $z = \pi_1^{\text{KIN}}(x)$ (for proving '$\Rightarrow$') and using eq. (3.1.1) with $\gamma(z) = \pi_1^{\text{DYN}}(y)$ (for '$\Leftarrow$'). Hence, $\pi_2^{\text{DYN}}$ fulfills eq. (3.1.1).

In particular, we then have $\eta \circ \pi_2^{\text{KIN}} = \pi_2^{\text{DYN}} \circ \gamma$. Thus, since $\mathcal{N}^{\text{KIN}}$, $\mathcal{N}^{\text{DYN}}$ and $\mathcal{P}^{\text{DYN}}$ are smooth finite dimensional manifolds, $\eta \circ \pi_2^{\text{KIN}}$ is smooth and $\gamma$ is surjective with surjective derivative at any point (def. A.1.3), the rank theorem implies [11, prop. 5.19] that $\pi_2^{\text{DYN}}$ is smooth.

Finally, we need to show that $\pi_2^{\text{DYN}}$ is a surjective map compatible with the symplectic structures. We have $\pi_2^{\text{DYN}} \langle \mathcal{N}^{\text{DYN}} \rangle = \pi_2^{\text{DYN}} \langle \pi_1^{\text{DYN}} \langle \mathcal{M}^{\text{DYN}} \rangle \rangle = \pi_3^{\text{DYN}} \langle \mathcal{M}^{\text{DYN}} \rangle = \mathcal{P}^{\text{DYN}}$. And, for any $w \in \mathcal{N}^{\text{DYN}}$, there exists $y \in \mathcal{M}^{\text{DYN}}$ with $\pi_1^{\text{DYN}}(y) = w$, so that for any $\upsilon \in T^*_{\pi_2^{\text{DYN}}(w)}(\mathcal{P}^{\text{DYN}})$:

$$[T_w \pi_2^{\text{DYN}}]\left(\underline{\upsilon \circ [T_w \pi_2^{\text{DYN}}]}\right) = [T_w \pi_2^{\text{DYN}}] \circ [T_y \pi_1^{\text{DYN}}] \left(\underline{\upsilon \circ [T_w \pi_2^{\text{DYN}}] \circ [T_y \pi_1^{\text{DYN}}]}\right)$$

$$= [T_y \pi_3^{\text{DYN}}]\left(\underline{\upsilon \circ [T_y \pi_3^{\text{DYN}}]}\right) = \underline{\upsilon},$$

therefore $\pi_2^{\text{DYN}}$ fulfills eq. (2.1.1). □

**Proof of prop. 3.12** Let $\eta \in \mathcal{L}$ and $\eta' \in \mathcal{L} \setminus \mathcal{L}'$, with $\eta' \succcurlyeq \eta$. Since $\mathcal{L}'$ is a cofinal part of $\mathcal{L}$, there exists $\eta'' \in \mathcal{L}'$ such that $\eta'' \succcurlyeq \eta' \succcurlyeq \eta$. Using lemma 3.13 ($\mathcal{M}_{\eta'}^{\text{KIN}}$ is finite dimensional, for $\eta' \in \mathcal{L} \setminus \mathcal{L}'$, so $\mathcal{M}_{\eta}^{\text{KIN}}$ is finite dimensional, for $\eta \preccurlyeq \eta'$), the reductions on $\mathcal{M}_{\eta'}^{\text{KIN}}$ and $\mathcal{M}_{\eta}^{\text{KIN}}$ are related by $\pi_{\eta' \to \eta}^{\text{KIN}}$.

Hence, using prop. 3.10, there exists an elementary reduction $(\mathcal{L}, \mathcal{M}, \widetilde{\pi}, \delta)^{\text{DYN}}$ of $(\mathcal{L}, \mathcal{M}^{\text{KIN}}, \pi^{\text{KIN}})$, where $\forall \eta \preccurlyeq \eta' \in \mathcal{L}$, $\widetilde{\pi}_{\eta' \to \eta}^{\text{KIN}} = \pi_{\eta' \to \eta}^{\text{KIN}}$. And by prop. 3.2, $\forall \eta \preccurlyeq \eta' \in \mathcal{L}'$, $\widetilde{\pi}_{\eta' \to \eta}^{\text{DYN}} = \pi_{\eta' \to \eta}^{\text{DYN}}$, which supplies the desired result. □

In practice, we will be interested in a kinematical projective structure that is a rendering, by a system of finite dimensional manifolds $\mathcal{M}_{\eta}^{\text{KIN}}$, of an infinite dimensional symplectic manifold $\mathcal{M}_{\infty}^{\text{KIN}}$ (def. 2.6). If the phase space reduction on $\mathcal{M}_{\infty}^{\text{KIN}}$ satisfies the (admittedly very restrictive) requirement that it projects as a closed dynamics on $\mathcal{M}_{\eta}^{\text{KIN}}$ for all $\eta$, we will get an elementary reduction of the kinematical projective structure, and the thus obtained dynamical projective system will automatically be a rendering of the physical phase space $\mathcal{M}_{\infty}^{\text{DYN}}$ (fig. 3.3).

Moreover, the map turning observables on the kinematical projective structure into observables on the dynamical structure coincides with the one that can be defined directly from the phase space reduction of $\mathcal{M}_{\infty}^{\text{KIN}}$ (identifying the observables on the projective structures with functions on $\mathcal{M}_{\infty}^{\text{KIN}}$ or $\mathcal{M}_{\infty}^{\text{DYN}}$, as described in def. 2.6). It cannot be too much emphasized that this is a crucial point, for a physical theory is more than just a space of states: it is also a *labeling* of the observables over this state space, that associates to the elementary observables a particular physical meaning. This labeling is the interface that allows us to make the connection between a given concrete measure protocol and an observable of the theory, between the experimental world and the mathematical formalism. Hence, from a physical point of view, a rendering of the physical phase space would be useless if we do not tell at the same time how the elementary observables of our theory are constructed in this rendering.

As already mentioned above, we have, dual to the translation of observables, the possibility of transporting dynamical states back to the kinematical theory (as projective families of orbits), and again this transport reflects the map $\delta_{\infty}^{-1} \langle \cdot \rangle$ that sends a point in $\mathcal{M}_{\infty}^{\text{DYN}}$ to an orbit in $\mathcal{M}_{\infty}^{\text{SHELL}}$. This is probably not needed when the constraints are there to implement dynamics, since, as soon as we have obtained the physical state space (and observables thereon!), the kinematical theory has played its role and can be discarded. However, the same mathematical formalism of imposing constraints can also describe the symmetry restriction of a theory. It has in this case an entirely different physical interpretation, and we are then not only interested in the symmetry restricted theory itself, but we also want to understand its states as special, symmetric states in the full theory (note that the constraints describing symmetry restriction being second class, we map a state on the restricted

Figure 3.3 – Elementary reduction and rendering

state to a state on the unrestricted side: orbits are in this case just single points).

**Proposition 3.14** Let $(\mathcal{L}, \mathcal{M}^{\text{KIN}}, \pi^{\text{KIN}})^{\downarrow}$ be a rendering of a (possibly infinite dimensional) symplectic manifold $\mathcal{M}_{\infty}^{\text{KIN}}$ (def. 2.6) and let $(\mathcal{M}_{\infty}^{\text{DYN}}, \mathcal{M}_{\infty}^{\text{SHELL}}, \delta_{\infty})$ be a phase space reduction of $\mathcal{M}_{\infty}^{\text{KIN}}$. We suppose that, for all $\eta \in \mathcal{L}$:

1. $\mathcal{M}_{\eta}$ is finite dimensional;

2. we are given a phase space reduction $(\mathcal{M}_{\eta}^{\text{DYN}}, \mathcal{M}_{\eta}^{\text{SHELL}}, \delta_{\eta})$ of $\mathcal{M}_{\eta}^{\text{KIN}}$ that is related by $\pi_{\infty \to \eta}^{\text{KIN}}$ to the reduction of $\mathcal{M}_{\infty}^{\text{KIN}}$.

Then, we have an elementary reduction $(\mathcal{L}, \mathcal{M}, \pi, \delta)^{\text{DYN}}$ of $(\mathcal{L}, \mathcal{M}^{\text{KIN}}, \pi^{\text{KIN}})^{\downarrow}$ and a rendering of $\mathcal{M}_{\infty}^{\text{DYN}}$ by $(\mathcal{L}, \mathcal{M}^{\text{DYN}}, \pi^{\text{DYN}})^{\downarrow}$ such that, for any $y_{\infty} \in \mathcal{M}_{\infty}^{\text{DYN}}$:

$$\widehat{\sigma}_{\downarrow}^{\text{KIN}}\left(\delta_{\infty}^{-1}\langle y_{\infty}\rangle\right) = \Delta \circ \sigma_{\downarrow}^{\text{DYN}}(y_{\infty}), \tag{3.14.1}$$

where $\widehat{\sigma}_{\downarrow}^{\text{KIN}} : \mathcal{P}(\mathcal{M}_{\infty}^{\text{KIN}}) \to \widehat{\mathcal{S}}_{(\mathcal{L},\mathcal{M}^{\text{KIN}},\pi^{\text{KIN}})}^{\downarrow}$, $\sigma_{\downarrow}^{\text{DYN}} : \mathcal{M}_{\infty}^{\text{DYN}} \to \mathcal{S}_{(\mathcal{L}',\mathcal{M}^{\text{DYN}},\pi^{\text{DYN}})}^{\downarrow}$ are defined in analogy to def. 2.6. Moreover, for any $f \in \mathcal{A}_{(\mathcal{L},\mathcal{M}^{\text{KIN}},\pi^{\text{KIN}})}^{\downarrow}$, we have:

$$\left(\beta_{\uparrow}^{\text{KIN}}(f)\right)^{\text{DYN}} = \beta_{\uparrow}^{\text{DYN}}(f^{\text{DYN}}), \tag{3.14.2}$$

where $\beta_{\uparrow}^{\text{KIN/DYN}} : \mathcal{A}_{(\mathcal{L},\mathcal{M}^{\text{KIN/DYN}},\pi^{\text{KIN/DYN}})}^{\downarrow} \to B(\mathcal{M}_{\infty}^{\text{KIN/DYN}})$ are defined in the same way as $\alpha_{\uparrow}^{\text{KIN/DYN}} : \mathcal{O}_{(\mathcal{L},\mathcal{M}^{\text{KIN/DYN}},\pi^{\text{KIN/DYN}})}^{\downarrow} \to C^{\infty}(\mathcal{M}_{\infty}^{\text{KIN/DYN}}, \mathbb{R})$ (def. 2.6).

**Proof** From prop. 3.12, we can complete the phase space reduction $(\mathcal{M}_{\infty}^{\text{DYN}}, \mathcal{M}_{\infty}^{\text{SHELL}}, \delta_{\infty})$ of $\mathcal{M}_{\infty}^{\text{KIN}}$ into an elementary reduction $(\mathcal{L} \sqcup \{\infty\}, \mathcal{M}, \pi, \delta)^{\text{DYN}}$ of $(\mathcal{L} \sqcup \{\infty\}, \mathcal{M}^{\text{KIN}}, \pi^{\text{KIN}})^{\downarrow}$. In particular, $(\mathcal{L} \sqcup \{\infty\}, \mathcal{M}^{\text{DYN}}, \pi^{\text{DYN}})^{\downarrow}$ is a projective system of phase spaces; in other words, $(\mathcal{L}, \mathcal{M}^{\text{DYN}}, \pi^{\text{DYN}})^{\downarrow}$ is a rendering of $\mathcal{M}_{\infty}^{\text{DYN}}$.

Eq. (3.14.1) and eq. (3.14.2) then follow by applying twice the corresponding results from prop. 3.11 (to go down from $\mathcal{L} \sqcup \{\infty\}$ to both $\mathcal{L}$ and $\{\infty\}$), together with:

$$\widehat{\sigma}_{\downarrow}^{\text{KIN}} = \widehat{\sigma}_{\mathcal{L} \sqcup \{\infty\} \to \mathcal{L}}^{\text{KIN}} \circ \widehat{\sigma}_{\mathcal{L} \sqcup \{\infty\} \to \{\infty\}}^{\text{KIN},-1} \quad \& \quad \sigma_{\downarrow}^{\text{DYN}} = \sigma_{\mathcal{L} \sqcup \{\infty\} \to \mathcal{L}}^{\text{DYN}} \circ \sigma_{\mathcal{L} \sqcup \{\infty\} \to \{\infty\}}^{\text{DYN},-1},$$

and:

$$\beta^{\text{KIN}}_{\uparrow} = \beta^{\text{KIN},-1}_{\mathcal{L} \sqcup \{\infty\} \leftarrow \{\infty\}} \circ \beta^{\text{KIN}}_{\mathcal{L} \sqcup \{\infty\} \leftarrow \mathcal{L}} \quad \& \quad \beta^{\text{DYN}}_{\uparrow} = \beta^{\text{DYN},-1}_{\mathcal{L} \sqcup \{\infty\} \leftarrow \{\infty\}} \circ \beta^{\text{DYN}}_{\mathcal{L} \sqcup \{\infty\} \leftarrow \mathcal{L}} \, .$$

□

## 3.2 Regularized reductions

We now turn to the general case, where, typically, the prerequisites of the previous section will *not* be satisfied.

Recall that, as underlined above (prop. 3.12), these prerequisites will become milder and milder if we look for elementary reductions only defined on smaller and smaller cofinal subsets of the label set $\mathcal{L}$: the argument is that it's easier to write closed dynamics over truncations of the theory if we consent to give up the coarsest truncations for lost and to only try to formulate such truncated dynamics in partial theories retaining enough elementary observables (and being thus able to exhibit finer properties of the states). On the other hand, for what we are interested in (namely, defining a projective structure for the dynamical theory, constructing on it the observables inherited from the kinematical theory, and, if need be, embedding its states in the initial projective structure), such an elementary restriction restricted to a cofinal part of $\mathcal{L}$ is all we need.

This observation motivates the following strategy: we will try to design an approximating scheme, indexed by a directed set $\mathcal{E}$, that approaches the exact constraints (unadapted to the projective structure) by approximate constraints, projecting well on the $\mathcal{M}^{\text{KIN}}_{\eta}$ at least for all $\eta \in \mathcal{L}^{\varepsilon}$ (where $\varepsilon \in \mathcal{E}$ parametrizes the level of approximation and $\mathcal{L}^{\varepsilon}$ is a cofinal part of $\mathcal{L}$ that depends on $\varepsilon$). We expect that the label subset $\mathcal{L}^{\varepsilon}$ will get smaller and smaller (yet remaining cofinal), since formulating more accurate approximations of the dynamics will require deeper and deeper knowledge of the properties of the states (such knowledge that is only accessible in partial theories with labels at the high end of $\mathcal{L}$).

As an illustration of this idea, suppose that $\mathcal{L}$ consists of all possible finite subsets of points on the real line (ordered by inclusion), take $\mathcal{E}$ to be the set of positive reals $\varepsilon$ and define $\mathcal{L}_{\varepsilon} \subset \mathcal{L}$ to select those subsets in which next neighbor points have a distance of at most $\varepsilon$. Thus a label $\eta \in \mathcal{L}$ will only qualify for belonging to $\mathcal{L}_{\varepsilon}$ if, given a real function $f$, it can provide an approximation $f|_{\eta}$ with at least a resolution of $\varepsilon$ (over the convex hull of $\eta$). As $\varepsilon$ gets smaller and smaller, we retain less and less labels $\eta$, yet $\mathcal{L}_{\varepsilon}$ will keep cofinal. Notice that in this example we would use on $\mathcal{E}$ the reverse order $\varepsilon \preccurlyeq \varepsilon' \Leftrightarrow \varepsilon \geqslant \varepsilon'$, because we think of the partial order on $\mathcal{E}$ in terms of coarser lattices being included in finer ones, rather than as an ordering of the lattice parameters (thus we will sometimes refer to the continuum limit as $\varepsilon = \infty$, in the sense of having an infinitely fine lattice, although in the present case $\varepsilon = 0$ would have been more intuitive).

To make it more precise what we mean by approaching the exact constraints, we want the approximation scheme to come with an additional input, namely a family of projections, going from the space of exact solutions of the dynamics $\mathcal{M}^{\text{DYN}}$ into each space of approximate solutions $\mathcal{M}^{\text{DYN},\varepsilon}$: it will tell us, for each level of approximation $\varepsilon \in \mathcal{E}$, how to map the exacts orbits in $\mathcal{M}^{\text{SHELL}}$ to their approximate versions in $\mathcal{M}^{\text{SHELL},\varepsilon}$. In other words, we will associate to each orbit in the exact constraint surface a family (indexed by $\mathcal{E}$) of orbits intended to approach it, thus setting the stage to formulate a notion of convergence (this point will be examined more closely in the second half of the present subsection).

Besides, it is sensible that the map from $\mathcal{M}^{\text{DYN}}$ to $\mathcal{M}^{\text{DYN},\varepsilon}$ does not retain all degrees of freedom, so that it only depends on the most distinctive properties of the dynamical states in $\mathcal{M}^{\text{DYN}}$: the approximation of an exact solution should drop those finest details, that can anyway not be handled correctly by the coarse dynamics underlying $\mathcal{M}^{\text{DYN},\varepsilon}$. More precisely, we will require that the family of approximated theories build on their own a projective system of phase spaces (with label set $\mathcal{E}$), and, in addition, we would like the approximating maps, bringing us from a finer approximated dynamical theory to a coarser one, to be expressible at the level of the truncated theories $\mathcal{M}_{\eta}^{\text{DYN},\varepsilon}$, so that we can assemble all $\mathcal{M}_{\eta}^{\text{DYN},\varepsilon}$ into a big projective system of phase spaces (whose label set will be a part of $\mathcal{E} \times \mathcal{L}$). The return of these quite restrictive requirements is that it supplies immediately a dynamical projective structure, where we can represent the dynamical states and start doing calculations with them, even before we have settled the question of convergence.

Clearly, we are assuming here that we are provided with some non-trivial input, that will have to come from a precise understanding of the system under study. The examples in [10], besides demonstrating that the procedure described here can indeed be put into practice in simple systems, also give some insights on how the needed input can be obtained, but it will require more extensive investigations to develop systematic ways of constructing suitable approximating schemes in the sense above.

**Proposition 3.15** Let $\mathcal{E}, \preccurlyeq$ and $\mathcal{L}, \preccurlyeq$ be preordered, directed sets and suppose there exists for all $\varepsilon \in \mathcal{E}$ a cofinal part $\mathcal{L}^{\varepsilon}$ of $\mathcal{L}$ such that:

$$\forall \varepsilon \preccurlyeq \varepsilon', \mathcal{L}^{\varepsilon} \supset \mathcal{L}^{\varepsilon'}.$$

We define $\mathcal{EL} := \{(\varepsilon, \eta) \mid \varepsilon \in \mathcal{E}, \eta \in \mathcal{L}^{\varepsilon}\}$ and equip it with the preorder:

$$\forall (\varepsilon, \eta), (\varepsilon', \eta') \in \mathcal{EL}, \quad (\varepsilon, \eta) \preccurlyeq (\varepsilon', \eta') \Leftrightarrow \left(\varepsilon \preccurlyeq \varepsilon' \ \& \ \eta \preccurlyeq \eta'\right).$$

Then $\mathcal{EL}, \preccurlyeq$ is directed.

**Proof** Let $(\varepsilon, \eta), (\varepsilon', \eta') \in \mathcal{EL}$. Since $\mathcal{E}$ and $\mathcal{L}$ are directed, there exist $\varepsilon'' \in \mathcal{E}$ and $\widetilde{\eta} \in \mathcal{L}$ such that $\varepsilon, \varepsilon' \preccurlyeq \varepsilon''$ and $\eta, \eta' \preccurlyeq \widetilde{\eta}$. $\mathcal{L}^{\varepsilon''}$ being cofinal in $\mathcal{L}$, there exists $\eta'' \in \mathcal{L}^{\varepsilon''} / \widetilde{\eta} \preccurlyeq \eta''$. $\square$

**Definition 3.16** Let $(\mathcal{L}, \mathcal{M}^{\text{KIN}}, \pi^{\text{KIN}})^{\downarrow}$ be a projective system of phase spaces. A regularized reduction of $(\mathcal{L}, \mathcal{M}^{\text{KIN}}, \pi^{\text{KIN}})^{\downarrow}$ is a sextuple:

$$\left(\mathcal{E}, (\mathcal{L}^{\varepsilon})_{\varepsilon \in \mathcal{E}}, \left(\mathcal{M}_{\eta}^{\text{DYN},\varepsilon}\right)_{(\varepsilon,\eta) \in \mathcal{EL}}, \left(\mathcal{M}_{\eta}^{\text{SHELL},\varepsilon}\right)_{(\varepsilon,\eta) \in \mathcal{EL}}, \left(\pi_{\eta' \to \eta}^{\text{DYN},\varepsilon' \to \varepsilon}\right)_{(\varepsilon,\eta) \preccurlyeq (\varepsilon',\eta')}, \left(\delta_{\eta}^{\varepsilon}\right)_{(\varepsilon,\eta) \in \mathcal{EL}}\right)$$

such that:

1. $\mathcal{E}$ is a directed set indexing a family $(\mathcal{L}^{\varepsilon})_{\varepsilon \in \mathcal{E}}$ of decreasing ($\forall \varepsilon \preccurlyeq \varepsilon', \mathcal{L}^{\varepsilon} \supset \mathcal{L}^{\varepsilon'}$), cofinal parts of $\mathcal{L}$ as in prop. 3.15;

2. $\forall \varepsilon \in \mathcal{E}, \left(\left(\mathcal{M}_{\eta}^{\text{DYN},\varepsilon}\right)_{\eta \in \mathcal{L}^{\varepsilon}}, \left(\mathcal{M}_{\eta}^{\text{SHELL},\varepsilon}\right)_{\eta \in \mathcal{L}^{\varepsilon}}, \left(\pi_{\eta' \to \eta}^{\text{DYN},\varepsilon \to \varepsilon}\right)_{\eta \preccurlyeq \eta'}, \left(\delta_{\eta}^{\varepsilon}\right)_{\eta \in \mathcal{L}^{\varepsilon}}\right)$ is an elementary reduction of $(\mathcal{L}^{\varepsilon}, \mathcal{M}^{\text{KIN}}, \pi^{\text{KIN}})^{\downarrow}$;

3. $\left(\mathcal{EL}, \left(\mathcal{M}_{\eta}^{\text{DYN},\varepsilon}\right)_{(\varepsilon,\eta) \in \mathcal{EL}}, \left(\pi_{\eta' \to \eta}^{\text{DYN},\varepsilon' \to \varepsilon}\right)_{(\varepsilon,\eta) \preccurlyeq (\varepsilon',\eta')}\right)$ is a projective system of phase spaces.

Whenever possible, we will use the shortened notation $(\mathcal{L}, \mathcal{M}, \pi, \delta)^{\text{DYN}, \mathcal{E}}$ instead of $\left(\mathcal{E}, (\mathcal{L}^\varepsilon)_{\varepsilon \in \mathcal{E}}, \left(\mathcal{M}_\eta^{\text{DYN}, \varepsilon}\right)_{(\varepsilon, \eta) \in \mathcal{EL}}, \left(\mathcal{M}_\eta^{\text{SHELL}, \varepsilon}\right)_{(\varepsilon, \eta) \in \mathcal{EL}}, \left(\pi_{\eta' \to \eta}^{\text{DYN}, \varepsilon \to \varepsilon}\right)_{(\varepsilon, \eta) \preccurlyeq (\varepsilon', \eta')}, \left(\delta_\eta^\varepsilon\right)_{(\varepsilon, \eta) \in \mathcal{EL}}\right)$.

At that point we have written a projective structure for the dynamical theory, but as we emphasized in the previous subsection, constructing the space of physical states is of little use if we do not prescribe how to define on it the observables inherited from the kinematical theory. As a first step in this direction, we will construct maps that transport kinematical observables into the dynamical theory *at some level of approximation* $\varepsilon$: given a particular kinematical observable, the dynamical observables constructed this way, for all possible $\varepsilon$, should be thought of as successive approximations of the exact dynamical version of this kinematical observable.

Moreover, we can check that these maps transform well under restriction of the label sets $\mathcal{L}$ and $\mathcal{E}$ (provided the label subsets $\mathcal{L}'$ and $\mathcal{E}'$ considered are such that we still have a regularized reduction after restricting ourselves to $\mathcal{E}'\mathcal{L}'$). We will make use of this result at the end of the present subsection, when we will consider how regularized reductions interact with renderings.

**Definition 3.17** We consider the same objects as in def. 3.16. For $\varepsilon \in \mathcal{E}$, we define $\Delta^\varepsilon : \mathcal{S}^\downarrow_{(\mathcal{EL}, \mathcal{M}^{\text{DYN}}, \pi^{\text{DYN}})} \to \widehat{\mathcal{S}}^\downarrow_{(\mathcal{L}, \mathcal{M}^{\text{KIN}}, \pi^{\text{KIN}})}$ as:

$$\Delta^\varepsilon := \widehat{\sigma}_{\mathcal{L} \to \mathcal{L}^\varepsilon}^{\text{KIN}, -1} \circ \Delta^\varepsilon_{\mathcal{L}^\varepsilon} \circ \sigma^{\text{DYN}}_{\mathcal{EL} \to \mathcal{L}^\varepsilon},$$

where $\sigma^{\text{DYN}}_{\mathcal{EL} \to \mathcal{L}^\varepsilon} : \mathcal{S}^\downarrow_{(\mathcal{EL}, \mathcal{M}^{\text{DYN}}, \pi^{\text{DYN}})} \to \mathcal{S}^\downarrow_{(\mathcal{L}^\varepsilon, \mathcal{M}^{\text{DYN}, \varepsilon}, \pi^{\text{DYN}, \varepsilon \to \varepsilon})}$ is defined as in prop. 2.5 (for the directed part $\{(\varepsilon, \eta) \mid \eta \in \mathcal{L}^\varepsilon\}$ of $\mathcal{EL}$), $\Delta^\varepsilon_{\mathcal{L}^\varepsilon} : \mathcal{S}^\downarrow_{(\mathcal{L}^\varepsilon, \mathcal{M}^{\text{DYN}, \varepsilon}, \pi^{\text{DYN}, \varepsilon \to \varepsilon})} \to \widehat{\mathcal{S}}^\downarrow_{(\mathcal{L}^\varepsilon, \mathcal{M}^{\text{KIN}}, \pi^{\text{KIN}})}$ is defined as in def. 3.8 (for the elementary reduction $(\mathcal{L}^\varepsilon, \mathcal{M}^\varepsilon, \pi^{\varepsilon \to \varepsilon}, \delta^\varepsilon)^\downarrow$ of $(\mathcal{L}^\varepsilon, \mathcal{M}^{\text{KIN}}, \pi^{\text{KIN}})^\downarrow$), and $\widehat{\sigma}^{\text{KIN}}_{\mathcal{L} \to \mathcal{L}^\varepsilon} : \widehat{\mathcal{S}}^\downarrow_{(\mathcal{L}, \mathcal{M}^{\text{KIN}}, \pi^{\text{KIN}})} \to \widehat{\mathcal{S}}^\downarrow_{(\mathcal{L}^\varepsilon, \mathcal{M}^{\text{KIN}}, \pi^{\text{KIN}})}$ is defined in analogy to prop. 2.5 (for the cofinal part $\mathcal{L}^\varepsilon$ of $\mathcal{L}$).

Similarly, we define $(\,\cdot\,)^\varepsilon : \mathcal{A}^\downarrow_{(\mathcal{L}, \mathcal{M}^{\text{KIN}}, \pi^{\text{KIN}})} \to \mathcal{A}^\downarrow_{(\mathcal{EL}, \mathcal{M}^{\text{DYN}}, \pi^{\text{DYN}})}$ as:

$$(\,\cdot\,)^\varepsilon := \beta^{\text{DYN}}_{\mathcal{EL} \leftarrow \mathcal{L}^\varepsilon} \circ (\,\cdot\,)^{\text{DYN}, \varepsilon} \circ \beta^{\text{KIN}, -1}_{\mathcal{L} \leftarrow \mathcal{L}^\varepsilon},$$

where $\beta^{\text{DYN}}_{\mathcal{EL} \leftarrow \mathcal{L}^\varepsilon} : \mathcal{A}^\downarrow_{(\mathcal{L}^\varepsilon, \mathcal{M}^{\text{DYN}, \varepsilon}, \pi^{\text{DYN}, \varepsilon \to \varepsilon})} \to \mathcal{A}^\downarrow_{(\mathcal{EL}, \mathcal{M}^{\text{DYN}}, \pi^{\text{DYN}})}$ and $\beta^{\text{KIN}}_{\mathcal{L} \leftarrow \mathcal{L}^\varepsilon} : \mathcal{A}^\downarrow_{(\mathcal{L}^\varepsilon, \mathcal{M}^{\text{KIN}}, \pi^{\text{KIN}})} \to \mathcal{A}^\downarrow_{(\mathcal{L}, \mathcal{M}^{\text{KIN}}, \pi^{\text{KIN}})}$ are defined as in analogy to prop. 2.5 (for the directed part $\{(\varepsilon, \eta) \mid \eta \in \mathcal{L}^\varepsilon\}$ of $\mathcal{EL}$ and the cofinal part $\mathcal{L}^\varepsilon$ of $\mathcal{L}$) and $(\,\cdot\,)^{\text{DYN}, \varepsilon} : \mathcal{A}^\downarrow_{(\mathcal{L}^\varepsilon, \mathcal{M}^{\text{DYN}, \varepsilon}, \pi^{\text{DYN}, \varepsilon \to \varepsilon})} \to \mathcal{A}^\downarrow_{(\mathcal{L}^\varepsilon, \mathcal{M}^{\text{KIN}}, \pi^{\text{KIN}})}$ is defined as in prop. 3.9 (for the elementary reduction $(\mathcal{L}^\varepsilon, \mathcal{M}^\varepsilon, \pi^{\varepsilon \to \varepsilon}, \delta^\varepsilon)^\downarrow$ of $(\mathcal{L}^\varepsilon, \mathcal{M}^{\text{KIN}}, \pi^{\text{KIN}})^\downarrow$).

We have for all $y \in \mathcal{S}^\downarrow_{(\mathcal{EL}, \mathcal{M}^{\text{DYN}}, \pi^{\text{DYN}})}$ and all $f \in \mathcal{A}^\downarrow_{(\mathcal{L}, \mathcal{M}^{\text{KIN}}, \pi^{\text{KIN}})}$:

$$f^\varepsilon(y) = f(\Delta^\varepsilon(y)).$$

**Proposition 3.18** Let $(\mathcal{L}, \mathcal{M}^{\text{KIN}}, \pi^{\text{KIN}})^\downarrow$ be a projective system of phase spaces and let $(\mathcal{L}, \mathcal{M}, \pi, \delta)^{\text{DYN}, \mathcal{E}}$ be a regularized reduction of $(\mathcal{L}, \mathcal{M}^{\text{KIN}}, \pi^{\text{KIN}})^\downarrow$. Let $\mathcal{L}'$ and $\mathcal{E}'$ be directed subsets of $\mathcal{L}$ and $\mathcal{E}$ respectively, such that, for all $\varepsilon \in \mathcal{E}'$, $\mathcal{L}'^\varepsilon := \mathcal{L}^\varepsilon \cap \mathcal{L}'$ is a cofinal part of $\mathcal{L}'$.

Then, $(\mathcal{L}', \mathcal{M}, \pi, \delta)^{\text{DYN}, \mathcal{E}'}$ is a regularized reduction of $(\mathcal{L}', \mathcal{M}^{\text{KIN}}, \pi^{\text{KIN}})^\downarrow$ and, for any $\varepsilon \in \mathcal{E}'$, we have:

$$\widehat{\sigma}^{\text{KIN}}_{\mathcal{L}\to\mathcal{L}'} \circ \Delta^{\varepsilon} = \Delta'^{\varepsilon} \circ \sigma^{\text{DYN}}_{\mathcal{E}\mathcal{L}\to\mathcal{E}'\mathcal{L}'}, \qquad (3.18.1)$$

where $\widehat{\sigma}^{\text{KIN}}_{\mathcal{L}\to\mathcal{L}'} : \widehat{\mathcal{S}}^{\downarrow}_{(\mathcal{L},\mathcal{M}^{\text{KIN}},\pi^{\text{KIN}})} \to \widehat{\mathcal{S}}^{\downarrow}_{(\mathcal{L}',\mathcal{M}^{\text{KIN}},\pi^{\text{KIN}})}$, $\sigma^{\text{DYN}}_{\mathcal{E}\mathcal{L}\to\mathcal{E}'\mathcal{L}'} : \mathcal{S}^{\downarrow}_{(\mathcal{E}\mathcal{L},\mathcal{M}^{\text{DYN}},\pi^{\text{DYN}})} \to \mathcal{S}^{\downarrow}_{(\mathcal{E}'\mathcal{L}',\mathcal{M}^{\text{DYN}},\pi^{\text{DYN}})}$ are defined in analogy to prop. 2.5, while $\Delta^{\varepsilon} : \mathcal{S}^{\downarrow}_{(\mathcal{E}\mathcal{L},\mathcal{M}^{\text{DYN}},\pi^{\text{DYN}})} \to \widehat{\mathcal{S}}^{\downarrow}_{(\mathcal{L},\mathcal{M}^{\text{KIN}},\pi^{\text{KIN}})}$ and $\Delta'^{\varepsilon} : \mathcal{S}^{\downarrow}_{(\mathcal{E}'\mathcal{L}',\mathcal{M}^{\text{DYN}},\pi^{\text{DYN}})} \to \widehat{\mathcal{S}}^{\downarrow}_{(\mathcal{L}',\mathcal{M}^{\text{KIN}},\pi^{\text{KIN}})}$ are defined as in def. 3.17.

In addition, for any $f \in \mathcal{A}^{\downarrow}_{(\mathcal{L}',\mathcal{M}^{\text{KIN}},\pi^{\text{KIN}})}$ and any $\varepsilon \in \mathcal{E}'$, we have:

$$\left(\beta^{\text{KIN}}_{\mathcal{L}\leftarrow\mathcal{L}'}(f)\right)^{\varepsilon} = \beta^{\text{DYN}}_{\mathcal{E}\mathcal{L}\leftarrow\mathcal{E}'\mathcal{L}'}(f^{\varepsilon}), \qquad (3.18.2)$$

where $\beta^{\text{KIN}}_{\mathcal{L}\leftarrow\mathcal{L}'} : \mathcal{A}^{\downarrow}_{(\mathcal{L}',\mathcal{M}^{\text{KIN}},\pi^{\text{KIN}})} \to \mathcal{A}^{\downarrow}_{(\mathcal{L},\mathcal{M}^{\text{KIN}},\pi^{\text{KIN}})}$ and $\beta^{\text{DYN}}_{\mathcal{L}\leftarrow\mathcal{L}'} : \mathcal{A}^{\downarrow}_{(\mathcal{E}'\mathcal{L}',\mathcal{M}^{\text{DYN}},\pi^{\text{DYN}})} \to \mathcal{A}^{\downarrow}_{(\mathcal{E}\mathcal{L},\mathcal{M}^{\text{DYN}},\pi^{\text{DYN}})}$ are defined in analogy to prop. 2.5, while $(\,\cdot\,)^{\varepsilon} : \mathcal{A}^{\downarrow}_{(\mathcal{L},\mathcal{M}^{\text{KIN}},\pi^{\text{KIN}})} \to \mathcal{A}^{\downarrow}_{(\mathcal{E}\mathcal{L},\mathcal{M}^{\text{DYN}},\pi^{\text{DYN}})}$ and $(\,\cdot\,)^{\varepsilon} : \mathcal{A}^{\downarrow}_{(\mathcal{L}',\mathcal{M}^{\text{KIN}},\pi^{\text{KIN}})} \to \mathcal{A}^{\downarrow}_{(\mathcal{E}'\mathcal{L}',\mathcal{M}^{\text{DYN}},\pi^{\text{DYN}})}$ are defined as in def. 3.17.

**Proof** $\mathcal{L}'^{\varepsilon}$ being a cofinal part of $\mathcal{L}'$ for all $\varepsilon \in \mathcal{E}'$ ensures that def. 3.16.1 is fullfiled. Moreover, $\mathcal{E}'\mathcal{L}'$ is then a directed subset of $\mathcal{E}\mathcal{L}$, hence def. 3.16.3 holds. Lastly, def. 3.16.2 follows from prop. 3.11, since, for any $\varepsilon \in \mathcal{E}'$, $\mathcal{L}'^{\varepsilon}$ is a directed part of $\mathcal{L}^{\varepsilon}$ (as a cofinal part of the directed set $\mathcal{L}'$).

Let $\varepsilon \in \mathcal{E}'$. We have $\mathcal{L} \supset \mathcal{L}'$, $\mathcal{L}^{\varepsilon} \supset \mathcal{L}'^{\varepsilon}$, hence:

$$\widehat{\sigma}^{\text{KIN}}_{\mathcal{L}'\to\mathcal{L}'^{\varepsilon}} \circ \widehat{\sigma}^{\text{KIN}}_{\mathcal{L}\to\mathcal{L}'} = \widehat{\sigma}^{\text{KIN}}_{\mathcal{L}^{\varepsilon}\to\mathcal{L}'^{\varepsilon}} \circ \widehat{\sigma}^{\text{KIN}}_{\mathcal{L}\to\mathcal{L}^{\varepsilon}}.$$

And, from $\mathcal{E}\mathcal{L} \supset \mathcal{E}'\mathcal{L}'$, $\mathcal{L}^{\varepsilon} \supset \mathcal{L}'^{\varepsilon}$ (identifying $\mathcal{L}^{\varepsilon}$ with the subset $\{(\varepsilon,\eta) \mid \eta \in \mathcal{L}^{\varepsilon}\}$ of $\mathcal{E}\mathcal{L}$ and $\mathcal{L}'^{\varepsilon}$ with the subset $\{(\varepsilon,\eta) \mid \eta \in \mathcal{L}'^{\varepsilon}\}$ of $\mathcal{E}'\mathcal{L}'$ as in def. 3.17), we also have:

$$\sigma^{\text{DYN}}_{\mathcal{L}^{\varepsilon}\to\mathcal{L}'^{\varepsilon}} \circ \sigma^{\text{DYN}}_{\mathcal{E}\mathcal{L}\to\mathcal{L}^{\varepsilon}} = \sigma^{\text{DYN}}_{\mathcal{E}'\mathcal{L}'\to\mathcal{L}'^{\varepsilon}} \circ \sigma^{\text{DYN}}_{\mathcal{E}\mathcal{L}\to\mathcal{E}'\mathcal{L}'}.$$

So, using the definition of $\Delta^{\varepsilon}$ and $\Delta'^{\varepsilon}$ from def. 3.17:

$$\widehat{\sigma}^{\text{KIN}}_{\mathcal{L}\to\mathcal{L}'} \circ \Delta^{\varepsilon} = \widehat{\sigma}^{\text{KIN}}_{\mathcal{L}\to\mathcal{L}'} \circ \widehat{\sigma}^{\text{KIN},-1}_{\mathcal{L}\to\mathcal{L}^{\varepsilon}} \circ \Delta^{\varepsilon}_{\mathcal{L}^{\varepsilon}} \circ \sigma^{\text{DYN}}_{\mathcal{E}\mathcal{L}\to\mathcal{L}^{\varepsilon}}$$

$$= \widehat{\sigma}^{\text{KIN},-1}_{\mathcal{L}'\to\mathcal{L}'^{\varepsilon}} \circ \widehat{\sigma}^{\text{KIN}}_{\mathcal{L}^{\varepsilon}\to\mathcal{L}'^{\varepsilon}} \circ \Delta^{\varepsilon}_{\mathcal{L}^{\varepsilon}} \circ \sigma^{\text{DYN}}_{\mathcal{E}\mathcal{L}\to\mathcal{L}^{\varepsilon}}$$

$$= \widehat{\sigma}^{\text{KIN},-1}_{\mathcal{L}'\to\mathcal{L}'^{\varepsilon}} \circ \Delta'^{\varepsilon}_{\mathcal{L}'^{\varepsilon}} \circ \sigma^{\text{DYN}}_{\mathcal{L}^{\varepsilon}\to\mathcal{L}'^{\varepsilon}} \circ \sigma^{\text{DYN}}_{\mathcal{E}\mathcal{L}\to\mathcal{L}^{\varepsilon}} \quad \text{(using prop. 3.11)}$$

$$= \widehat{\sigma}^{\text{KIN},-1}_{\mathcal{L}'\to\mathcal{L}'^{\varepsilon}} \circ \Delta'^{\varepsilon}_{\mathcal{L}'^{\varepsilon}} \circ \sigma^{\text{DYN}}_{\mathcal{E}'\mathcal{L}'\to\mathcal{L}'^{\varepsilon}} \circ \sigma^{\text{DYN}}_{\mathcal{E}\mathcal{L}\to\mathcal{E}'\mathcal{L}'}$$

$$= \Delta'^{\varepsilon} \circ \sigma^{\text{DYN}}_{\mathcal{E}\mathcal{L}\to\mathcal{E}'\mathcal{L}'}.$$

Similarly, we have:

$$(\,\cdot\,)^{\varepsilon} \circ \beta^{\text{KIN}}_{\mathcal{L}\leftarrow\mathcal{L}'} = \beta^{\text{DYN}}_{\mathcal{E}\mathcal{L}\leftarrow\mathcal{E}'\mathcal{L}'} \circ (\,\cdot\,)^{\varepsilon}.$$

$\square$

Now, we would like to give a precise definition of the convergence we have been hinting at repeatedly above. To ascertain convergence is crucial for ensuring that we will get consistent predictions when refining the level of approximation at which we are conducting the calculations.

Our unchanged goal is to make it possible to transport kinematical observables over to the dynamical theory, not only in an approximated fashion, but in such a way that we faithfully realize the transport prescribed by the exact dynamics we are trying to implement (and, if the constraints we are considering describe a symmetry restriction, we are also interested in the correct embedding of the symmetric states in the full theory; as we already underlined in the previous subsection, the map between observables and the one between states are the two dual aspects of the bond between the initial or kinematical theory and the restricted or dynamical one). Additionally, we would like to be able to investigate the properties of this correct dynamics (at a certain level of precision) by making use of the approximated dynamics, where calculations will probably be more tractable.

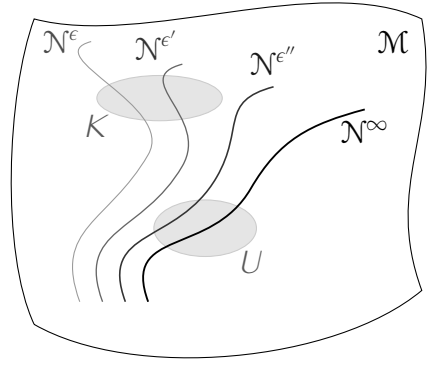

Figure 3.4 – Convergence of a net of orbits

The straightforward course is to obtain the correct transport map as the limit of the net of approximated maps introduced previously. As illustrated in fig. 3.4, we begin by defining, given a family of orbits in a manifold, a notion of convergence, which is adjusted to our method for transposing kinematical observables into dynamical ones (as explained in appendix A, this method is itself motivated by taking the indicator functions as model observables).

**Definition 3.19** Let $\mathcal{M}$ be a finite dimensional manifold and let $(\mathcal{N}^\varepsilon)_{\varepsilon \in \mathcal{E}}$ be a net of subsets of $\mathcal{M}$. We say that the net $(\mathcal{N}^\varepsilon)_{\varepsilon \in \mathcal{E}}$ converges to the subset $\mathcal{N}^\infty$ if:

1. $\forall U$ open set $\subset \mathcal{M}$ such that $\mathcal{N}^\infty \cap U \neq \varnothing$, $\exists \varepsilon \in \mathcal{E} \,/\, \forall \varepsilon' \succcurlyeq \varepsilon$, $\mathcal{N}^{\varepsilon'} \cap U \neq \varnothing$;

2. and $\forall K$ compact set $\subset \mathcal{M}$ such that $\mathcal{N}^\infty \cap K = \varnothing$, $\exists \varepsilon \in \mathcal{E} \,/\, \forall \varepsilon' \succcurlyeq \varepsilon$, $\mathcal{N}^{\varepsilon'} \cap K = \varnothing$.

**Proposition 3.20** Let $\mathcal{M}$ and $(\mathcal{N}^\varepsilon)_{\varepsilon \in \mathcal{E}}$ be as in def. 3.19 and let $f \in C_o^\infty(\mathcal{M}, \mathbb{R})$ (the space of compactly supported, smooth, real-valued functions on $\mathcal{M}$). We define:

$$f^\varepsilon := \sup\{f(x) \mid x \in \mathcal{N}^\varepsilon\} \quad \& \quad f^\infty := \sup\{f(x) \mid x \in \mathcal{N}^\infty\}.$$

Then, $\lim_{\varepsilon \in \mathcal{E}, \preccurlyeq} f^\varepsilon = f^\infty$.

**Proof** Let $\delta > 0$. We choose $x \in \mathcal{N}^\infty$ such that $f(x) > f^\infty - \delta/2$. $f$ is smooth, so there exists an open neighborhood $U$ of $x$ such that $\forall x' \in U$, $f(x') > f^\infty - \delta$. From def. 3.19.1, there exists $\varepsilon_1$ such that $\forall \varepsilon' \succcurlyeq \varepsilon_1$, $\mathcal{N}^{\varepsilon'} \cap U \neq \varnothing$. Hence, $\forall \varepsilon' \succcurlyeq \varepsilon_1$, $f^{\varepsilon'} > f^\infty - \delta$.

Let $K := \{x \in \mathcal{M} \mid f(x) \geqslant f^\infty + \delta\}$. Since $f$ is compactly supported, $K$ is compact. We have $K \cap \mathcal{N}^\infty = \varnothing$, so, from def. 3.19.2, there exists $\varepsilon_2$ such that $\forall \varepsilon' \succcurlyeq \varepsilon_2$, $\mathcal{N}^{\varepsilon'} \cap K = \varnothing$. Hence, $\forall \varepsilon' \succcurlyeq \varepsilon_2$, $f^{\varepsilon'} \leqslant f^\infty + \delta$.

$\mathcal{E}$ being directed, there exists $\varepsilon \succcurlyeq \varepsilon_1, \varepsilon_2$. Then, $\forall \varepsilon' \succcurlyeq \varepsilon$, $\left|f^\infty - f^{\varepsilon'}\right| \leqslant \delta$. $\square$

For any state over the dynamical projective system, and any $\eta \in \mathcal{L}$, the framework laid at the beginning of the present subsection allows to construct a net of approximated orbits representing this dynamical state in $\mathcal{M}_\eta^{\text{KIN}}$ (def. 3.17). Thus we can define the space $\mathcal{R}$ of all dynamical states such that, for all $\eta$, the net of their approached projections on $\mathcal{M}_\eta^{\text{KIN}}$ converges in the sense above.

Hopefully, $\mathcal{R}$ will be dense in the space of all dynamical states, but we do not require both spaces to coincide. It is in fact not really surprising that formulating the exact dynamics may require to consider only states that are well-behaved enough (we can view this prescription on the same footing as, for example, the routine requirement for fields to be smooth so that we can describe their dynamics by partial differential equations).

Reciprocally, we can also associate to any (sufficiently regular) kinematical observable its corresponding exact dynamical version, as an observable defined on $\mathcal{R}$. At this point we can comment on the issue raised at the beginning of section 2, namely that even the Poisson-algebra generated by finitely many observables could be too complicated to be represented on a finite dimensional symplectic manifold. We had argued that this problem should not arise when looking at kinematical observables, yet it might (and generically will) occur for the dynamical observables. By defining a dynamical observable as the limit of a family of imperfect estimations, we escape this difficulty. On the one hand, each such estimation can be expressed over a sufficiently big partial theory, while keeping the partial theories finite dimensional, because the regularization allows us to keep under control the algebra generated by finitely many of these approximated versions of the observables. On the other hand, an exact dynamical observable, being a limit, is allowed to depend on the full projective state $\left(y_\eta^\varepsilon\right)_{(\varepsilon, \eta) \in \mathcal{E}\mathcal{L}} \in \mathcal{R}$.

**Definition 3.21** Let $(\mathcal{L}, \mathcal{M}^{\text{KIN}}, \pi^{\text{KIN}})^\downarrow$ be a projective system of finite dimensional phase spaces and let $(Y^\varepsilon)_{\varepsilon \in \mathcal{E}}$ be a net in $\widehat{\mathcal{S}}^\downarrow_{(\mathcal{L}, \mathcal{M}^{\text{KIN}}, \pi^{\text{KIN}})}$. We say that the net $(Y^\varepsilon)_{\varepsilon \in \mathcal{E}}$ converges to the element $Y^\infty \in \widehat{\mathcal{S}}^\downarrow_{(\mathcal{L}, \mathcal{M}^{\text{KIN}}, \pi^{\text{KIN}})}$ iff:

1. $\forall \eta \in \mathcal{L}$, the net $(Y_\eta^\varepsilon)_{\varepsilon \in \mathcal{E}}$ converges to the subset $Y_\eta^\infty$ of $\mathcal{M}_\eta^{\text{KIN}}$ in the sense of def. 3.19.

If $(\mathcal{L}, \mathcal{M}, \pi, \delta)^{\text{DYN}, \mathcal{E}}$ is a regularized reduction of $(\mathcal{L}, \mathcal{M}^{\text{KIN}}, \pi^{\text{KIN}})^\downarrow$, we say that the regularization converges on a subset $\mathcal{R}$ of $\mathcal{S}^\downarrow_{(\mathcal{E}\mathcal{L}, \mathcal{M}^{\text{DYN}}, \pi^{\text{DYN}})}$ iff:

2. $\forall y \in \mathcal{R}$, the net $(\Delta^\varepsilon(y))_{\varepsilon \in \mathcal{E}}$ converges in $\widehat{\mathcal{S}}^\downarrow_{(\mathcal{L}, \mathcal{M}^{\text{KIN}}, \pi^{\text{KIN}})}$.

**Proposition 3.22** Let $(\mathcal{L}, \mathcal{M}^{\text{KIN}}, \pi^{\text{KIN}})^\downarrow$ be a projective system of finite dimensional phase spaces. We define:

$$\mathcal{A}^{o,\downarrow}_{(\mathcal{L}, \mathcal{M}^{\text{KIN}}, \pi^{\text{KIN}})} = \left\{ f \in \mathcal{A}^\downarrow_{(\mathcal{L}, \mathcal{M}^{\text{KIN}}, \pi^{\text{KIN}})} \,\Big|\, \exists \eta \in \mathcal{L}, \exists f_\eta \in f \,/\, f_\eta \in C_o^\infty(\mathcal{M}_\eta^{\text{KIN}}, \mathbb{R}) \right\}.$$

If the net $(Y^\varepsilon)_{\varepsilon \in \mathcal{E}}$ in $\widehat{\mathcal{S}}^\downarrow_{(\mathcal{L}, \mathcal{M}^{\text{KIN}}, \pi^{\text{KIN}})}$ converges to the element $Y^\infty \in \widehat{\mathcal{S}}^\downarrow_{(\mathcal{L}, \mathcal{M}^{\text{KIN}}, \pi^{\text{KIN}})}$, then, for all $f \in \mathcal{A}^{o,\downarrow}_{(\mathcal{L}, \mathcal{M}^{\text{KIN}}, \pi^{\text{KIN}})}$, the net $(f(Y^\varepsilon))_{\varepsilon \in \mathcal{E}}$ converges to $f(Y^\infty)$.

If $(\mathcal{L}, \mathcal{M}, \pi, \delta)^{\text{DYN}, \mathcal{E}}$ is a regularized reduction of $(\mathcal{L}, \mathcal{M}^{\text{KIN}}, \pi^{\text{KIN}})^\downarrow$ such that the regularization converges on $\mathcal{R} \subset \mathcal{S}^\downarrow_{(\mathcal{E}\mathcal{L}, \mathcal{M}^{\text{DYN}}, \pi^{\text{DYN}})}$, then, for all $f \in \mathcal{A}^{o,\downarrow}_{(\mathcal{L}, \mathcal{M}^{\text{KIN}}, \pi^{\text{KIN}})}$, we can define an application $f^{\text{DYN}}$ on $\mathcal{R}$ by:

$$\forall y \in \mathcal{R}, f^{\text{DYN}}(y) := \lim_{\varepsilon \in \mathcal{E}, \preccurlyeq} f^{\varepsilon}(y).$$

**Proof** Let $f \in \mathcal{A}^{o,\downarrow}_{(\mathcal{L}, \mathcal{M}^{\text{KIN}}, \pi^{\text{KIN}})}$ and let $\eta \in \mathcal{L}$, $f_\eta \in f$ such that $f_\eta \in C_o^\infty(\mathcal{M}^{\text{KIN}}_\eta, \mathbb{R})$. For all $\varepsilon \in \mathcal{E}$, we have $f(Y^\varepsilon) = \sup_{Y^\varepsilon_\eta} f_\eta$ (for we can choose any representative of $f$ to evaluate $f$ on $Y^\varepsilon$). Now, the net $(Y^\varepsilon_\eta)_{\varepsilon \in \mathcal{E}}$ converges to the subset $Y^\infty_\eta$ of $\mathcal{M}^{\text{KIN}}_\eta$, hence, using prop. 3.20:

$$\lim_{\varepsilon \in \mathcal{E}, \preccurlyeq} f(Y^\varepsilon) = \sup_{Y^\infty_\eta} f_\eta = f(Y^\infty).$$

Now, if $y \in \mathcal{R}$, we have from def. 3.21 that the net $(\Delta^\varepsilon(y))_{\varepsilon \in \mathcal{E}}$ converges in $\widehat{\mathcal{S}}^{\downarrow}_{(\mathcal{L}, \mathcal{M}^{\text{KIN}}, \pi^{\text{KIN}})}$. Hence, the net $(f(\Delta^\varepsilon(y)))_{\varepsilon \in \mathcal{E}}$ converges, but we have $\forall \varepsilon \in \mathcal{E}, f(\Delta^\varepsilon(y)) = f^\varepsilon(y)$ (def. 3.17). □

Finally, we want to discuss how renderings (def. 2.6) can be incorporated in this procedure, and more specifically, how regularized reductions can be used to mirror phase space reductions in infinite dimensional symplectic manifolds. Given a rendering of some infinite dimensional symplectic manifold $\mathcal{M}^{\text{KIN}}_\infty$, and a phase space reduction thereof, we will aim at constructing a regularized reduction whose dynamical projective system renders the dynamical phase space $\mathcal{M}^{\text{DYN}}_\infty$. Additionally, we will require that the regularization converges (at least) on the dynamical states that are identified, through this rendering, with points in $\mathcal{M}^{\text{DYN}}_\infty$, and that, for any such state, the family of orbits reflecting it in the kinematical structure can be identified with the corresponding orbit in $\mathcal{M}^{\text{KIN}}_\infty$.

Then, besides being provided with a rendering of the dynamical theory, this last point will ensure that the maps linking the kinematical side and the dynamical one are appropriately intertwined by the identifications arising from the renderings on both sides.

In prop. 3.24, we formulate more concise assumptions that are sufficient to bring forth this optimal setting. As illustrated in fig. 3.5, it relies on the successive approximations of the dynamics being formulated as phase space reductions of $\mathcal{M}^{\text{KIN}}_\infty$, and the thus defined dynamical phase spaces $\mathcal{M}^{\text{DYN},\varepsilon}_\infty$ building a rendering of the exact dynamical theory (denoted by $\mathcal{M}^{\text{DYN},\infty}_\infty$).

**Proposition 3.23** Let $(\mathcal{L}, \mathcal{M}^{\text{KIN}}, \pi^{\text{KIN}})^{\downarrow}$ be a projective system of finite dimensional phase spaces and let $(\mathcal{L}, \mathcal{M}, \pi, \delta)^{\text{DYN}, \mathcal{E}}$ be a regularized reduction of $(\mathcal{L}, \mathcal{M}^{\text{KIN}}, \pi^{\text{KIN}})^{\downarrow}$. Assume that we have a symplectic manifold $\mathcal{M}^{\text{KIN}}_\infty$ and a phase space reduction $(\mathcal{M}^{\text{DYN}}_\infty, \mathcal{M}^{\text{SHELL}}_\infty, \delta_\infty)$ of $\mathcal{M}^{\text{KIN}}_\infty$ such that:

1. we have a rendering of $\mathcal{M}^{\text{KIN}}_\infty$ by $(\mathcal{L}, \mathcal{M}^{\text{KIN}}, \pi^{\text{KIN}})^{\downarrow}$ and of $\mathcal{M}^{\text{DYN}}_\infty$ by $(\mathcal{E}\mathcal{L}, \mathcal{M}^{\text{DYN}}, \pi^{\text{DYN}})^{\downarrow}$;

2. for all $y$ in $\mathcal{M}^{\text{DYN}}_\infty$, the net $\left(\Delta^\varepsilon \circ \sigma^{\text{DYN}}_\downarrow(y)\right)_{\varepsilon \in \mathcal{E}}$ converges in $\widehat{\mathcal{S}}^{\downarrow}_{(\mathcal{L}, \mathcal{M}^{\text{KIN}}, \pi^{\text{KIN}})}$ to $\widehat{\sigma}^{\text{KIN}}_\downarrow \left(\delta_\infty^{-1} \langle \{y\} \rangle\right)$, where $\sigma^{\text{DYN}}_\downarrow : \mathcal{M}^{\text{DYN}}_\infty \to \mathcal{S}^{\downarrow}_{(\mathcal{E}\mathcal{L}, \mathcal{M}^{\text{DYN}}, \pi^{\text{DYN}})}$ is defined as in def. 2.6 and $\widehat{\sigma}^{\text{KIN}}_\downarrow : \mathcal{P}(\mathcal{M}^{\text{KIN}}_\infty) \to \widehat{\mathcal{S}}^{\downarrow}_{(\mathcal{L}, \mathcal{M}^{\text{KIN}}, \pi^{\text{KIN}})}$ is defined in a similar way.

Then, the regularization converges on $\mathcal{R} := \text{Im}\, \sigma^{\text{DYN}}_\downarrow$ and for all $y \in \mathcal{M}^{\text{DYN}}_\infty$, for all $f \in \mathcal{A}^{o,\downarrow}_{(\mathcal{L}, \mathcal{M}^{\text{KIN}}, \pi^{\text{KIN}})}$, we have:

$$f^{\text{DYN}} \circ \sigma^{\text{DYN}}_\downarrow(y) = \left(\beta^{\text{KIN}}_\uparrow(f)\right)^{\text{DYN}}(y).$$

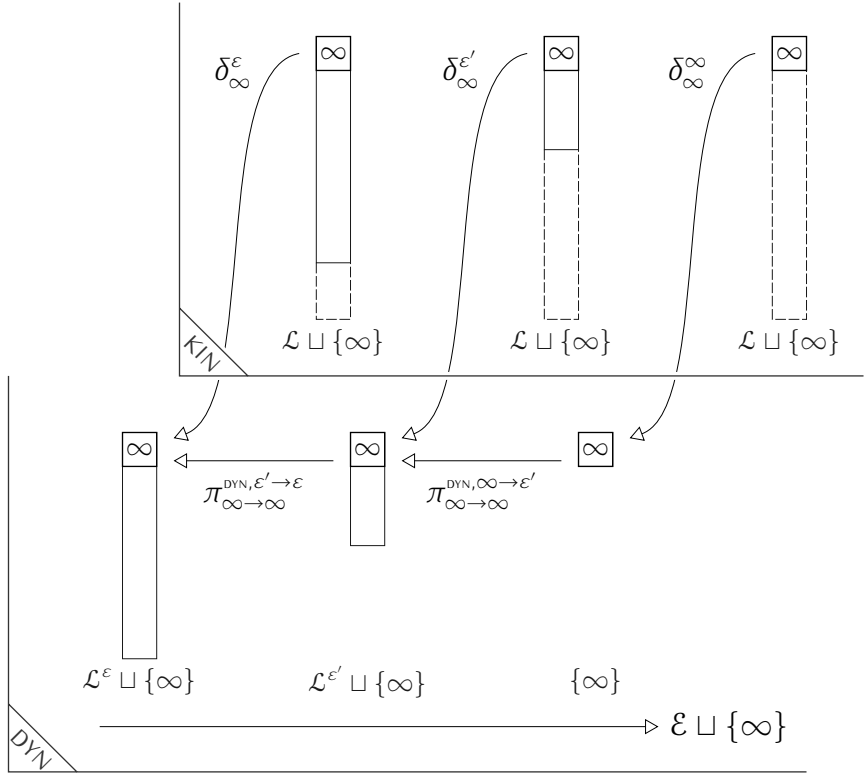

Figure 3.5 – Regularized reduction and rendering

**Proof** Let $y \in \mathcal{M}_\infty^{\text{DYN}}$ and $f \in \mathcal{A}_{(\mathcal{L}, \mathcal{M}^{\text{KIN}}, \pi^{\text{KIN}})}^{o, \downarrow}$. We have, using prop. 3.22:

$$f^{\text{DYN}} \circ \sigma_\downarrow^{\text{DYN}}(y) = \lim_{\varepsilon \in \mathcal{E}, \preccurlyeq} f^\varepsilon \circ \sigma_\downarrow^{\text{DYN}}(y) = \lim_{\varepsilon \in \mathcal{E}, \preccurlyeq} f \circ \Delta^\varepsilon \circ \sigma_\downarrow^{\text{DYN}}(y) = f \circ \widehat{\sigma}_\downarrow^{\text{KIN}} \left( \delta_\infty^{-1} \langle \{y\} \rangle \right)$$

$$= \beta_\uparrow^{\text{KIN}}(f) \left( \delta_\infty^{-1} \langle \{y\} \rangle \right) = \sup_{\delta_\infty^{-1} \langle \{y\} \rangle} \beta_\uparrow^{\text{KIN}}(f) = \left( \beta_\uparrow^{\text{KIN}}(f) \right)^{\text{DYN}}(y).$$

$\square$

**Proposition 3.24** Let $(\mathcal{L}, \mathcal{M}^{\text{KIN}}, \pi^{\text{KIN}})^\downarrow$ be a projective system of finite dimensional phase spaces yielding a rendering of a symplectic manifold $\mathcal{M}_\infty^{\text{KIN}}$. Let $\mathcal{E}$ be a directed preordered set and assume that:

1. for any $\varepsilon \in \mathcal{E} \sqcup \{\infty\}$, we have a phase space reduction $(\mathcal{M}_\infty^{\text{DYN},\varepsilon}, \mathcal{M}_\infty^{\text{SHELL},\varepsilon}, \delta_\infty^\varepsilon)$ of $\mathcal{M}_\infty^{\text{KIN}}$;

2. for any $\varepsilon \in \mathcal{E}$, we have a cofinal subset $\mathcal{L}^\varepsilon$ of $\mathcal{L}$ and an elementary reduction
$$\left( \left( \mathcal{M}_\eta^{\text{DYN},\varepsilon} \right)_{\eta \in \mathcal{L}^\varepsilon \sqcup \{\infty\}}, \left( \mathcal{M}_\eta^{\text{SHELL},\varepsilon} \right)_{\eta \in \mathcal{L}^\varepsilon \sqcup \{\infty\}}, \left( \pi_{\eta' \to \eta}^{\text{DYN},\varepsilon \to \varepsilon} \right)_{\eta \preccurlyeq \eta'}, \left( \delta_\eta^\varepsilon \right)_{\eta \in \mathcal{L}^\varepsilon \sqcup \{\infty\}} \right) \text{ of } (\mathcal{L}^\varepsilon \sqcup \{\infty\}, \mathcal{M}^{\text{KIN}}, \pi^{\text{KIN}})^\downarrow,$$
arising from $(\mathcal{M}_\infty^{\text{DYN},\varepsilon}, \mathcal{M}_\infty^{\text{SHELL},\varepsilon}, \delta_\infty^\varepsilon)$;

3. we have a rendering of $\mathcal{M}_\infty^{\text{DYN},\infty}$ by $(\mathcal{E}, \mathcal{M}_\infty^{\text{DYN}}, \pi_{\infty \to \infty}^{\text{DYN}})^\downarrow$;

4. for any $\varepsilon \preccurlyeq \varepsilon'$, $\mathcal{L}^{\varepsilon'} \subset \mathcal{L}^\varepsilon$ and, for any $\eta \in \mathcal{L}^{\varepsilon'}$, we have a projection $\pi_{\eta \to \eta}^{\text{DYN},\varepsilon' \to \varepsilon} : \mathcal{M}_\eta^{\text{DYN},\varepsilon'} \to \mathcal{M}_\eta^{\text{DYN},\varepsilon}$, compatible with the symplectic structures, and such that $\pi_{\infty \to \eta}^{\text{DYN},\varepsilon \to \varepsilon} \circ \pi_{\infty \to \infty}^{\text{DYN},\varepsilon' \to \varepsilon} = \pi_{\eta \to \eta}^{\text{DYN},\varepsilon' \to \varepsilon} \circ \pi_{\infty \to \eta}^{\text{DYN},\varepsilon' \to \varepsilon'}$.

Then, defining $\widetilde{\mathcal{L}} := \mathcal{L} \sqcup \{\infty\}$ and $\widetilde{\mathcal{E}} := \mathcal{E} \sqcup \{\infty\}$ (extending the preorders in such a way that $\infty$ is a greatest element), we can complete this input to build a regularized reduction $\left( \widetilde{\mathcal{L}}, \mathcal{M}, \pi, \delta \right)^{\text{DYN}, \widetilde{\mathcal{E}}}$ of $\left(\widetilde{\mathcal{L}}, \mathcal{M}^{\text{KIN}}, \pi^{\text{KIN}}\right)^{\downarrow}$.

If we moreover have:

5. for any $y \in \mathcal{M}^{\text{DYN},\infty}_{\infty}$, the net $\left(\widehat{\sigma}^{\text{KIN}}_{\downarrow}\left(\delta^{\varepsilon,-1}_{\infty}\left\langle\pi^{\text{DYN},\infty\to\varepsilon}_{\infty\to\infty}(y)\right\rangle\right)\right)_{\varepsilon\in\mathcal{E}}$ converges in $\widehat{\mathcal{S}}^{\downarrow}_{(\mathcal{L},\mathcal{M}^{\text{KIN}},\pi^{\text{KIN}})}$ to $\widehat{\sigma}^{\text{KIN}}_{\downarrow}\left(\delta^{\infty,-1}_{\infty}\langle y\rangle\right)$;

then the hypotheses of prop. 3.23 are fulfilled.

**Proof** For any $\varepsilon \in \mathcal{E}$ we define $\widetilde{\mathcal{L}}^{\varepsilon} := \mathcal{L}^{\varepsilon} \sqcup \{\infty\}$ and we additionally define $\widetilde{\mathcal{L}}^{\infty} := \{\infty\}$. With this def. 3.16.1 is fulfilled.

For any $\varepsilon \in \mathcal{E}$, def. 3.16.2 comes from assumption 3.24.2, and for $\infty \in \widetilde{\mathcal{E}}$, it reduces to $(\mathcal{M}^{\text{DYN},\infty}_{\infty}, \mathcal{M}^{\text{SHELL},\infty}_{\infty}, \delta^{\infty}_{\infty})$ being a phase space reduction of $\mathcal{M}^{\text{KIN}}_{\infty}$, which has been assumed in 3.24.1.

For any $\varepsilon \preccurlyeq \varepsilon' \in \widetilde{\mathcal{E}}$ and any $\eta \in \widetilde{\mathcal{L}}^{\varepsilon}$, $\eta' \in \widetilde{\mathcal{L}}^{\varepsilon'}$ with $\eta \preccurlyeq \eta'$, we define $\pi^{\text{DYN},\varepsilon'\to\varepsilon}_{\eta'\to\eta} := \pi^{\text{DYN},\varepsilon\to\varepsilon}_{\eta'\to\eta} \circ \pi^{\text{DYN},\varepsilon'\to\varepsilon}_{\eta'\to\eta'}$. From assumption 3.24.2, $\pi^{\text{DYN},\varepsilon\to\varepsilon}_{\eta'\to\eta} : \mathcal{M}^{\text{DYN},\varepsilon}_{\eta'} \to \mathcal{M}^{\text{DYN},\varepsilon}_{\eta}$ is a projection compatible with the symplectic structures, and from assumption 3.24.4 (or 3.24.3 if $\eta' = \infty$), $\pi^{\text{DYN},\varepsilon'\to\varepsilon}_{\eta'\to\eta'} : \mathcal{M}^{\text{DYN},\varepsilon'}_{\eta'} \to \mathcal{M}^{\text{DYN},\varepsilon}_{\eta'}$ is a projection compatible with the symplectic structures. Hence, $\pi^{\text{DYN},\varepsilon'\to\varepsilon}_{\eta'\to\eta} : \mathcal{M}^{\text{DYN},\varepsilon'}_{\eta'} \to \mathcal{M}^{\text{DYN},\varepsilon}_{\eta}$ is a projection compatible with the symplectic structures.

Let $(\varepsilon,\eta) \preccurlyeq (\varepsilon',\eta') \preccurlyeq (\varepsilon'',\eta'') \in \widetilde{\mathcal{E}\mathcal{L}}$. We have $\eta',\eta'' \in \mathcal{L}^{\varepsilon'}$, hence, using points 3.24.2 and 3.24.4:

$$\pi^{\text{DYN},\varepsilon\to\varepsilon}_{\eta'\to\eta'} \circ \pi^{\text{DYN},\varepsilon'\to\varepsilon}_{\eta''\to\eta''} \circ \pi^{\text{DYN},\varepsilon'\to\varepsilon'}_{\infty\to\eta''} = \pi^{\text{DYN},\varepsilon\to\varepsilon}_{\infty\to\eta'} \circ \pi^{\text{DYN},\varepsilon'\to\varepsilon}_{\infty\to\infty} = \pi^{\text{DYN},\varepsilon'\to\varepsilon}_{\eta'\to\eta'} \circ \pi^{\text{DYN},\varepsilon'\to\varepsilon'}_{\eta''\to\eta''} \circ \pi^{\text{DYN},\varepsilon'\to\varepsilon'}_{\infty\to\eta''}.$$

Since $\pi^{\text{DYN},\varepsilon'\to\varepsilon'}_{\infty\to\eta''}$ is surjective, we then have:

$$\pi^{\text{DYN},\varepsilon\to\varepsilon}_{\eta'\to\eta'} \circ \pi^{\text{DYN},\varepsilon'\to\varepsilon}_{\eta''\to\eta''} = \pi^{\text{DYN},\varepsilon'\to\varepsilon}_{\eta'\to\eta'} \circ \pi^{\text{DYN},\varepsilon'\to\varepsilon'}_{\eta''\to\eta'},$$

hence, using once more from 3.24.2:

$$\pi^{\text{DYN},\varepsilon'\to\varepsilon}_{\eta'\to\eta} \circ \pi^{\text{DYN},\varepsilon''\to\varepsilon'}_{\eta''\to\eta'} = \pi^{\text{DYN},\varepsilon\to\varepsilon}_{\eta'\to\eta} \circ \pi^{\text{DYN},\varepsilon\to\varepsilon}_{\eta''\to\eta'} \circ \pi^{\text{DYN},\varepsilon'\to\varepsilon}_{\eta''\to\eta''} \circ \pi^{\text{DYN},\varepsilon''\to\varepsilon'}_{\eta''\to\eta''}$$

$$= \pi^{\text{DYN},\varepsilon\to\varepsilon}_{\eta''\to\eta} \circ \pi^{\text{DYN},\varepsilon'\to\varepsilon}_{\eta''\to\eta''} \circ \pi^{\text{DYN},\varepsilon''\to\varepsilon'}_{\eta''\to\eta''}. \tag{3.24.1}$$

Now, using repeatedly 3.24.4, together with 3.24.3, we have:

$$\pi^{\text{DYN},\varepsilon'\to\varepsilon}_{\eta''\to\eta''} \circ \pi^{\text{DYN},\varepsilon''\to\varepsilon'}_{\eta''\to\eta''} \circ \pi^{\text{DYN},\varepsilon''\to\varepsilon''}_{\infty\to\eta''} = \pi^{\text{DYN},\varepsilon\to\varepsilon}_{\infty\to\eta''} \circ \pi^{\text{DYN},\varepsilon'\to\varepsilon}_{\infty\to\infty} \circ \pi^{\text{DYN},\varepsilon''\to\varepsilon'}_{\infty\to\infty}$$

$$= \pi^{\text{DYN},\varepsilon''\to\varepsilon}_{\eta''\to\eta''} \circ \pi^{\text{DYN},\varepsilon''\to\varepsilon''}_{\infty\to\eta''}$$

and, since $\pi^{\text{DYN},\varepsilon''\to\varepsilon''}_{\infty\to\eta''}$ is surjective:

$$\pi^{\text{DYN},\varepsilon'\to\varepsilon}_{\eta''\to\eta''} \circ \pi^{\text{DYN},\varepsilon''\to\varepsilon'}_{\eta''\to\eta''} = \pi^{\text{DYN},\varepsilon''\to\varepsilon}_{\eta''\to\eta''}. \tag{3.24.2}$$

Combining eq. (3.24.1) and eq. (3.24.2), we get:

$$\pi^{\text{DYN},\varepsilon'\to\varepsilon}_{\eta'\to\eta} \circ \pi^{\text{DYN},\varepsilon''\to\varepsilon'}_{\eta''\to\eta'} = \pi^{\text{DYN},\varepsilon\to\varepsilon}_{\eta''\to\eta} \circ \pi^{\text{DYN},\varepsilon''\to\varepsilon}_{\eta''\to\eta''} = \pi^{\text{DYN},\varepsilon''\to\varepsilon}_{\eta''\to\eta},$$

therefore $\left(\widetilde{\mathcal{E}\mathcal{L}}, \mathcal{M}^{\text{DYN}}, \pi^{\text{DYN}}\right)^{\downarrow}$ is a projective system of phase spaces, so def. 3.16.3 holds.

Thus, using prop. 3.18 with assumption 3.24.2, $(\mathcal{L},\mathcal{M},\pi,\delta)^{\text{DYN},\mathcal{E}}$ is a regularized reduction of $(\mathcal{L}, \mathcal{M}^{\text{KIN}}, \pi^{\text{KIN}})^{\downarrow}$, while $(\mathcal{EL}, \mathcal{M}^{\text{DYN}}, \pi^{\text{DYN}})^{\downarrow}$ is a rendering of $\mathcal{M}^{\text{DYN},\infty}_{\infty}$.

We now assume that assumption 3.24.5 holds. Using eq. (3.18.1) for $\mathcal{EL} \subset \widetilde{\mathcal{EL}}$, we have:

$$\widehat{\sigma}^{\text{KIN}}_{\widetilde{\mathcal{L}} \to \mathcal{L}} \circ \widetilde{\Delta}^{\varepsilon} = \Delta^{\varepsilon} \circ \sigma^{\text{DYN}}_{\widetilde{\mathcal{EL}} \to \mathcal{EL}},$$

and using it for $\widetilde{\mathcal{E}}\{\infty\} \subset \widetilde{\mathcal{EL}}$:

$$\widehat{\sigma}^{\text{KIN}}_{\widetilde{\mathcal{L}} \to \{\infty\}} \circ \widetilde{\Delta}^{\varepsilon} = \delta^{\varepsilon,-1}_{\infty} \langle \cdot \rangle \circ \sigma^{\text{DYN}}_{\widetilde{\mathcal{E}}\{\infty\} \to \{(\varepsilon,\infty)\}} \circ \sigma^{\text{DYN}}_{\widetilde{\mathcal{EL}} \to \widetilde{\mathcal{E}}\{\infty\}}$$

$$= \delta^{\varepsilon,-1}_{\infty} \langle \cdot \rangle \circ \sigma^{\text{DYN}}_{\widetilde{\mathcal{EL}} \to \{(\varepsilon,\infty)\}},$$

therefore:

$$\Delta^{\varepsilon} \circ \sigma^{\text{DYN}}_{\downarrow} = \Delta^{\varepsilon} \circ \sigma^{\text{DYN}}_{\widetilde{\mathcal{EL}} \to \mathcal{EL}} \circ \sigma^{\text{DYN},-1}_{\widetilde{\mathcal{EL}} \to \{(\infty,\infty)\}}$$

$$= \widehat{\sigma}^{\text{KIN}}_{\widetilde{\mathcal{L}} \to \mathcal{L}} \circ \widehat{\sigma}^{\text{KIN},-1}_{\widetilde{\mathcal{L}} \to \{\infty\}} \circ \delta^{\varepsilon,-1}_{\infty} \langle \cdot \rangle \circ \sigma^{\text{DYN}}_{\widetilde{\mathcal{EL}} \to \{(\varepsilon,\infty)\}} \circ \sigma^{\text{DYN},-1}_{\widetilde{\mathcal{EL}} \to \{(\infty,\infty)\}}$$

$$= \widehat{\sigma}^{\text{KIN}}_{\downarrow} \circ \delta^{\varepsilon,-1}_{\infty} \langle \cdot \rangle \circ \pi^{\text{DYN},\infty \to \varepsilon}_{\infty \to \infty}.$$

Hence, for all $y \in \mathcal{M}^{\text{DYN},\infty}_{\infty}$, the net $\left(\Delta^{\varepsilon} \circ \sigma^{\text{DYN}}_{\downarrow}(y)\right)_{\varepsilon \in \mathcal{E}}$ converges in $\widehat{\mathcal{S}}^{\downarrow}_{(\mathcal{L}, \mathcal{M}^{\text{KIN}}, \pi^{\text{KIN}})}$ to $\widehat{\sigma}^{\text{KIN}}_{\downarrow} \left(\delta^{\infty,-1}_{\infty} \langle y \rangle\right)$.
□

# 4 Outlook

At this point, the question that remains open is how to construct a rendering of $\mathcal{M}^{\text{DYN},\infty}$ by a net of reduced phase spaces $\mathcal{M}^{\text{DYN},\epsilon}$, arising from constraint surfaces approaching $\mathcal{M}^{\text{SHELL},\infty}$. In other words, we are lacking systematic recipes for setting up regularization schemes in the sense of the procedure just described.

Among the tools that are at our disposal is the gauge fixing/unfixing trick (taken from [1], where it was however used in a completely different context), that would consist in first partially gauge fixing (prop. A.8) the original phase space reduction, and then gauge unfixing it in a slightly different direction: thus we would deform the orbits (in the view of improving their projectability), and get an approximation of the dynamics that should be satisfactory in some neighborhood of the common gauge fixing surface (this technique is the one used in [10, section 3]). Another option, that might be in particular relevant when the gauge orbits are infinite dimensional, could be to drastically gauge fix them, before progressively lifting the gauge fixing conditions, thus approaching a given orbit by an increasing net of submanifolds inside it. In both cases, we get a natural symplectomorphism between $\mathcal{M}^{\text{DYN},\infty}$ and each $\mathcal{M}^{\text{DYN},\epsilon}$, so we probably want to combine such methods with projections from $\mathcal{M}^{\text{DYN},\infty}$ into symplectic submanifolds of it, to drop the degrees of freedom that are disproportionately accurate at a given level of approximation.

Also, there is presumably some link between the regularization procedure we are considering and various concepts developed in the context of Loop Quantum Gravity (often within a Lagrangian setting), exploring the interplay between discretization, coarse graining, diffeomorphism invariance, and the continuum limit [12]. Studying more precisely how these approaches are related to the strategy proposed here could in particular help incorporate renormalization group ideas into the picture.

## Acknowledgements

This work has been financially supported by the Université François Rabelais, Tours, France.

This research project has been supported by funds to Emerging Field Project "Quantum Geometry" from the FAU Erlangen-Nuernberg within its Emerging Fields Initiative.

# A Appendix: Classical constrained systems

To fix the notations and definitions, we summarize here some facts about constrained classical systems. We recall how a reduced phase space arises from a constraint surface in a symplectic manifold [20, section 1.7], we introduce a notion of transport of observables to translate kinematical observables into dynamical ones (this facility is the main object of the physical discussion in section 3), and give a very brief account of partial gauge fixing [13].

When considering a constraint surface $\mathcal{M}^{\text{SHELL}}$ in a symplectic manifold $\mathcal{M}^{\text{KIN}}$ the pullback of the symplectic structure $\Omega_{\text{KIN}}$ does not, in general, define a symplectic structure on $\mathcal{M}^{\text{SHELL}}$: there might be directions in the tangent space of $\mathcal{M}^{\text{SHELL}}$ on which this pullback vanishes. These directions correspond to the gauge flow generated by first class constraints, and the gauge orbits need to be quotiented out in order to get a reduced phase space $\mathcal{M}^{\text{DYN}}$ with a non-degenerate symplectic structure $\Omega_{\text{DYN}}$.

Except for the first few definitions (which are tailored to match the needs of some results in section 3), this appendix focuses on finite dimensional manifolds: this is anyway the point of the formalism presented in the main text that we aim at describing a field theory in such a way that we can work mostly within the context of finite dimensional manifolds.

In this appendix all manifolds will be smooth manifolds, all maps between them will be smooth and all submanifolds will be regular (ie. embedded) submanifolds. Where infinite dimensional manifolds are considered, these are Banach-modeled smooth manifolds, and symplectic structures on them are always strong symplectic structures [2, chap. VII].

**Definition A.1** Let $\mathcal{M}^{\text{KIN}}$ be a (possibly infinite dimensional) smooth symplectic manifold (with symplectic structure $\Omega_{\text{KIN}}$). A phase space reduction of $\mathcal{M}^{\text{KIN}}$ is a triple $(\mathcal{M}^{\text{DYN}}, \mathcal{M}^{\text{SHELL}}, \delta)$ such that:

1. $\mathcal{M}^{\text{SHELL}}$ is a submanifold of $\mathcal{M}^{\text{KIN}}$ and $\mathcal{M}^{\text{DYN}}$ is a symplectic manifold (with symplectic structure $\Omega_{\text{DYN}}$);

2. $\delta : \mathcal{M}^{\text{SHELL}} \to \mathcal{M}^{\text{DYN}}$ is a surjective map and, for all $y \in \mathcal{M}^{\text{DYN}}$, $\delta^{-1}\langle y \rangle$ is connected;

3. for all $x \in \mathcal{M}^{\text{SHELL}}$, $\text{Im}(T_x \delta) = T_{\delta(x)}(\mathcal{M}^{\text{DYN}})$    &    $\Omega_{\text{KIN},x}\big|_{T_x(\mathcal{M}^{\text{SHELL}})} = [\delta^* \Omega_{\text{DYN}}]_x$.

For any bounded real-valued function $f$ on $\mathcal{M}^{\text{KIN}}$, we define a corresponding dynamical observable on $\mathcal{M}^{\text{DYN}}$ by mapping to a point $y$ in the reduced phase space the supremum of $f$ on the corresponding orbit $\delta^{-1}\langle y \rangle$. The motivation for this definition is that we regard indicator functions as the most fundamental observables: with the transport of observables defined this way, the indicator function of some region in $\mathcal{M}^{\text{KIN}}$ is mapped into the indicator function on the space of orbits that characterize whether a given orbit crosses this region or not. In other words, the dynamical observable related to the indicator function of some region of $\mathcal{M}^{\text{KIN}}$ will tell us whether the dynamical state of the system allows it to be measured in that region.

Note that there can be relations between the dynamical observables $f_1^{\text{DYN}}, \ldots, f_k^{\text{DYN}}$ arising from functionally independent kinematical observables $f_1, \ldots, f_k$, or to state this more precisely we can have dependencies:

$$\text{Im}\,(f_1^{\text{DYN}} \times \ldots \times f_k^{\text{DYN}}) \neq (\text{Im}\,f_1^{\text{DYN}}) \times \ldots \times (\text{Im}\,f_k^{\text{DYN}}),$$

although the corresponding kinematical observables were independent:

$$\text{Im}\,(f_1 \times \ldots \times f_k) = (\text{Im}\,f_1) \times \ldots \times (\text{Im}\,f_k).$$

This is a crucial observation, since, indeed, the dynamical content of theory lies in such functional relations emerging between observables that were kinematically independent.

**Definition A.2** Let $(\mathcal{M}^{\text{DYN}}, \mathcal{M}^{\text{SHELL}}, \delta)$ be a phase space reduction of $\mathcal{M}^{\text{KIN}}$. We denote by $B(\mathcal{M}^{\text{KIN}})$ the space of bounded, real-valued, functions on $\mathcal{M}^{\text{KIN}}$. For all $f \in B(\mathcal{M}^{\text{KIN}})$, we define $f^{\text{DYN}} \in B(\mathcal{M}^{\text{DYN}})$ by:

$$\forall y \in \mathcal{M}^{\text{DYN}},\; f^{\text{DYN}}(y) := \sup\,\{f(x)\,|\,x \in \delta^{-1}\langle y \rangle\}. \tag{A.2.1}$$

For the rest of this appendix all manifolds will be *finite dimensional* manifolds.

**Definition A.3** Let $\mathcal{M}^{\text{KIN}}$ be a smooth, finite dimensional, symplectic manifold (with symplectic structure $\Omega_{\text{KIN}}$). A pre-reduction of $\mathcal{M}^{\text{KIN}}$ is a triple $(\mathcal{M}^{\text{DYN}}, \mathcal{M}^{\text{SHELL}}, \delta)$ such that:

1. $\mathcal{M}^{\text{SHELL}}$ is a submanifold of $\mathcal{M}^{\text{KIN}}$ and $\mathcal{M}^{\text{DYN}}$ is a manifold;

2. the restriction of $\Omega_{\text{KIN}}$ to $T(\mathcal{M}^{\text{SHELL}})$ is of constant rank, thus defining a foliation $K(\mathcal{M}^{\text{SHELL}})$ by

$$\forall x \in \mathcal{M}^{\text{SHELL}},\; K_x(\mathcal{M}^{\text{SHELL}}) := \left\{ v \in T_x(\mathcal{M}^{\text{SHELL}}) \,\Big|\, \Omega_{\text{KIN},x}(v,\,\cdot)\big|_{T_x(\mathcal{M}^{\text{SHELL}})} = 0 \right\} \subset T_x(\mathcal{M}^{\text{SHELL}});$$

3. $\delta : \mathcal{M}^{\text{SHELL}} \to \mathcal{M}^{\text{DYN}}$ is a surjective map and $\forall x \in \mathcal{M}^{\text{SHELL}}$, $\text{Im}(T_x \delta) = T_{\delta(x)}(\mathcal{M}^{\text{DYN}})$;

4. $\forall y \in \mathcal{M}^{\text{DYN}}$, $\delta^{-1}\langle y \rangle$ is a leaf of the foliation $K(\mathcal{M}^{\text{SHELL}})$.

**Proposition A.4** Let $\mathcal{M}^{\text{KIN}}$ be a smooth, finite dimensional, symplectic manifold (with symplectic structure $\Omega_{\text{KIN}}$) and let $(\mathcal{M}^{\text{DYN}}, \mathcal{M}^{\text{SHELL}}, \delta)$ be a phase space reduction of $\mathcal{M}^{\text{KIN}}$. Then, $(\mathcal{M}^{\text{DYN}}, \mathcal{M}^{\text{SHELL}}, \delta)$ is a pre-reduction of $\mathcal{M}^{\text{KIN}}$ and we have:

$$\forall x \in \mathcal{M}^{\text{SHELL}},\quad K_x(\mathcal{M}^{\text{SHELL}}) = \text{Ker}\,T_x \delta = T_x\left(\delta^{-1}\langle \delta(x) \rangle\right). \tag{A.4.1}$$

**Proof** Defs. A.3.1 and A.3.3 are directly implied by def. A.1.

Let $y \in \mathcal{M}^{\text{DYN}}$ and $x \in \delta^{-1}\langle y \rangle$. Since $\delta$ has surjective derivative at each point, we have as an implication of the rank theorem [11, theorem 5.22] that $\delta^{-1}\langle y\rangle$ is a submanifold of $\mathcal{M}^{\text{SHELL}}$ with tangent space $\operatorname{Ker} T_x\delta \subset T_x(\mathcal{M}^{\text{SHELL}})$ at $x$. Now, from def. A.1.3, together with the non-degeneracy of $\Omega_{\text{DYN}}$ (for $\mathcal{M}^{\text{DYN}}$ is a symplectic manifold), we have:

$$\forall x' \in \delta^{-1}\langle y\rangle, \quad \operatorname{Ker} T_{x'}\delta = K_{x'}(\mathcal{M}^{\text{SHELL}}).$$

Hence, $K(\mathcal{M}^{\text{SHELL}})$ has constant dimension, so def. A.3.2 is fulfilled.

Additionaly, by maximality of the leaves, the connected submanifold $\delta^{-1}\langle y\rangle$ is included in the leaf of the foliation $K(\mathcal{M}^{\text{SHELL}})$ that goes through $x$. Reciprocally, since the leaf that goes through $x$ is connected, and has tangent space $K_{x'}(\mathcal{M}^{\text{SHELL}}) = \operatorname{Ker} T_{x'}\delta$ at any point, $\delta$ is constant on it, hence it is included in $\delta^{-1}\langle y\rangle$. Thus, def. A.3.4 is fulfilled. □

**Proposition A.5** Let $\mathcal{M}^{\text{KIN}}$ be a smooth, finite dimensional, symplectic manifold and let $(\mathcal{M}^{\text{DYN}}, \mathcal{M}^{\text{SHELL}}, \delta)$ be a pre-reduction of $\mathcal{M}^{\text{KIN}}$. Then, there exists a symplectic structure $\Omega_{\text{DYN}}$ on $\mathcal{M}^{\text{DYN}}$ such that $(\mathcal{M}^{\text{DYN}}, \mathcal{M}^{\text{SHELL}}, \delta)$ is a phase space reduction of $\mathcal{M}^{\text{KIN}}$.

**Proof** What we need to prove is that there exists a symplectic structure $\Omega_{\text{DYN}}$ on $\mathcal{M}^{\text{DYN}}$ such that:

$$\forall x \in \mathcal{M}^{\text{SHELL}}, \quad \Omega_{\text{KIN},x}|_{T_x(\mathcal{M}^{\text{SHELL}})} = [\delta^* \Omega_{\text{DYN}}]_x.$$

The others points in def. A.1 are immediately fulfilled (in particular, for any $y \in \mathcal{M}^{\text{DYN}}$, $\delta^{-1}\langle y\rangle$ is connected as a leaf of a foliation).

Let $x \in \mathcal{M}^{\text{SHELL}}$ and let $y := \delta(x)$. Since $\delta$ has surjective derivative at each point, there exist by the rank theorem [11, theorem 5.13] open neighborhoods $U$ of $x$ in $\mathcal{M}^{\text{SHELL}}$, $V$ of $y$ in $\mathcal{M}^{\text{DYN}}$ and $W$ of $0$ in $\mathcal{R}^{s-d}$ (with $s := \dim \mathcal{M}^{\text{SHELL}}$ and $d := \dim \mathcal{M}^{\text{DYN}}$), and a diffeomorphism $\varphi : V \times W \to U$ such that:

$$\forall y' \in V, \forall z' \in W, \quad \delta \circ \varphi(y', z') = y'.$$

For any $y', z' \in V \times W$, we define $\Omega_{\text{DYN},y'}^{\varphi,z'}$ by:

$$\forall v, v' \in T_{y'}(\mathcal{M}^{\text{DYN}}), \quad \Omega_{\text{DYN},y'}^{\varphi,z'}(v, v') = \Omega_{\text{KIN},\varphi(y',z')}\left(T_{y',z'}\varphi(v,0), T_{y',z'}\varphi(v',0)\right).$$

Then, setting $x' := \varphi(y', z')$, $\Omega_{\text{DYN},y'}^{\varphi,z'}$ satisfies:

$$\forall u, u' \in T_{x'}(\mathcal{M}^{\text{SHELL}}), \quad \Omega_{\text{KIN},x'}(u, u') = \Omega_{\text{DYN},y'}^{\varphi,z'}\left(T_{x'}\delta(u), T_{x'}\delta(u')\right),$$

for we have from def. A.3.4:

$$\left\{T_{y',z'}\varphi(0, w) \,|\, w \in T_{z'}(W)\right\} = T_{x'}\left(\delta^{-1}\langle y'\rangle\right) = K_{x'}(\mathcal{M}^{\text{SHELL}}). \tag{A.5.1}$$

Let $\widetilde{Y}, \widetilde{Y}'$ be vector fields on $V$ and $\widetilde{Z}$ be a vector field on $W$. Defining $Y := \varphi_*\left(\widetilde{Y}, 0\right)$, $Y' := \varphi_*\left(\widetilde{Y}', 0\right)$, and $Z := \varphi_*\left(0, \widetilde{Z}\right)$, we have $[Y, Z] = [Y', Z] = 0$ and, from eq. (A.5.1):

$$\Omega_{\text{KIN}}(Z, \cdot)|_{T(\mathcal{M}^{\text{SHELL}})} = 0.$$

Hence, we get, for any $y', z' \in V \times W$:

$$d\Omega_{\text{KIN}}\left(Y, Y', Z\right)_{\varphi(y',z')} = 0 \quad \text{(by definition of a symplectic form)}$$

$$= Z\left(\Omega_{\text{KIN}}(Y, Y')\right)_{\varphi(y',z')}$$

$$= \widetilde{Z}_{z'}\left(z'' \mapsto \Omega^{\varphi,z''}_{\text{DYN},y'}\left(\widetilde{Y}_{y'}, \widetilde{Y}'_{y'}\right)\right).$$

Now, for any $x \in \mathcal{M}^{\text{SHELL}}$, we define $\Omega^x_{\text{DYN}} : T_{\delta(x)}(\mathcal{M}^{\text{DYN}}) \times T_{\delta(x)}(\mathcal{M}^{\text{DYN}}) \to \mathbb{R}$ by:

$$\forall v, w \in T_x(\mathcal{M}^{\text{SHELL}}),\ \Omega^x_{\text{DYN}}\left(T_x\delta(v), T_x\delta(w)\right) = \Omega_{\text{KIN},x}(v, w).$$

That such an $\Omega^x_{\text{DYN}}$ exists is established by the previous discussion and, since $\operatorname{Im} T_x\delta = T_{\delta(x)}(\mathcal{M}^{\text{DYN}})$, it is moreover unique. Thus, $\Omega^x_{\text{DYN}}$ is well-defined.

The previous argument also shows that, for any vector fields $Y, Y'$ on $\mathcal{M}^{\text{DYN}}$, $x \mapsto \Omega^x_{\text{DYN}}\left(Y_{\delta(x)}, Y'_{\delta(x)}\right)$ is smooth and satisfies:

$$\forall x \in \mathcal{M}^{\text{SHELL}},\ \forall w \in T_x\left(\delta^{-1}\langle\delta(x)\rangle\right),\ T_x\left(x' \mapsto \Omega^{x'}_{\text{DYN}}\left(Y_{\delta(x')}, Y'_{\delta(x')}\right)\right)(w) = 0.$$

The level sets of $\delta$ being connected, as underlined above, this allows us to define a smooth differential 2-form $\Omega_{\text{DYN}}$ satisfying:

$$\forall x \in \mathcal{M}^{\text{SHELL}},\ \Omega_{\text{KIN},x}\big|_{T_x(\mathcal{M}^{\text{SHELL}})} = [\delta^* \Omega_{\text{DYN}}]_x.$$

Lastly, for any $x \in \mathcal{M}^{\text{SHELL}}$, we also have:

$$[\delta^* d\Omega_{\text{DYN}}]_x = d\Omega_{\text{KIN},x}\big|_{T_x(\mathcal{M}^{\text{SHELL}})} = 0,$$

and, from eq. (A.5.1):

$$\operatorname{Ker} T_x\delta = K_x(\mathcal{M}^{\text{SHELL}}).$$

Thus, $T_x\delta$ being surjective, $\Omega_{\text{DYN}}$ is closed and non-degenerate, so it is indeed a symplectic form on $\mathcal{M}^{\text{DYN}}$. $\square$

**Proposition A.6** Let $\mathcal{M}^{\text{KIN}}$ be a smooth, finite dimensional, symplectic manifold and let $(\mathcal{M}^{\text{DYN},1}, \mathcal{M}^{\text{SHELL}}, \delta_1)$ and $(\mathcal{M}^{\text{DYN},2}, \mathcal{M}^{\text{SHELL}}, \delta_2)$ be two phase space reductions of $\mathcal{M}^{\text{KIN}}$ arising from the same submanifold $\mathcal{M}^{\text{SHELL}}$ of $\mathcal{M}^{\text{KIN}}$. Then there exists a unique map $\psi : \mathcal{M}^{\text{DYN},1} \to \mathcal{M}^{\text{DYN},2}$ such that $\delta_2 = \psi \circ \delta_1$. Moreover, $\psi$ is a symplectomorphism.

**Proof** From def. A.3.4 $\delta_2$ is constant on the level sets of $\delta_1$ and reciprocally, hence, as a consequence [11, prop. 5.21] of the rank theorem (using that both $\delta_1$ and $\delta_2$ are surjective and have surjective derivative at each point, from def. A.3.3), there exists a unique diffeomorphism $\psi : \mathcal{M}^{\text{DYN},1} \to \mathcal{M}^{\text{DYN},2}$ such that $\delta_2 = \psi \circ \delta_1$.

In particular, for $x \in \mathcal{M}^{\text{SHELL}}$ (with $y := \delta_1(x)$), we have $T_x\delta_2 = T_y\psi \circ T_x\delta_1$, so that, using def. A.1.3:

$$[\delta_1^* \Omega_{\text{DYN},1}]_x = \Omega_{\text{KIN},x}\big|_{T_x(\mathcal{M}^{\text{SHELL}})} = [\delta_1^* \psi^* \Omega_{\text{DYN},2}]_x.$$

Since $T_x\delta_1$ and $\delta_1$ are surjective, $\psi$ is a symplectomorphism. $\square$

**Proposition A.7** Let $\mathcal{M}^{\text{KIN}}$ be a smooth, finite dimensional, symplectic manifold and let $(\mathcal{M}^{\text{DYN}}, \mathcal{M}^{\text{SHELL}}, \delta)$ be a phase space reduction of $\mathcal{M}^{\text{KIN}}$. Let $f, g$ and $\{f, g\}_{\text{KIN}} \in C^\infty(\mathcal{M}^{\text{KIN}}, \mathbb{R}) \cap B(\mathcal{M}^{\text{KIN}})$, and assume that:

$$\forall x \in \mathcal{M}^{\text{SHELL}},\ X_{f,x}, X_{g,x} \in T_x(\mathcal{M}^{\text{SHELL}}),$$

where the Hamiltonian vector field $X_f := \underline{df}$ is defined by $\Omega_{\text{KIN}}(X_f, \cdot) = df$.

Then $f^{\text{DYN}}, g^{\text{DYN}} \in C^\infty(\mathcal{M}^{\text{DYN}}, \mathbb{R})$ and $\{f^{\text{DYN}}, g^{\text{DYN}}\}_{\text{DYN}} = \left(\{f, g\}_{\text{KIN}}\right)^{\text{DYN}}$.

**Proof** For all $x \in \mathcal{M}^{\text{SHELL}}$, $X_{f,x} \in T_x(\mathcal{M}^{\text{SHELL}})$, hence $df_x\langle K_x(\mathcal{M}^{\text{SHELL}})\rangle = \Omega_{\text{KIN},x}\left(X_{f,x}, K_x(\mathcal{M}^{\text{SHELL}})\right) \subset$

$\Omega_{\text{KIN},x}(T_x(\mathcal{M}^{\text{SHELL}}), K_x(\mathcal{M}^{\text{SHELL}})) = \{0\}$. The same holds for $g$. Therefore, $f$ and $g$ are constant on the leaves of the foliation $K(\mathcal{M}^{\text{SHELL}})$ on $\mathcal{M}^{\text{SHELL}}$. As a consequence [11, prop. 5.20] of the rank theorem (using defs. A.3.3 and A.3.4), there exist smooth maps $\widetilde{f}$ and $\widetilde{g} : \mathcal{M}^{\text{DYN}} \to \mathbb{R}$ such that $f|_{\mathcal{M}^{\text{SHELL}}} = \widetilde{f} \circ \delta$ and $g|_{\mathcal{M}^{\text{SHELL}}} = \widetilde{g} \circ \delta$. Hence, $f^{\text{DYN}} = \widetilde{f}$ and $g^{\text{DYN}} = \widetilde{g}$.

In addition, we have for any $x \in \mathcal{M}^{\text{SHELL}}$ (with $y := \delta(x)$):

$$[\delta^* \, df^{\text{DYN}}]_x = df_x|_{T_x(\mathcal{M}^{\text{SHELL}})}$$

$$= \Omega_{\text{KIN},x}|_{T_x(\mathcal{M}^{\text{SHELL}})} \left( X_{f,x}; \cdot \right) \text{ (using } X_{f,x} \in T_x(\mathcal{M}^{\text{SHELL}}))$$

$$= \Omega_{\text{DYN},y} \left( T_x\delta \left( X_{f,x} \right); T_x\delta(\cdot) \right) \text{ (using def. A.1.3).}$$

Thus, $T_x\delta$ being surjective, $X_{f^{\text{DYN}},y} = T_x\delta \left( X_{f,x} \right)$ and, similarly $X_{g^{\text{DYN}},y} = T_x\delta \left( X_{g,x} \right)$.

Hence, we have:

$$\{f^{\text{DYN}}, g^{\text{DYN}}\}_{\text{DYN},y} = \Omega_{\text{DYN},y} \left( X_{g^{\text{DYN}},y}, X_{f^{\text{DYN}},y} \right)$$

$$= \Omega_{\text{DYN},y} \left( T_x\delta \left( X_{g,x} \right), T_x\delta \left( X_{f,x} \right) \right)$$

$$= \Omega_{\text{KIN},x} \left( X_{g,x}, X_{f,x} \right) \text{ (using def. A.1.3 and } X_{f,x}, X_{g,x} \in T_x(\mathcal{M}^{\text{SHELL}}))$$

$$= \{f, g\}_{\text{KIN},x}.$$

Since this holds for all $x \in \delta^{-1}\langle y \rangle$, this implies in particular that $\{f, g\}_{\text{KIN}}$ is constant on $\delta^{-1}\langle y \rangle$. Therefore $\left(\{f, g\}_{\text{KIN}}\right)^{\text{DYN}}(y) = \{f, g\}_{\text{KIN},x} = \{f^{\text{DYN}}, g^{\text{DYN}}\}_{\text{DYN},y}$. □

**Proposition A.8** Let $(\mathcal{M}^{\text{DYN}}, \mathcal{M}^{\text{SHELL}}, \delta)$ be a phase space reduction of $\mathcal{M}^{\text{KIN}}$ and $\mathcal{M}^{\text{FIX}}$ a submanifold of $\mathcal{M}^{\text{SHELL}}$ such that:

1. for all $x \in \mathcal{M}^{\text{FIX}}$, $T_x(\mathcal{M}^{\text{SHELL}}) = T_x(\mathcal{M}^{\text{FIX}}) + K_x(\mathcal{M}^{\text{SHELL}})$;

2. the intersection of a leaf of the foliation $K(\mathcal{M}^{\text{SHELL}})$ with $\mathcal{M}^{\text{FIX}}$ is not void and is connected;

We define $\delta^{\text{FIX}} : \mathcal{M}^{\text{FIX}} \to \mathcal{M}^{\text{DYN}}$ by $\delta^{\text{FIX}} := \delta|_{\mathcal{M}^{\text{FIX}}}$. Then, $(\mathcal{M}^{\text{DYN}}, \mathcal{M}^{\text{FIX}}, \delta^{\text{FIX}})$ is a phase space reduction of $\mathcal{M}^{\text{KIN}}$.

Moreover, if $f \in C^\infty(\mathcal{M}^{\text{KIN}}, \mathbb{R}) \cap B(\mathcal{M}^{\text{KIN}})$ and $\forall x \in \mathcal{M}^{\text{SHELL}}$, $X_{f,x} \in T_x(\mathcal{M}^{\text{SHELL}})$, we have $f^{\text{DYN}} = f^{\text{FIX}}$ where:

$$\forall y \in \mathcal{M}^{\text{DYN}}, \, f^{\text{DYN}}(y) := \sup\left\{ f(x) \mid x \in \delta^{-1}\langle y \rangle \right\} \quad \text{(def. A.2)}$$

and $f^{\text{FIX}}(y) := \sup\left\{ f(x) \mid x \in \delta^{\text{FIX},-1}\langle y \rangle \right\}$.

**Proof** *Statements A.1.1 & A.1.2.* $\mathcal{M}^{\text{FIX}}$ is a submanifold of $\mathcal{M}^{\text{SHELL}}$, and $\mathcal{M}^{\text{SHELL}}$ is a submanifold of $\mathcal{M}^{\text{KIN}}$, hence $\mathcal{M}^{\text{FIX}}$ is a submanifold of $\mathcal{M}^{\text{KIN}}$. $\mathcal{M}^{\text{DYN}}$ is a symplectic manifold.

The level sets of $\delta^{\text{FIX}}$ are the intersection with $\mathcal{M}^{\text{FIX}}$ of the leaves of the foliation $K(\mathcal{M}^{\text{SHELL}})$ (using def. A.3.4), hence from assumption A.8.2, $\delta^{\text{FIX}}$ is surjective and its level sets are connected.

*Statement A.1.3.* Let $x \in \mathcal{M}^{\text{FIX}}$. We have $T_x\delta^{\text{FIX}} = T_x\delta|_{T_x(\mathcal{M}^{\text{FIX}})}$, hence:

$$T_x\delta^{\text{FIX}} \langle T_x(\mathcal{M}^{\text{FIX}}) \rangle = T_x\delta \langle T_x(\mathcal{M}^{\text{FIX}}) \rangle = T_x\delta \langle T_x(\mathcal{M}^{\text{FIX}}) + \text{Ker } T_x\delta \rangle$$

$$= T_x \delta \langle T_x(\mathfrak{M}^{\text{FIX}}) + K_x(\mathfrak{M}^{\text{SHELL}}) \rangle = T_x \delta \langle T_x(\mathfrak{M}^{\text{SHELL}}) \rangle = T_{\delta^{\text{FIX}}(x)}(\mathfrak{M}^{\text{DYN}}) \text{ (using assumption A.8.1, eq. (A.4.1) and def. A.1.3 for the phase space reduction } (\mathfrak{M}^{\text{DYN}}, \mathfrak{M}^{\text{SHELL}}, \delta)).$$

Next, we have:

$$\Omega_{\text{KIN},x}|_{T_x(\mathfrak{M}^{\text{FIX}})} = [\delta^* \Omega_{\text{DYN}}]_x|_{T_x(\mathfrak{M}^{\text{FIX}})} \text{ (using def. A.1.3 for the phase space reduction } (\mathfrak{M}^{\text{DYN}}, \mathfrak{M}^{\text{SHELL}}, \delta) \text{ and } T_x(\mathfrak{M}^{\text{FIX}}) \subset T_x(\mathfrak{M}^{\text{SHELL}}))$$

$$= \left[\left(\delta|_{\mathfrak{M}^{\text{FIX}}}\right)^* \Omega_{\text{DYN}}\right]_x$$

$$= [\delta^{\text{FIX},*} \Omega_{\text{DYN}}]_x.$$

*Observables.* Let $f \in C^\infty(\mathfrak{M}^{\text{KIN}}, \mathbb{R}) \cap B(\mathfrak{M}^{\text{KIN}})$ with $\forall x \in \mathfrak{M}^{\text{SHELL}}, X_{f,x} \in T_x(\mathfrak{M}^{\text{SHELL}})$. From the proof of prop. A.7, $f$ is then constant on the leaves of the foliation $K(\mathfrak{M}^{\text{SHELL}})$ on $\mathfrak{M}^{\text{SHELL}}$. Therefore, $f^{\text{DYN}} = f^{\text{FIX}}$.
□

# B References